\documentclass[preprint,superscriptaddress,amsmath,amssymb,aps,prd,nofootinbib]{revtex4}
\usepackage{epsfig,graphicx,color,appendix}
\begin{document}
\begin{flushright}
 KIAS-P13021
\end{flushright}
\def\CP{{\it CP}~}
\def\C{{\it C}~}

\title{\mbox{}\\[10pt]
 Spontaneous CP Violation in $A_4$ Flavor Symmetry\\
 and Leptogenesis}

\author{Y. H. Ahn}\email{yhahn@kias.re.kr}
\affiliation{School of Physics, KIAS, Seoul 130-722, Korea}
\author{Sin Kyu Kang} \email{skkang@seoultech.ac.kr}
\affiliation{
School of Liberal Arts, Seoul-Tech, Seoul 139-743, Korea}
\affiliation{
PITT-PACC, Department of Physics and Astronomy, University of Pittsburgh, Pittsburgh, PA 15260, USA}
\author{C. S. Kim} \email{cskim@yonsei.ac.kr}
\affiliation{ Department of Physics \& IPAP, Yonsei University, Seoul, 120-479, Korea}
\begin{abstract}
\noindent
We propose a simple renormalizable model for the spontaneous CP violation based on $SU(2)_{L}\times U(1)_{Y}\times A_{4}$ symmetry in a radiative seesaw mechanism, which can be guaranteed by an extra $Z_{2}$ symmetry.
In our model CP is spontaneously broken at high energies, after breaking of flavor symmetry, by a complex vacuum expectation value of $A_{4}$-triplet and gauge singlet scalar field.
We show that the spontaneously generated CP phase  could become a natural source of leptogenesis,
and also investigate CP violation at low energies in the lepton sector and show how the CP phases in PMNS could be arisen through spontaneous symmetry breaking mechanism.

As a numerical study, interestingly, we show that the normal mass hierarchy favors relatively large values of $\theta_{13}$, large deviations from maximality of $\theta_{23}<\pi/4$ and Dirac-CP phase $0^{\circ}\leq\delta_{CP}\leq50^{\circ}$ and $300^{\circ}\leq\delta_{CP}\leq360^{\circ}$. For the inverted hierarchy case,
 the experimentally  measured values of $\theta_{13}$  favors $\theta_{23}>\pi/4$ and discrete values of $\delta_{CP}$ around $100^{\circ},135^{\circ},255^{\circ}$ and $300^{\circ}$. Finally, with a successful leptogenesis our numerical results give more predictive values on the Dirac CP phase: for the normal mass hierarchy $1^{\circ}\lesssim\delta_{CP}\lesssim10^{\circ}$ and for inverted one $\delta_{CP}\sim100^{\circ},135^{\circ},300^{\circ}$.
\end{abstract}

\maketitle %
\section{Introduction}

CP violation plays a crucial role in our understanding of the observed baryon asymmetry of the Universe (BAU)~\cite{Farrar:1993hn}.
This is because the preponderance of matter over antimatter in the observed Universe cannot be generated from an equal amounts of matter and antimatter unless CP is broken as  shown by Sakharov (1967), who pointed out that in addition to CP violation baryon-number violation, C (charge-conjugation) violation, and a departure from thermal equilibrium are all necessary to successfully achieve a net baryon asymmetry in early Universe.
In the Standard Model (SM)
CP symmetry is violated due to a complex phase in the Cabibbo-Kobayashi-Maskawa (CKM) matrix \cite{CKM}. However, since the extent of CP violation in the SM is not enough for achieving the observed BAU,  we need new source of CP violation for successful BAU.
On the other hand, CP violations in the lepton sector are imperative if the BAU could be realized through leptogenesis.
So, any hint or observation of the leptonic CP violation  can strengthen our belief in leptogenesis.

The violation of the CP symmetry is a crucial ingredient of any dynamical mechanism which intends to explain both low energy CP violation and the baryon asymmetry.
Renormalizable gauge theories are based on the spontaneous symmetry breaking mechanism, and it is natural to have the spontaneous CP violation (SCPV)~\cite{Lee:1973iz, Branco:1999fs}
 as an integral part of that mechanism. Determining all
possible sources of CP violation is a fundamental challenge for high energy physics. In theoretical and economical viewpoints, the spontaneous CP breaking necessary to generate the baryon asymmetry and leptonic CP violation at low energies brings us to a common source which comes from the phase of the scalar field responsible for the spontaneous CP breaking at a high energy scale.

Under $SU(2)\times U(1)$, we propose a simple renormalizable model for the SCPV based on $A_{4}$ flavor symmetry\footnote{E. Ma and G. Rajasekaran~\cite{Ma:2001dn} have introduced for the first time the $A_{4}$ symmetry to avoid the mass degeneracy of $\mu$ and $\tau$ under a $\mu$--$\tau$ symmetry~\cite{mutau}.} in a radiative seesaw mechanism~\cite{loopnu}, which can be guaranteed by an auxiliary $Z_{2}$ symmetry.
The main theoretical challenge for our work is three fold: First, we investigate CP violation in the lepton sector and show how CP phases in Pontecorvo-Maki-Nakagawa-Sakata (PMNS)~\cite{PDG} can be brought in through spontaneous symmetry breaking mechanism. Second, we show that the model we propose can provide a nice explanation to the smallness of neutrino masses and to the mild hierarchy of neutrino masses.
Third, we discuss how to link between leptonic mixing and leptogenesis through the SCPV. CP symmetry is spontaneously broken at high energies, after breaking of $A_{4}$ flavor symmetry, by a complex vacuum expectation value (VEV) of $A_{4}$-triplet and gauge singlet scalar filed $\chi$, which is introduced by heavy neutrino.
The spontaneously generated CP phase could become a natural source of leptogenesis, and
bring into low energy CP violation as well in the lepton sector through the CP phases in PMNS matrix.
Due to the auxiliary $Z_{2}$ symmetry, there are three implications: (i) The usual seesaw mechanism does not operate any more, and thus light neutrino masses cannot be generated at tree level and can be generated through one loop diagram, thanks to the quartic scalar interactions. (ii) The vacuum alignment problem\footnote{Such stability problems can be naturally solved, for instance, in the presence of extra dimensions or in supersymmetric dynamical completions~\cite{A4, vacuum}.}, which arises in the presence of two $A_{4}$-triplet scalar fields, can be naturally solved by putting the neutral Higgs VEVs to be zero. And, (iii) there can be a natural dark matter candidate which is the $Z_{2}$-odd neutral components of scalar field.

The work we propose is different from the previous works~\cite{Ma:2001dn, Ahn:2012tv, Ahn:2011yj, A4, vacuum} in using $A_{4}$ flavor symmetry, where (i) the $A_{4}$ flavor symmetry is spontaneously broken, and thereby a CP breaking phase is generated spontaneously, and (ii) the neutrino Yukawa coupling constants do not have all the same magnitude. Our model can naturally explain the measured value of $\theta_{13}$ and thereby mild hierarchy of neutrino masses, and can also provide a possibility for low energy CP violation in neutrino oscillations with a renormalizable Lagrangian. The seesaw mechanism, besides explaining of smallness of the measured neutrino masses,  has another appealing feature: generating the observed baryon asymmetry in our Universe by means of  leptogenesis~\cite{review}. Since the conventional $A_{4}$ models realized with type-I~\cite{type1_seesaw} or -III seesaw~\cite{type3_seesaw} and a tree-level Lagrangian lead to an exact tri-bi-maximal (TBM) and vanishing leptonic CP-asymmetries responsible for leptogenesis (due to the proportionality  of the $Y^{\dag}_{\nu}Y_{\nu}$ combination of the Dirac neutrino Yukawa matrix $Y_{\nu}$ to the unit matrix), physicists  usually introduce soft-breaking terms or higher-dimensional operators with many parameters, in order to explain the non-zero $\theta_{13}$ as well as the non-vanishing CP-asymmetries.

Our model is based on a renormalizable $SU(2)_L\times U(1)_Y\times A_{4}\times Z_{2}$ Lagrangian in a radiative seesaw framework with minimal Yukawa couplings, and gives rise to a non-degenerate Dirac neutrino Yukawa matrix and a unique CP-phase which arises spontaneously. This opens the possibility of explaining the non-zero value of $\theta_{13}\simeq9^{\circ}$  and large deviations from maximality of atmospheric mixing angle $\theta_{23}$, still maintaining large neutrino mixing angle $\theta_{12}$; furthermore, this allows an economic and dynamical way to achieve low energy CP violation in neutrino oscillations as well as high energy CP violation for leptogenesis. In addition, auxiliary symmetry guarantees the smallness of neutrino masses and a dark matter candidate, and after the breaking of the $A_{4}$ flavor symmetry makes their connection under the $A_4$ symmetry flavored.

This paper is organized as follows. In the next section, we lay down the particle content
and the field representations under the $A_4$  flavor symmetry with an auxiliary $Z_{2}$ symmetry in our model, as well as construct Higgs scalar and Yukawa Lagrangian. In Sec.~III, we discuss how to realize the spontaneous breaking of CP symmetry.
In Sec.~IV, we consider the phenomenology of neutrino at low energy, and in Sec.~V we study numerical analysis. In Sec. VI we show possible leptogenesis and its link with low energy observables
We give our conclusions in Sec.~VII, and in Appendix A we outline the minimization of the scalar potential and the vacuum alignments.

\section{The model}

Gauge invariance does not restrict the flavor structure of Yukawa interactions. As a result, particle masses and mixings are generally undetermined and arbitrary in a gauge theory. We extend the SM by introducing  a right-handed Majorana neutrinos $N_R$ which are $A_4$ triplet and $SU(2)_L$ singlet
and two kinds of extra scalar fields, $SU(2)_L$ doublet scalars $\eta$ and
a $SU(2)_L$ singlet scalar $\chi$, which are $A_4$ triplets.
Note that $\eta$ is distinguished from the SM Higgs doublet $\Phi$ because $\Phi$ is $A_4$ singlet.
Thus, the scalar fields in this model can be presented as follows;
 \begin{eqnarray}
  \Phi =
\begin{pmatrix} \varphi^{+} \\ \varphi^{0} \end{pmatrix},
\qquad
\eta_j =
\begin{pmatrix}
  \eta^{+}_j \\
  \eta^{0}_j
\end{pmatrix} ,
\qquad
\chi_j,
\qquad
j=1,2,3.
  \label{Higgs}
 \end{eqnarray}
We impose $A_{4}$ flavor symmetry for leptons and scalars, and force CP to be invariant at the Lagrangian level which implies that all the parameters appearing in the Lagrangian are real. So, the extended Higgs sector can spontaneously break CP through a phase in the VEV of the singlet scalar field.
In addition to $A_4$ symmetry, we introduce an extra auxiliary $Z_{2}$ symmetry so that: (i) a light neutrino mass can be generated via one loop diagram, (ii) vacuum alignment problem which occurs in the presence of two $A_{4}$-triplet can be naturally solved, and (iii) there can be a good dark matter candidate.

The representations of the field content of the model under $SU(2)\times U(1)\times A_{4}\times Z_2$ are summarized in Table \ref{reps} :
\begin{widetext}
\begin{center}
\begin{table}[h]
\caption{\label{reps} Representations of the fields under $A_{4}\times Z_{2}$ and $SU(2)_{L}\times U(1)_{Y}$.}
\begin{ruledtabular}
\begin{tabular}{ccccccccccccc}
Field &$L_{e},L_{\mu},L_{\tau}$&$e_R,\mu_R,\tau_R$&$N_{R}$&$\chi$&$\Phi$&$\eta$\\
\hline
$A_4$&$\mathbf{1}$, $\mathbf{1^\prime}$, $\mathbf{1^{\prime\prime}}$&$\mathbf{1}$, $\mathbf{1^\prime}$, $\mathbf{1^{\prime\prime}}$&$\mathbf{3}$&$\mathbf{3}$&$\mathbf{1}$&$\mathbf{3}$\\
$Z_2$&$+$&$+$&$-$&$+$&$+$&$-$\\
$SU(2)_L\times U(1)_Y$&$(2,-1)$&$(1,-2)$&$(1,0)$&$(1,0)$&$(2,1)$&$(2,1)$\\
\end{tabular}
\end{ruledtabular}
\end{table}
\end{center}
\end{widetext}
In particular, the CP invariance associated with $A_4$ triplet fields  can be clarified by the non-trivial transformation~\cite{Holthausen:2012dk}
\begin{align}
\psi\rightarrow U\psi^{\ast}=\psi~,
\end{align}
where $\psi=N_{R}, \chi, \eta$ and
\begin{align}
U={\left(\begin{array}{ccc}
 1 &  0 &  0 \\
 0 & 0 & 1 \\
 0 & 1 & 0
 \end{array}\right)}.
\end{align}
With the help of this CP transformation, one can easily show that the Lagrangian we introduced is CP invariant when all the couplings and mass parameters are real.

The most general renormalizable scalar potential for the Higgs fields $\Phi, \eta$ and $\chi$ invariant under $SU(2)_{L}\times U(1)_{Y}\times A_{4}\times Z_{2}$ is given as
 \begin{eqnarray}
  V=V(\eta)+V(\Phi)+V(\chi)+V(\eta\Phi)+V(\eta\chi)+V(\Phi\chi)~,
 \end{eqnarray}
where
 \begin{eqnarray}
V(\eta) &=& \mu^{2}_{\eta}(\eta^{\dag}\eta)_{\mathbf{1}}+\lambda^{\eta}_{1}(\eta^{\dag}\eta)_{\mathbf{1}}(\eta^{\dag}\eta)_{\mathbf{1}}+\lambda^{\eta}_{2}
            (\eta^{\dag}\eta)_{\mathbf{1^\prime}}(\eta^{\dag}\eta)_{\mathbf{1^{\prime\prime}}}+\lambda^{\eta}_{3}(\eta^{\dag}\eta)_{\mathbf{3}_{s}}(\eta^{\dag}\eta)_{\mathbf{3}_{s}}\nonumber\\
  &+&\lambda^{\eta}_{4}(\eta^{\dag}\eta)_{\mathbf{3}_{a}}(\eta^{\dag}\eta)_{\mathbf{3}_{a}}+\lambda^{\eta}_{5}\left\{(\eta^{\dag}\eta)_{\mathbf{3}_{s}}(\eta^{\dag}\eta)_{\mathbf{3}_{a}}+\text{h.c.}\right\}~,\\
V(\Phi) &=& \mu^{2}_{\Phi}(\Phi^{\dag}\Phi)+\lambda^{\Phi}(\Phi^{\dag}\Phi)^{2}~,
 \end{eqnarray}
 \begin{eqnarray}
V(\chi) &=& \mu^{2}_{\chi}\left\{(\chi\chi)_{\mathbf{1}}+(\chi^{\ast}\chi^{\ast})_{\mathbf{1}}\right\}+m^{2}_{\chi}(\chi\chi^{\ast})_{\mathbf{1}}+\lambda^{\chi}_{1}\left\{(\chi\chi)_{\mathbf{1}}(\chi\chi)_{\mathbf{1}}+(\chi^{\ast}\chi^{\ast})_{\mathbf{1}}(\chi^{\ast}\chi^{\ast})_{\mathbf{1}}\right\}\nonumber\\
&+&\lambda^{\chi}_{2}
            \left\{(\chi\chi)_{\mathbf{1}^\prime}(\chi\chi)_{\mathbf{1}^{\prime\prime}}+(\chi^{\ast}\chi^{\ast})_{\mathbf{1}^\prime}(\chi^{\ast}\chi^{\ast})_{\mathbf{1}^{\prime\prime}}\right\}\nonumber\\
&+&\tilde{\lambda}^{\chi}_{2} \left\{(\chi^{\ast}\chi)_{\mathbf{1}^\prime}(\chi\chi)_{\mathbf{1}^{\prime\prime}}+(\chi^{\ast}\chi)_{\mathbf{1}^{\prime\prime}}(\chi^{\ast}\chi^{\ast})_{\mathbf{1}^{\prime}}\right\}\nonumber\\
  &+&\lambda^{\chi}_{3}\left\{(\chi\chi)_{\mathbf{3}_{s}}(\chi\chi)_{\mathbf{3}_{s}}+(\chi^{\ast}\chi^{\ast})_{\mathbf{3}_{s}}(\chi^{\ast}\chi^{\ast})_{\mathbf{3}_{s}}\right\}+\tilde{\lambda}^{\chi}_{3}(\chi^{\ast}\chi)_{\mathbf{3}_{s}}\left\{(\chi\chi)_{\mathbf{3}_{s}}+(\chi^{\ast}\chi^{\ast})_{\mathbf{3}_{s}}\right\}\nonumber\\
  &+&\lambda^{\chi}_{4}\left\{(\chi^\ast\chi)_{\mathbf{3}_{a}}(\chi\chi)_{\mathbf{3}_{s}}+(\chi\chi^{\ast})_{\mathbf{3}_{a}}(\chi^{\ast}\chi^{\ast})_{\mathbf{3}_{s}}\right\}\nonumber\\
  &+&\xi^{\chi}_{1}\left\{\chi(\chi\chi)_{\mathbf{3}_{s}}+\chi^{\ast}(\chi^{\ast}\chi^{\ast})_{\mathbf{3}_{s}}\right\}+\tilde{\xi}^{\chi}_{1}\left\{\chi(\chi^{\ast}\chi^{\ast})_{\mathbf{3}_{s}}+\chi^{\ast}(\chi\chi)_{\mathbf{3}_{s}}\right\}~,
 \end{eqnarray}
 \begin{eqnarray}
V(\eta\Phi) &=& \lambda^{\eta\Phi}_{1}(\eta^{\dag}\eta)_{\mathbf{1}}(\Phi^{\dag}\Phi)
  +\lambda^{\eta\Phi}_{2}(\eta^{\dag}\Phi)(\Phi^{\dag}\eta)+\lambda^{\eta\Phi}_{3}\left\{(\eta^{\dag}\Phi)(\eta^{\dag}\Phi)+h.c\right\}\nonumber\\
  &+&\lambda^{\eta\Phi}_{4}\left\{(\eta^{\dag}\eta)_{\mathbf{3}_{s}}(\eta^{\dag}\Phi)+\text{h.c.}\right\}+\lambda^{\eta\Phi}_{5}\left\{(\eta^{\dag}\eta)_{\mathbf{3}_{a}}(\eta^{\dag}\Phi)+\text{h.c.}\right\}~,\\
V(\Phi\chi) &=& \lambda^{\Phi\chi}(\Phi^{\dag}\Phi)\left\{(\chi\chi)_{\mathbf{1}}+(\chi^{\ast}\chi^{\ast})_{\mathbf{1}}\right\}~,\\
V(\eta\chi) &=& \lambda^{\eta\chi}_{1}(\eta^{\dag}\eta)_{\mathbf{1}}\left\{(\chi\chi)_{\mathbf{1}}+(\chi^{\ast}\chi^{\ast})_{\mathbf{1}}\right\}+\lambda^{\eta\chi}_{2}\left\{(\eta^{\dag}\eta)_{\mathbf{1}^{\prime}}(\chi\chi)_{\mathbf{1}^{\prime\prime}}
  +(\eta^{\dag}\eta)_{\mathbf{1}^{\prime\prime}}(\chi^{\ast}\chi^{\ast})_{\mathbf{1}^{\prime}}\right\}\nonumber\\
  &+&\lambda^{\eta\chi}_{3}(\eta^{\dag}\eta)_{\mathbf{3}_{s}}\left\{(\chi\chi)_{\mathbf{3}_{s}}+(\chi^{\ast}\chi^{\ast})_{\mathbf{3}_{s}}\right\}+\lambda^{\eta\chi}_{4}\left\{(\eta^{\dag}\eta)_{\mathbf{3}_{a}}(\chi\chi)_{\mathbf{3}_{s}}+h.c.\right\}\nonumber\\
  &+&\xi^{\eta\chi}_{1}(\eta^{\dag}\eta)_{\mathbf{3}_{s}}\left\{\chi+\chi^{\ast}\right\}+\xi^{\eta\chi}_{2}\left\{(\eta^{\dag}\eta)_{\mathbf{3}_{a}}\chi+h.c.\right\}~.
\label{potential}
\end{eqnarray}
Here, $\mu_{\eta},\mu_{\Phi},\mu_{\chi}$, $m_{\chi}$, $\xi^{\chi}_{1}$, $\tilde{\xi}^{\chi}_{1}$, $\xi^{\eta\chi}_{1}$ and $\xi^{\eta\chi}_{2}$ have a mass dimension,
whereas $\lambda^{\eta}_{1,...,5}$, $\lambda^{\Phi}$, $\lambda^{\chi}_{1,...,4}$, $\tilde{\lambda}^{\chi}_{2,3}$, $\lambda^{\eta\Phi}_{1,...,5}$, $\lambda^{\eta\chi}_{1,...,4}$ and $\lambda^{\Phi\chi}$ are all dimensionless.
In $V(\eta\Phi)$, the usual mixing term $\Phi^{\dag}\eta$ and $\Phi^{\dag}\eta\chi$ are forbidden by the $A_{4}\times Z_{2}$ symmetry.

With the field content and the symmetries specified in Table~\ref{reps}, the relevant renormalizable Lagrangian for the neutrino and charged lepton sectors
invariant under $SU(2)\times U(1)\times A_{4}\times Z_2$
is given by
 \begin{eqnarray}
 -{\cal L}_{\rm Yuk} &=& y^{\nu}_{1}\bar{L}_{e}(\tilde{\eta}N_{R})_{{\bf 1}}+y^{\nu}_{2}\bar{L}_{\mu}(\tilde{\eta}N_{R})_{{\bf 1}'}+y^{\nu}_{3}\bar{L}_{\tau}(\tilde{\eta}N_{R})_{{\bf 1}''}\nonumber\\
 &+&\frac{M}{2}(\overline{N^{c}_{R}}N_{R})_{{\bf 1}}+\frac{\lambda_{\chi}}{2}(\overline{N^{c}_{R}}N_{R})_{{\bf 3}_{s}} \chi\nonumber\\
 &+& y_{e}\bar{L}_{e}\Phi~e_{R}+y_{\mu}\bar{L}_{\mu}\Phi~\mu_{R}+y_{\tau}\bar{L}_{\tau}\Phi~ \tau_{R}
 +h.c~,
 \label{lagrangian}
 \end{eqnarray}
where $\tilde{\eta}\equiv i\tau_{2}\eta^{\ast}$ with the Pauli matrix $\tau_{2}$.
Here, $L_{e,\nu,\tau}$ and $e_{R},\mu_{R},\tau_{R}$ denote left handed lepton $SU(2)_L$ doublets and right handed lepton $SU(2)_L$ singlets, respectively.
In the above Lagrangian,
mass terms of the charged leptons are given by the diagonal form  because the Higgs scalar $\Phi$ and the charged lepton fields are assigned to be $A_4$ singlet.
The heavy neutrinos $N_{Ri}$ acquire a bare mass $M$ as well as a mass induced by a vacuum of electroweak singlet scalar  $\chi$ assigned to be $A_4$ triplet.
While the standard Higgs scalar $\Phi^0$ gets a VEV $v=(2\sqrt{2}G_{F})^{-1/2}=174$ GeV, the neutral component of scalar doublet $\eta$ would not acquire a nontrivial VEV
because $\eta$ has odd parity of $Z_2$ as assigned in Table~\ref{reps}
and the auxiliary $Z_{2}$ symmetry is  exactly conserved even after electroweak symmetry breaking;
 \begin{eqnarray}
  \langle\eta^{0}_{i}\rangle=0~,~(i=1,2,3)~,\qquad\langle\Phi^{0}\rangle=\upsilon_{\Phi}\neq0~.
 \label{vev1}
 \end{eqnarray}
Therefore, the neutral component of scalar doublet $\eta$ can be a good dark matter candidate, and the usual seesaw mechanism does not operate because the neutrino Yukawa interactions cannot generate masses. However, the light Majorana neutrino mass matrix can be generated radiatively through one-loop with the help of the Yukawa interaction $\bar{L}_{L}N_{R}\tilde{\eta}$ in the Lagrangian,
 which will be discussed more in detail in Sec. III.
Even though there exist interaction terms of the two $A_{4}$-triplet Higgs scalars $\chi,\eta$ in Higgs potential, there are no conflicts in vacuum stability because the $\eta$ fields do not have VEV.
In our model, the $A_4$ flavor symmetry is spontaneously broken by $A_4$ triplet scalars $\chi$, and thereby a CP breaking phase is generated spontaneously.
From the condition of the global minima of the scalar potential, we can obtain a vacuum alignment of the fields $\chi$.

\section{Spontaneous CP violation}

While CP symmetry is conserved at the Lagrangian level because all the parameters are assumed to be real, in our model  it can be spontaneously broken when the scalar singlet $\chi$ acquires a complex VEV.
Now let us discuss how to realize the spontaneous breaking of CP symmetry.

\subsection{Minimization of the neutral scalar potential}

After the breaking of the flavor and electroweak symmetry, we can find minimum configuration of the Higgs potential by taking as follow;
 \begin{eqnarray}
  \langle\Phi\rangle &=&
  {\left(\begin{array}{c}
  0 \\
  v_{\Phi}e^{i\theta}
 \end{array}\right)}~,\quad
 \langle\eta_{j}\rangle =0~,\quad
 \langle\chi_{1}\rangle= v_{\chi_{1}}e^{i\phi_{1}}~,\quad\langle\chi_{2}\rangle=v_{\chi_{2}}e^{i\phi_{2}}~,\quad\langle\chi_{3}\rangle=v_{\chi_{3}}e^{i\phi_{3}}~,
  \label{neuvevs}
 \end{eqnarray}
with $j=1-3$, where $v,v_{\chi_{1,2,3}}$ are real and positive, and $\phi_{1,2,3}$ are physically meaningful phases.
Since $\theta$ is not physical observable, we can set $\theta=0$ without loss of generality.
Then, we get seven minimization conditions for four VEVs and three phases.
By requiring that   the derivatives of $V$ with respect to each component of the scalar fields $\Phi$, $\chi_{i}$ and $\phi_{i}$ are vanished at $\langle\eta_{i}\rangle=0~(i=1,2,3)$
we can obtain the vacuum configurations as follows:
 \begin{eqnarray}
  \upsilon^{2}_{\chi_{i}}&=&-\frac{m^{2}_{\chi}+2(\mu^{2}_{\chi}+v^{2}_{\Phi}\lambda^{\Phi\chi})\cos2\phi_{i}}
  {4(\tilde{\lambda}^{\chi}_{2}\cos2\phi_{i}+(\lambda^{\chi}_{1}+\lambda^{\chi}_{2})\cos4\phi_{i})}\neq0~,\qquad\langle\chi_{j}\rangle=\langle\chi_{k}\rangle=0~, \label{vevchi}\\
  v^{2}_{\Phi}&=&\frac{-\mu^{2}_{\Phi}-2v^{2}_{\chi_i}\lambda^{\Phi\chi}\cos2\phi_{i}}{2\lambda^{\Phi}}~~~\text{for}~\langle\chi\rangle=v_{\chi_{i}}e^{i\phi_{i}}a_i~\nonumber
  \label{vevPhi1}
 \end{eqnarray}
where $i,j,k=1,2,3~(i\neq j\neq k)$,
$a_1=(1,0,0)$, $a_2=(0,1,0)$, $a_3=(0,0,1)$ and $\upsilon_{\chi_{i}}$ is real.
With the vacuum alignment of $\chi$ fields, Eq.~(\ref{vevchi}), minimal condition with respect to $\phi_{i}$ is given as
 \begin{eqnarray}
  -\frac{1}{4}\frac{\partial V}{\partial \phi_{i}}\Big|&=&  v^{2}_{\chi}\left\{v^{2}_{\Phi}\lambda^{\Phi\chi}+\mu^{2}_{\chi}+v^{2}_{\chi_i}\left(\tilde{\lambda}^{\chi}_{2}+4(\lambda^{\chi}_{1}+\lambda^{\chi}_{2})\cos2\phi_{i}\right)\right\}\sin2\phi_{i}=0~,
 \end{eqnarray}
and $\frac{\partial V}{\partial \phi_{j}}\Big|= \frac{\partial V}{\partial \phi_{k}}\Big|=0$ is automatically satisfied with $i,j,k=1,2,3$ $(i\neq j\neq k)$.

If we consider, as an example, the vacuum alignment $\langle\chi\rangle=v_{\chi}e^{i\phi}(1,0,0)$ and $\langle\Phi\rangle=v_{\Phi}$ where $v_{\chi}\equiv v_{\chi_{1}}$ and $\phi\equiv \phi_{1}$, the scalar potential can be written as\footnote{If we assume the $\chi$ VEV is very heavy and decouples from the theory at an energy scale much higher than electroweak scale, the scalar potential is roughly given as $V_{0}\simeq m^{2}_{\chi}v^{2}_{\chi}+2v^{2}_{\chi}(\mu^{2}_{\chi}+\tilde{\lambda}^{\chi}_{2}v^{2}_{\chi})\cos2\phi+2(\lambda^{\chi}_{1}+\lambda^{\chi}_{2})v^{4}_{\chi}\cos4\phi$.}
 \begin{eqnarray}
  V_{0}=v^{4}_{\Phi}\lambda^{\Phi}+v^{2}_{\Phi}\mu^{2}_{\Phi}+m^{2}_{\chi}v^{2}_{\chi}+2v^{2}_{\chi}(v^{2}_{\Phi}\lambda^{\Phi\chi}+\mu^{2}_{\chi}+\tilde{\lambda}^{\chi}_{2}v^{2}_{\chi})\cos2\phi+2(\lambda^{\chi}_{1}+\lambda^{\chi}_{2})v^{4}_{\chi}\cos4\phi~.
 \end{eqnarray}
In our scenario, we assume that $v_{\chi}$ is larger than $v_{\Phi}$.
Depending on the values of $\phi$, the vacuum configurations are given by:\\
{\bf (i)} for $\phi=0, \pm\pi$
 \begin{eqnarray}
  \upsilon^{2}_{\chi}=-\frac{m^{2}_{\chi}+2(\mu^{2}_{\chi}+v^{2}_{\Phi}\lambda^{\Phi\chi})}
  {4(\tilde{\lambda}^{\chi}_{2}+\lambda^{\chi}_{1}+\lambda^{\chi}_{2})}~,\qquad v^{2}_{\Phi}=\frac{-\mu^{2}_{\Phi}-2v^{2}_{\chi}\lambda^{\Phi\chi}}{2\lambda^{\Phi}}~,
  \label{VP1}
 \end{eqnarray}
{\bf (ii)} for $\phi=\pm\pi/2$
 \begin{eqnarray}
  \upsilon^{2}_{\chi}=\frac{m^{2}_{\chi}-2(\mu^{2}_{\chi}+v^{2}_{\Phi}\lambda^{\Phi\chi})}
  {4(\tilde{\lambda}^{\chi}_{2}-\lambda^{\chi}_{1}-\lambda^{\chi}_{2})}~,\qquad v^{2}_{\Phi}=\frac{-\mu^{2}_{\Phi}+2v^{2}_{\chi}\lambda^{\Phi\chi}}{2\lambda^{\Phi}}~,
  \label{VP2}
 \end{eqnarray}
{\bf (iii)} for $\cos2\phi=-\frac{v^{2}_{\Phi}\lambda^{\Phi\chi}+\mu^{2}_{\chi}+v^{2}_{\chi}\tilde{\lambda}^{\chi}_{2}}{4v^{2}_{\chi}(\lambda^{\chi}_{1}+\lambda^{\chi}_{2})}$
 \begin{eqnarray}
  \upsilon^{2}_{\chi}=\frac{2m^{2}_{\chi}(\lambda^{\chi}_{1}+\lambda^{\chi}_{2})-\tilde{\lambda}^{\chi}_{2}(v^{2}_{\Phi}\lambda^{\Phi\chi}+\mu^{2}_{\chi})}{\tilde{\lambda}^{\chi2}_{2}+8(\lambda^{\chi}_{1}+\lambda^{\chi}_{2})^{2}}~,\quad v^{2}_{\Phi}=\frac{(\mu^{2}_{\chi}+\tilde{\lambda}^{\chi}_{2}v^{2}_{\chi})\lambda^{\Phi\chi}-2\mu^{2}_{\Phi}(\lambda^{\chi}_{1}+\lambda^{\chi}_{2})}{4\lambda^{\Phi}(\lambda^{\chi}_{1}+\lambda^{\chi}_{2})-\lambda^{\Phi\chi2}}~.
  \label{VP3}
 \end{eqnarray}
In the first case (i) the vacuum configurations do not violate CP, while the second (ii) and third case (iii) lead not only to the the spontaneous breaking of the CP symmetry but also to a non-trivial CP violating phase in the one loop diagrams relevant for leptogenesis.

Let us examine which case corresponds to the global minimum of the potential in a wide region of the parameter space.
Imposing the parameter conditions, $m^{2}_{\chi}<0, \mu^{2}_{\Phi}<0, \lambda^{\Phi}>0$ and $\lambda^{\chi}_{1,2}<0$, into Eqs.~(\ref{VP1}-\ref{VP3}),
the vacuum configurations of each case become
we obtain for the case (i)
 \begin{eqnarray}
  V_{0}=-\lambda^{\Phi}v^{4}_{\Phi}-\frac{(m^{2}_{\chi}+2\mu^{2}_{\chi})^{2}-4v^{4}_{\Phi}\lambda^{\Phi\chi2}}{8(\lambda^{\chi}_{1}+\lambda^{\chi}_{2}+\tilde{\lambda}^{\chi}_{2})}~,\qquad \phi=0,\pm\pi~,
 \end{eqnarray}
for the case (ii)
 \begin{eqnarray}
  V_{0}=-\lambda^{\Phi}v^{4}_{\Phi}-\frac{(m^{2}_{\chi}-2\mu^{2}_{\chi})^{2}-4v^{4}_{\Phi}\lambda^{\Phi\chi2}}{8(\lambda^{\chi}_{1}+\lambda^{\chi}_{2}-\tilde{\lambda}^{\chi}_{2})}~,\qquad \phi=\pm\frac{\pi}{2}~,
 \end{eqnarray}
for the case (iii), we obtain
 \begin{eqnarray}
  v^{2}_{\chi}=\frac{m^{2}_{\chi}}{4(\lambda^{\chi}_{1}+\lambda^{\chi}_{2})}~,\qquad v^{2}_{\Phi}=-\frac{\mu^{2}_{\Phi}}{2\lambda^{\Phi}}~,\qquad {\rm for}~\phi=\pm\frac{\pi}{4}~,
 \end{eqnarray}
 leading to
 \begin{eqnarray}
  V_{0}=\frac{m^{4}_{\chi}}{8(\lambda^{\chi}_{1}+\lambda^{\chi}_{2})}-\frac{\mu^{4}_{\Phi}}{4\lambda^{\Phi}}~.
 \label{mini}
 \end{eqnarray}
The third case corresponds to the absolute minimum of the potential.
As shown in Appendix, it is also guaranteed that we are at a minimum by showing the eigenvalues of the neutral Higgs boson mass matrices and requiring that they are all positive.

\subsection{The lepton mass matrices and a CP phase}

After the scalar fields get VEVs, the Yukawa interactions in Eq.~(\ref{lagrangian}) and the charged gauge interactions in a weak eigenstate basis can be written as
 \begin{eqnarray}
 -{\cal L} &=& \frac{1}{2}\overline{N^{c}_{R}}M_{R}N_{R}+\overline{\ell_{L}}m_{\ell}\ell_{R}
 +\overline{\nu_{L}}Y_{\nu}\hat{\eta}N_{R}+\frac{g}{\sqrt{2}}W^{-}_{\mu}\overline{\ell_{L}}\gamma^{\mu}\nu_{L}+h.c~,
 \label{lagrangianA}
 \end{eqnarray}
where $\hat{\eta}={\rm Diag.}(\tilde{\eta}_{1},\tilde{\eta}_{2},\tilde{\eta}_{3})$.
In particular, thanks to the vacuum alignment given in Eqs.~(\ref{vevchi},\ref{vevPhi1}), $\langle\chi\rangle=v_{\chi}e^{i\phi}(1,0,0)$ and $\langle\Phi\rangle=v_{\Phi}$, the right-handed Majorana neutrino mass matrix and the charged lepton mass matrix are given by
 \begin{eqnarray}
 M_{R}={\left(\begin{array}{ccc}
 M &  0 &  0 \\
 0 &  M &  \lambda^{s}_{\chi}v_{\chi} e^{i\phi} \\
 0 &  \lambda^{s}_{\chi}v_{\chi} e^{i\phi} &  M
 \end{array}\right)}~,\qquad m_{\ell}=v_{\Phi}{\left(\begin{array}{ccc}
 y_{e} & 0 & 0 \\
 0 & y_{\mu} & 0 \\
 0 & 0 &  y_{\tau}
 \end{array}\right)}~.
 \label{MR}
 \end{eqnarray}
We note that the vacuum alignment given in Eq.~(\ref{vevchi}) implies that the $A_{4}$ symmetry is spontaneously broken to its residual symmetry $Z_{2}$
in the heavy neutrino sector since $(1,0,0)$ is invariant under the generator $S$ presented in Eq.~(\ref{generator}).
In addition, one can easily see that the neutrino Yukawa matrix is given as follows;
 \begin{eqnarray}
 Y_{\nu}=\sqrt{3}{\left(\begin{array}{ccc}
 y^{\nu}_{1} &  0 &  0 \\
 0 & y^{\nu}_{2} & 0 \\
 0 & 0 & y^{\nu}_{3}
 \end{array}\right)}U^{\dag}_{\omega}~,\qquad{\rm with}~~
 U_{\omega}=\frac{1}{\sqrt{3}}{\left(\begin{array}{ccc}
 1 &  1 &  1 \\
 1 &  \omega^{2} &  \omega \\
 1 &  \omega &  \omega^{2}
 \end{array}\right)}~.
 \label{yukawaNu}
 \end{eqnarray}
For our convenience, let us take the basis where heavy Majorana neutrino and charged lepton mass matrices are diagonal.
Rotating the basis with the help of a unitary matrix $U_{R}$,
 \begin{eqnarray}
 N_{R}\rightarrow U^{\dag}_{R}N_{R}~,
 \label{basis}
 \end{eqnarray}
the right-handed Majorana mass matrix $M_{R}$ becomes a diagonal matrix $\widehat{M}_{R}$ with real and positive mass eigenvalues $M_{1}=aM, M_{2}=M$ and $M_{3}=bM$,
\begin{eqnarray}
 \widehat{M}_{R}=U^{T}_{R}M_{R}U_{R}=MU^{T}_{R}{\left(\begin{array}{ccc}
 1 &  0 &  0 \\
 0 &  1 & \kappa e^{i\phi} \\
 0 &  \kappa e^{i\phi} &  1
 \end{array}\right)}U_{R}={\left(\begin{array}{ccc}
 aM & 0 & 0 \\
 0 & M & 0 \\
 0 & 0 & bM
 \end{array}\right)}~,
 \label{MR1}
 \end{eqnarray}
 where $\kappa=\lambda^{s}_{\chi}v_{\chi}/M$. We find $a=\sqrt{1+\kappa^{2}+2\kappa\cos\phi}$, $b=\sqrt{1+\kappa^{2}-2\kappa\cos\phi}$, and the diagonalizing matrix
\begin{eqnarray}
  U_{R} = \frac{1}{\sqrt{2}} {\left(\begin{array}{ccc}
  0  &  \sqrt{2}  &  0 \\
  1 &  0  &  -1 \\
  1 &  0  &  1
  \end{array}\right)}{\left(\begin{array}{ccc}
  e^{i\frac{\psi_1}{2}}  &  0  &  0 \\
  0  &  1  &  0 \\
  0  &  0  &  e^{i\frac{\psi_2}{2}}
  \end{array}\right)}~,
  \label{URN}
\end{eqnarray}
with the phases
\begin{eqnarray}
 \psi_1 = \tan^{-1} \Big( \frac{-\kappa\sin\phi}{1+\kappa\cos\phi} \Big)
 ~~~{\rm and}~~~ \psi_2 = \tan^{-1} \Big( \frac{\kappa\sin\phi}{1-\kappa\cos\phi} \Big)~.
\label{alphs_beta}
\end{eqnarray}
The phases $\psi_{1,2}$ go to $0$ or $\pi$ as the magnitude of $\kappa$ defined in Eq.~(\ref{MR1}) decreases.
Due to the rotation (\ref{basis}), the neutrino Yukawa matrix $Y_{\nu}$ gets modified to
 \begin{eqnarray}
 \tilde{Y}_{\nu} &=& Y_{\nu}U_{R}~, \nonumber \\
    &=& P_{\nu}^{\dag}~{\rm Diag.}(|y^{\nu}_{1}|,|y^{\nu}_{2}|,|y^{\nu}_{3}|)U^{\dag}_{\omega}U_{R}~.
 \label{YnuT}
 \end{eqnarray}

We perform basis rotations from weak to mass eigenstates in the leptonic sector,
 \begin{eqnarray}
 \ell_{L}\rightarrow P^{\ast}_{\nu}\ell_{L}~,\qquad \ell_{R}\rightarrow P^{\ast}_{\nu}\ell_{R}~,\qquad\nu_{L}\rightarrow U^{\dag}_{\nu}P^{\ast}_{\nu}\nu_{L}
 \label{basisR}
 \end{eqnarray}
where $P_{\ell}$ and $P_{\nu}$ are phase matrices and $U_{\nu}$ is a diagonalizing matrix of light neutrino mass matrix. Then, from the charged current term in Eq.~(\ref{lagrangianA}) we obtain the lepton mixing matrix $U_{\rm PMNS}$ as
 \begin{eqnarray}
 U_{\rm PMNS}=P^{\ast}_{\ell}P_{\nu}U_{\nu}~.
 \end{eqnarray}
The matrix $U_{\rm PMNS}$ can be written in terms of three mixing angles and three CP-odd phases (one for the Dirac neutrino and two for the Majorana neutrino) as follows \cite{PDG}
 \begin{eqnarray}
  U_{\rm PMNS}={\left(\begin{array}{ccc}
   c_{13}c_{12} & c_{13}s_{12} & s_{13}e^{-i\delta_{CP}} \\
   -c_{23}s_{12}-s_{23}c_{12}s_{13}e^{i\delta_{CP}} & c_{23}c_{12}-s_{23}s_{12}s_{13}e^{i\delta_{CP}} & s_{23}c_{13}  \\
   s_{23}s_{12}-c_{23}c_{12}s_{13}e^{i\delta_{CP}} & -s_{23}c_{12}-c_{23}s_{12}s_{13}e^{i\delta_{CP}} & c_{23}c_{13}
   \end{array}\right)}Q_{\nu}~,
 \label{PMNS}
 \end{eqnarray}
where $s_{ij}\equiv \sin\theta_{ij}$ and $c_{ij}\equiv \cos\theta_{ij}$, and $Q_{\nu}={\rm Diag.}(e^{-i\varphi_{1}/2},e^{-i\varphi_{2}/2},1)$.

It is important to notice that the phase matrix $P_{\nu}$ can be rotated away by choosing the matrix $P_{\ell}=P_\nu$, i.e.\ by an appropriate redefinition of the left-handed charged lepton fields,
which is always possible. Hence, we can take the eigenvalues $y^{\nu}_1$, $y^{\nu}_2$, and $y^{\nu}_3$ of $Y_{\nu}$ to be real and positive without  loss of generality.
The Yukawa matrix $Y_{\nu}$ can then be written as
 \begin{eqnarray}
 Y_{\nu}=y^{\nu}_{3}\sqrt{3}{\left(\begin{array}{ccc}
 y_{1} &  0 &  0 \\
 0 & y_{2} & 0 \\
 0 & 0 & 1
 \end{array}\right)}U^{\dag}_{\omega},
  \label{Ynu}
 \end{eqnarray}
where $y_{1}=|y^{\nu}_{1}/y^{\nu}_{3}|, y_{2}=|y^{\nu}_{2}/y^{\nu}_{3}|$, and $U_\omega$ is given in Eq.~(\ref{yukawaNu}).

Concerning CP violation, we notice that the CP phases $\psi_1,\psi_2$ in the scalar potential only take part in low-energy CP violation, as can be seen from Eqs.~(\ref{MR1}-\ref{Ynu}). The source of CP-violation relevant for leptogenesis originates from the neutrino Yukawa matrix $\widetilde{Y}_{\nu}=Y_{\nu}U_{R}$ and its combination, $H\equiv\widetilde{Y}^{\dag}_{\nu}\widetilde{Y}_{\nu}= U^{\dag}_{R}Y^{\dag}_{\nu}Y_{\nu}U_{R}$, which is
 \begin{eqnarray}
   H=3|y^{\nu}_{3}|^{2}\left(\begin{array}{ccc}
  \frac{1+4y^{2}_{1}+y^{2}_{2}}{2} & \frac{e^{-i\frac{\psi_{1}}{2}}}{\sqrt{2}}(2y^{2}_{1}-y^{2}_{2}-1) & \frac{i\sqrt{3}e^{i\frac{\psi_{21}}{2}}}{2}(y^{2}_{2}-1) \\
  \frac{e^{i\frac{\psi_{1}}{2}}}{\sqrt{2}}(2y^{2}_{1}-y^{2}_{2}-1) & 1+y^{2}_{1}+y^{2}_{2} & -i\sqrt{\frac{3}{2}}e^{i\frac{\psi_{2}}{2}}(y^{2}_{2}-1) \\
  -\frac{i\sqrt{3}e^{-i\frac{\psi_{21}}{2}}}{2}(y^{2}_{2}-1) & i\sqrt{\frac{3}{2}}e^{-i\frac{\psi_{2}}{2}}(y^{2}_{2}-1) & \frac{3}{2}(1+y^{2}_{2})
  \end{array}\right) ,
 \label{YnuYnu}
 \end{eqnarray}
where $\psi_{ij}\equiv\psi_{i}-\psi_{j}$.
 However, in the limit $|y^{\nu}_1|=|y^{\nu}_2|=|y^{\nu}_3|$ , i.e.\ $y_{1,2}\rightarrow1$, the off-diagonal entries of $H$  vanish, and thus leptogenesis can not be realized because of no CP violation.
 In our model, baryogenesis via leptogenesis, and non-zero $\theta_{13}\simeq9^{\circ}$ while keeping large mixing angles ($\theta_{23},\theta_{12}$)~\cite{Data} are achievable only when
 the neutrino Yukawa couplings $y^{\nu}_1$, $y^{\nu}_2$, and $y^{\nu}_3$ are non-degenerate.
  We see that all ${\rm Im}[H_{ij}]$ and $\tilde{Y}_{\nu}$ itself depend on the phases $\psi_{1,2}$ which are functions of $\phi$ and $\kappa$. Therefore, the origins of a low energy CP violation in neutrino oscillation and/or a high energy CP violation in leptogenesis are the non-degeneracy of the neutrino Yukawa couplings and a non-zero phase $\phi$ generated from spontaneous breaking of symmetry.

\section{Low energy neutrino mass matrix}

\begin{figure}[t]
\begin{center}
\includegraphics*[width=0.4\textwidth]{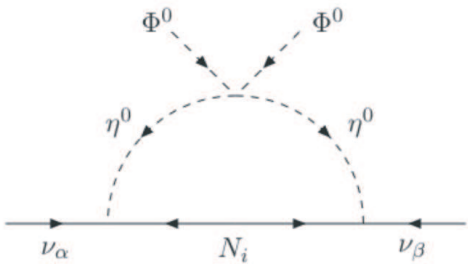}
\caption{\label{Fig1} One-loop generation of light neutrino masses.}
\end{center}
\end{figure}
In the present model, the light neutrino mass matrix can be generated through one loop diagram, shown in Fig.~\ref{Fig1}, which is similar to the scenario presented in ~\cite{loopnu, Ahn:2010cc}.
After electroweak symmetry breaking,
the light neutrino masses in the flavor basis, where the charged lepton mass matrix is real and diagonal, are written as
 \begin{eqnarray}
   (m_{\nu})_{\alpha\beta}&=&\sum_{i}\frac{\Delta m^{2}_{\eta_i}}{16\pi^{2}}\frac{(\tilde{Y}_{\nu})_{\alpha i}(\tilde{Y}_{\nu})_{\beta i}}{M_{i}}f\left(\frac{M^{2}_{i}}{\bar{m}^{2}_{\eta_i}}\right),~~{\rm for}~\Delta m^{2}_{\eta_{i}}\ll \bar{m}^{2}_{\eta_i}~,
   \label{lownu1}
 \end{eqnarray}
where
 \begin{eqnarray}
   f(z_{i})&=&\frac{z_{i}}{1-z_{i}}\left[1+\frac{z_{i}\ln z_{i}}{1-z_{i}}\right]~,\quad\Delta m^{2}_{\eta_{i}}\equiv|m^{2}_{R_i}-m^{2}_{I_i}|=4v^{2}\lambda^{\Phi\eta}_{3}~,
 \label{lambda}
 \end{eqnarray}
with $z_{i}=M^{2}_{i}/\bar{m}^{2}_{\eta_i}$ and $\bar{m}^{2}_{\eta_i}\equiv(m^{2}_{R_i}+m^{2}_{I_i})/2$. The explicit expressions for $\bar{m}^{2}_{\eta_i}$ are presented in the Appendix.
Here, $m_{R_i}(m_{I_i})$ is the mass of the field component $\eta^{0}_{R_i}(\eta^{0}_{I_i})$ and $m^{2}_{R_i(I_i)}=\bar{m}^{2}_{\eta_i}\pm\Delta m^{2}_{\eta_i}/2$ where the subscripts $R$ and $I$ indicate real and imaginary component, respectively. With $\tilde{M}_{R}={\rm Diag}(M_{r1},M_{r2},M_{r3})$ and $M_{ri}\equiv M_{i}f^{-1}(z_{i})$, the above formula Eq.~(\ref{lownu1}) can be expressed as
 \begin{eqnarray}
  m_{\nu} &=& \frac{v^2_{\Phi}\lambda^{\Phi\eta}_{3}}{4\pi^{2}}\tilde{Y}_{\nu}\tilde{M}^{-1}_{R}\tilde{Y}^{T}_{\nu}
  = U_{\rm PMNS}~{\rm Diag.}(m_{1},m_{2},m_{3}) U^{T}_{\rm PMNS} \nonumber\\
  &=& m_{0}{\left(\begin{array}{ccc}
  Ay^{2}_{1} & By_{1}y_{2} & By_{1} \\
  By_{1}y_{2} & Dy^{2}_{2} & Gy_{2}  \\
  By_{1} & Gy_{2} & D
 \end{array}\right)}~,
 \label{radseesaw1}
 \end{eqnarray}
where $m_{i}(i=1,2,3)$ are the light neutrino mass eigenvalues,  $y_{1(2)}=y^{\nu}_{1(2)}/y^{\nu}_3$, and
 \begin{eqnarray} A&=&f(z_{2})+\frac{2e^{i\psi_{1}}f(z_{1})}{a}~,~~\qquad \qquad \qquad \quad B=f(z_{2})-\frac{e^{i\psi_{1}}f(z_{1})}{a}~,\nonumber\\
  D&=&f(z_{2})+\frac{e^{i\psi_{1}}f(z_{1})}{2a}-\frac{3e^{i\psi_{2}}f(z_{3})}{2b}~,\qquad
  m_{0}= \frac{v^2_{\Phi}|y^{\nu}_{3}|^{2}\lambda^{\Phi\eta}_{3}}{4\pi^{2}M}~,\nonumber\\
  G&=&f(z_{2})+\frac{e^{i\psi_{1}}f(z_{1})}{2a}+\frac{3e^{i\psi_{2}}f(z_{3})}{2b}~.
 \label{entries}
 \end{eqnarray}
It is worthwhile to notice that in the case of $y_{2}=1$ the mass matrices given by Eq.~(\ref{radseesaw1}) get to have $\mu-\tau$ symmetry~\cite{mutau} leading to $\theta_{13}=0$ and $\theta_{23}=-\pi/4$. Moreover, in the case of $y_{1}, y_{2}=1$, the mass matrices give rise to TBM angles and masses, respectively,
 \begin{eqnarray}
  \theta_{13}&=&0~, \qquad\qquad\theta_{23}=-\frac{\pi}{4}~,\qquad\qquad\theta_{12}=\sin^{-1}\left(\frac{1}{\sqrt{3}}\right)~,\nonumber\\
  m_{1}&=&3m_{0}\frac{f(z_{1})}{a}e^{i\psi_{1}}~,\quad m_{2}=3m_{0}f(z_{2})~,\quad m_{3}=3m_{0}\frac{f(z_{3})}{b}e^{i(\psi_{2}+\pi)}~,
 \label{masseigen}
  \end{eqnarray}
indicating that neutrino masses are divorced from mixing angles.
However, in order to accommodate recent neutrino data including the observations of non-zero $\theta_{13}$, the parameters $y_{1,2}$ should be lifted from unit maintaining the Yukawa neutrino couplings being mild hierarchy\footnote{With the lift of $y_{1,2}$ from unit, the heavy neutrino mass relation given by Eq.~(\ref{MR1}) guarantees the mild hierarchy of light neutrino mass.}. Interestingly, due to the loop function $f(z_{i})$ which has a scale dependence, contrary to the usual seesaw~\cite{Ahn:2012cg}, the mixing parameters can have various behaviors and predictions depending on a scale of dark matter mass once a successful leptogenesis scale is fixed, which can be named as ``flavored dark matter" and will be shown as examples in Sec.~V.

To see how neutrino mass matrix given by Eq.(\ref{radseesaw1}) can lead to the deviations of neutrino mixing angles from their TBM values,
 we first introduce three small quantities $\epsilon_{i},~(i=1-3)$ which are responsible for the deviations of the $\theta_{jk}$ from their TBM values ;
 \begin{eqnarray}
  \theta_{23}=-\frac{\pi}{4}+\epsilon_{1}~, \qquad\theta_{13}=\epsilon_{2}~, \qquad\theta_{12}= \sin^{-1}\left(\frac{1}{\sqrt{3}}\right)+\epsilon_{3}~.
 \end{eqnarray}
Then, the PMNS mixing matrix keeping unitarity up to order of $\epsilon_{i}$ can be written as
 \begin{eqnarray}
 U_{\rm PMNS}&=&{\left(\begin{array}{ccc}
 \frac{\sqrt{2}-\epsilon_{3}}{\sqrt{3}} &  \frac{1+\epsilon_{3}\sqrt{2}}{\sqrt{3}} &  \epsilon_{2}e^{-i\delta_{CP}} \\
 -\frac{1+\epsilon_{1}+\epsilon_{3}\sqrt{2}}{\sqrt{6}}+\frac{\epsilon_{2}e^{i\delta_{CP}}}{\sqrt{3}} &  \frac{\sqrt{2}+\epsilon_{1}\sqrt{2}-\epsilon_{3}}{\sqrt{6}}+\frac{\epsilon_{2}e^{i\delta_{CP}} }{\sqrt{6}} &  \frac{-1+\epsilon_{1}}{\sqrt{2}} \\
 \frac{-1+\epsilon_{1}+\epsilon_{3}\sqrt{2}}{\sqrt{6}}-\frac{\epsilon_{2}}{\sqrt{3}}e^{i\delta_{CP}} &  \frac{\sqrt{2}-\epsilon_{3}-\sqrt{2}\epsilon_{1}}{\sqrt{6}}-\frac{\epsilon_{2}}{\sqrt{6}}e^{i\delta_{CP}} &  \frac{1+\epsilon_{1}}{\sqrt{2}}
 \end{array}\right)}Q_{\nu}
 +{\cal O}(\epsilon^{2}_{i})~.
 \label{Unu}
 \end{eqnarray}

The small deviation $\epsilon_{1}$ from the maximality of the atmospheric mixing angle $\theta_{23}$ is expressed in terms of the parameters in Eq.~(\ref{MMele}) presented in Appendix B as
 \begin{eqnarray}
  \tan\epsilon_{1}=\frac{R(1+y_{2})-S(y_{2}-1)}{R(y_{2}-1)-S(1+y_{2})}~.
 \label{Atmdevi}
 \end{eqnarray}
In the limit of $y_{1},y_{2}\rightarrow1$, $\epsilon_1$ goes to zero (or equivalently $\theta_{23}\rightarrow-\pi/4$) due to $R,S\rightarrow0$.
The reactor angle $\theta_{13}$ and the Dirac-CP phase $\delta_{CP}$ are expressed as
 \begin{eqnarray}
  \tan2\theta_{13}&=&\frac{y_{1}|s_{23}(P-Q)y_{2}+c_{23}(P+Q)-3i\{s_{23}(R+S)y_{2}+c_{23}(R-S)\}|}{\Psi_{3}-y^{2}_{1}\tilde{A}}~,\nonumber\\
  \tan\delta_{CP}&=&3\frac{(R-S)^{2}+y^{2}_{2}(R+S)^{2}}{(P+Q)(R-S)-y^{2}_{2}(P-Q)(R+S)}~,
 \label{DiracCP}
 \end{eqnarray}
where the parameters $P, Q, R, S$ and $\tilde{A}$ are given in Eq.~(\ref{MMele}) in Appendix B.
In the limit of  $y_{1},y_{2}\rightarrow1$, the parameters $Q,R,S$ go to zero, which in turn leads to  $\theta_{13}\rightarrow0$ and $\delta_{CP}\rightarrow0$ as expected.
Finally, the solar mixing angle is given by
 \begin{eqnarray}
  \tan2\theta_{12}=y_{1}\frac{y_{2}c_{23}(P-Q)-s_{23}(P+Q)}{c_{13}(\Psi_{2}-\Psi_{1})}~.
 \label{sol1}
 \end{eqnarray}
 One can easily check $\theta_{12}$ is recovered to be $\sin^{-1}(1/\sqrt{3})$ in the limit $y_{1},y_{2}\rightarrow1$.

The expressions of the squared-mass eigenvalues of the three light neutrinos are given by
 \begin{eqnarray}
    m^{2}_{1}&=&m^{2}_{0}\left\{s^{2}_{12}\Psi_{1}+c^{2}_{12}\Psi_{2}-y_{1}\frac{y_{2}c_{23}(P-Q)-s_{23}(P+Q)}{2c_{13}}\sin2\theta_{12}\right\}~,\nonumber\\
    m^{2}_{2}&=&m^{2}_{0}\left\{c^{2}_{12}\Psi_{1}+s^{2}_{12}\Psi_{2}+y_{1}\frac{y_{2}c_{23}(P-Q)-s_{23}(P+Q)}{2c_{13}}\sin2\theta_{12}\right\}~,\nonumber\\
    m^{2}_{3}&=&m^{2}_{0}\Big\{c^{2}_{13}\Psi_{3}
    +y^{2}_{1}\tilde{A}s^{2}_{13}+\frac{y_{1}\sin2\theta_{13}}{2}\Big[c_{23}\left((Q+P)\cos\delta_{CP}+3(R-S)\sin\delta_{CP}\right)\nonumber\\
    &+&s_{23}y_{2}\left((P-Q)\cos\delta_{CP}+3(R+S)\sin\delta_{CP}\right)\Big]\Big\}~.
 \label{eigenvalueGen}
 \end{eqnarray}
Note here that, when $y_{1,2}=1$, the mixing angles are reduced to TBM and independent to the mass eigenvalues, which means Eqs.~(\ref{Atmdevi}-\ref{eigenvalueGen}) do not work at all.
Since the parameters participating in mixing angles $(\theta_{12},\theta_{23},\theta_{13},\delta_{\rm CP})$ are simultaneously involved in mass-squared differences $(m^{2}_{2}-m^{2}_{1}, |m^{2}_{3}-m^{2}_{1}|)$, the case giving TBM values (or $y_{1,2}\rightarrow1$) may not be obtained if not $y_{1,2}=1$ (see, normal mass hierarchy case in Sec. V).

Actually, in the limiting case of $y_{1,2}\rightarrow1$ the combination of $\tilde{Y}^{\dag}_{\nu}\tilde{Y}_{\nu}$ is proportional to unit matrix (see Eq.~(\ref{YnuYnu})) and their deviations are responsible for non-zero $\theta_{13}$.
As will be shown later, a successful leptogenesis can be achieved when $M\geq10^{10}$ GeV and $y^{\nu}_{3}\geq0.01$ because of mild hierarchy of neutrino Yukawa couplings.
 Depending on the mass scale of the scalar field $\eta^{0}$ which can be a good dark matter candidate,
the parameter $M_{ri}$ in Eq.~(\ref{radseesaw1}) can be simplified in the following limit cases as
 \begin{eqnarray}
  M_{ri}\simeq\left\{
  \begin{array}{ll}
    M_{i}\left[\ln z_{i}-1\right]^{-1}, & \hbox{for $z_{i}\gg1$} \\
    2M_{i}, & \hbox{for $z_{i}\rightarrow1$} \\
    M_{i}z^{-1}_{i}, & \hbox{for $z_{i}\ll1$~.}
  \end{array}
\right.
 \end{eqnarray}
 Also, from the mass spectrum given in Eq.~(\ref{Higgsmass2}) and Eq.~(\ref{Higgsmass4}), we consider two plausible and simple scenarios as shown below.
\vskip 0.4cm
{\bf Case-I.}  $\bar{m}^{2}_{\eta_{1}}\simeq \bar{m}^{2}_{\eta_{2}}\simeq \bar{m}^{2}_{\eta_{3}}\sim{\cal O}(v^{2}_{\Phi})$ :
\vskip 0.5cm
 This case can be realized when $\lambda^{\eta\chi}_{2}\rightarrow0$ and $\lambda^{\eta\chi}_{1}\cos2\phi\rightarrow0$, leading to
 \begin{eqnarray}
  \phi\in[0,2\pi]~,\quad \bar{m}^{2}_{\eta_{1}}\simeq \bar{m}^{2}_{\eta_{2}}\simeq \bar{m}^{2}_{\eta_{3}}\simeq \mu^{2}_{\eta}+v^{2}_{\Phi}\lambda^{\eta\Phi}_{12}~,
 \end{eqnarray}
and corresponding to $z_{i}\gg1$. Since the light neutrino masses Eq.~(\ref{masseigen}) contain $3m_{0}f(z_{i})$ which is order of $0.01$ eV for hierarchical case, we have $f(z_{i})\sim{\cal O}(10)$
for $\bar{m}_{\eta}\sim{\cal O}(100{\rm GeV})$ and $M=10^{10}$ GeV, and then the quartic coupling $\lambda^{\Phi\eta}_{3}$ should be order of 0.1 and $10^{-5}$ for $y^{\nu}_{3}=0.01$ and $y^{\nu}_{3}=1$, respectively.

Since all new particles $\eta^{\pm}, \eta^{0}_{R}, \eta^{0}_{I}$ carry a $Z_{2}$ odd quantum number and only couple to Higgs boson and electroweak gauge bosons of the standard model, they can be produced in pairs through the standard model gauge bosons $W^{\pm}, Z$ or $\gamma$. Once produced, $\eta^{\pm}$ will decay into $\eta^{0}_{R,I}$ and a virtual $W^{\pm}$, then $\eta^{0}_{I}$ subsequently becomes $\eta^{0}_{R}+Z$-boson, which will decay a quark-antiquark or lepton-antilepton pair.
Here the mass hierarchy $m_{\eta^{\pm}}>m_{\eta^{0}_{I}}>m_{\eta^{0}_{R}}$ is assumed. That is, the stable $\eta^{0}_{R}$ appears as missing energy in the decays of $\eta^{\pm}\rightarrow\eta^{0}_{I}l^{\pm}\nu$ with the subsequent decay $\eta^{0}_{I}\rightarrow\eta^{0}_{R}l^{\pm}l^{\mp}$, which can be compared to the direct decay $\eta^{\pm}\rightarrow\eta^{0}_{R}l^{\pm}\nu$ to extract the masses of the respective particles.
Therefore, probing a signal of scalar particle $\eta$ in collider can be a search of the dark matter candidate\footnote{Here we will not consider the relic abundance of dark matter compatible with observation.}.

\vskip 0.4cm
{\bf Case-II.} For $\bar{m}^{2}_{\eta_{1}}\simeq \bar{m}^{2}_{\eta_{2}}\simeq \bar{m}^{2}_{\eta_{3}}\sim{\cal O}(v^{2}_{\chi})$ \footnote{More generally, as can be seen in Eq.~(\ref{Higgsmass4}),  $\bar{m}^{2}_{\eta_{1}}\neq \bar{m}^{2}_{\eta_{2}}\neq \bar{m}^{2}_{\eta_{3}}\sim{\cal O}(v^{2}_{\chi})$.}:
\vskip 0.5cm
It can be realized when $\lambda^{\eta\chi}_{2}\rightarrow0$ and $\phi\neq\pm\pi/4,\pm3\pi/4$, giving
 \begin{eqnarray}
  \phi\in[0,2\pi]~,\quad \bar{m}^{2}_{\eta_{1}}\simeq \bar{m}^{2}_{\eta_{2}}\simeq \bar{m}^{2}_{\eta_{3}}\simeq 2v^{2}_{\chi}\lambda^{\eta\chi}_{1}\cos2\phi~.
 \end{eqnarray}
Assuming $\mu^{2}_{\eta}+v^{2}_{\Phi}\lambda^{\eta\Phi}_{12}\sim{\cal O}(v^{2}_{\Phi})$ and $v_{\chi}\gg v_{\Phi}$, it can lead to $f(z_{i})\simeq1-10$, but much milder than Case-I.

\section{Numerical Analysis}

Now we perform a numerical analysis using the linear algebra tools in Ref.~\cite{Antusch:2005gp}.
The Daya Bay and RENO experiments have accomplished the measurement of three mixing angles
$\theta_{12}, \theta_{23}$, and $\theta_{13}$ from three kinds of neutrino oscillation experiments. The most recent analysis based on global fits~\cite{GonzalezGarcia:2012sz} of neutrino oscillations enters into a new phase of precise determinations of the neutrino mixing angles and mass-squared differences, indicating that the TBM mixing for the three flavors of leptons should be modified. Their allowed ranges at $1\sigma$ $(3\sigma)$ from global fits are given by
 \begin{eqnarray}
  &&\theta_{13}=8.66^{\circ+0.44^{\circ}~(+1.30^{\circ})}_{~-0.46^{\circ}~(-1.47^{\circ})}~,\qquad\delta_{\rm CP}=300^{\circ+66^{\circ}~~(+60^{\circ})}_{~-138^{\circ}~(-300^{\circ})}~,\qquad\theta_{12}=33.36^{\circ+0.81^{\circ}~(+2.53^{\circ})}_{~-0.78^{\circ}~(-1.27^{\circ})}~,\nonumber\\
  &&\theta_{23}=40.0^{\circ+2.1^{\circ}}_{~-1.5^{\circ}}\oplus50.4^{\circ+1.3^{\circ}}_{~-1.3^{\circ}}~{1\sigma},\quad\left(35.8^{\circ}\thicksim54.8^{\circ}~{3\sigma}\right)~,\nonumber\\
  &&\Delta m^{2}_{\rm Sol}[10^{-5}{\rm eV}^{2}]=7.50^{+0.18~(+0.59)}_{-0.19~(-0.50)}~,~\Delta m^{2}_{\rm Atm}[10^{-3}{\rm eV}^{2}]=\left\{\begin{array}{ll}
                2.473^{+0.070~(+0.222)}_{-0.067~(-0.197)}~, & \hbox{NMH} \\
                2.427^{+0.042~(+0.185)}_{-0.065~(-0.222)}~, & \hbox{IMH}~
                                  \end{array},
                                \right.
 \label{expnu}
 \end{eqnarray}
where $\Delta m^{2}_{\rm Sol}\equiv m^{2}_{2}-m^{2}_{1}$, $\Delta m^{2}_{\rm Atm}\equiv m^{2}_{3}-m^{2}_{1}$ for the normal mass hierarchy (NMH), and  $\Delta m^{2}_{\rm Atm}\equiv |m^{2}_{3}-m^{2}_{2}|$ for the inverted mass hierarchy (IMH).
Note here that the $3\sigma$ data for the oscillation parameters ($\theta_{23}, \theta_{12}, \Delta m^{2}_{\rm Sol}, \Delta m^{2}_{\rm Atm}$) except for $\theta_{13}$ and $\delta_{CP}$ are used to predict the values of model parameters in our numerical analysis.
For $\theta_{13}$ and $\delta_{CP}$, we scan the regions $\theta_{13}<12^{\circ}$ and $0\lesssim \delta_{CP} \lesssim 360^{\circ}$.
The mass matrix in Eq.~(\ref{radseesaw1}) contains 10 parameters: $y^{\nu}_{3}, M, \lambda^{\Phi\eta}_{3}, z_{1}, z_{2}, z_{3}, y_{1}, y_{2}, \kappa, \phi$. The first four ($y^{\nu}_{3}$, $M$, $\lambda^{\Phi\eta}_{3}, z_{2}$) contribute to the overall scale of  neutrino scale parameter given by $m_{0}f(z_{2})$. The next six ($y_1,y_2,\kappa,\phi, z_{1}, z_{3}$) are responsible for the deviations from TBM, the CP phases and corrections to the masse eigenvalues. Actually, the three parameters ($z_{1},z_{2},z_{3}$) can be determined by the mass scale of dark matter, as can be seen from Eq.~(\ref{Higgsmass4}). The determination of neutrino masses and mixing parameters in our numerical analysis requires to fix a leptogenesis scale $M$.

In Table \ref{bench}, we present the benchmark points for the unknown parameters $M, y^{\nu}_3, \lambda^{\Phi\eta}$ and $\bar{m}_{\eta_1}$.
Such a choice presented in Table \ref{bench} makes the parameters $z_{1}$ and $z_{3}$ no longer arbitrary. The choice of  $M=10^{10} (10^{11})$ GeV, $y^{\nu}_{3}=0.01$ leads to a successful leptogenesis, as can be seen in Sec. IV.
We take $\bar{m}_{\eta_i}$ to be degenerate for the sake of simplicity.
\begin{widetext}
\begin{center}
\begin{table}[h]
\caption{\label{bench} Benchmark points for the unknown parameters.}
\begin{ruledtabular}
\begin{tabular}{c|cccc}
hierarchy& $M$(GeV) &$y^{\nu}_{3}$ &$\lambda^{\Phi\eta}_3$&$\bar{m}_{\eta_i}$(GeV)\\
\hline
NH (Case-I)&$ 10^{10}$&0.01&0.1 & 500 \\
NH (Case-II)&$ 10^{10}$&0.01&0.4 & $10^8$ \\
IH (Case-I)&$ 10^{11}$&0.32& $10^{-3}$ & 100 \\
IH (Case-II)&$ 10^{11}$&0.01&  5 & $10^{9} $
\end{tabular}
\end{ruledtabular}
\end{table}
\end{center}
\end{widetext}

It is worthwhile to note that the neutrino masses are sensitive to the combination $m_{0}f(z_{2})=\lambda^{\Phi\eta}_{3}v^{2}_{\Phi}|y^{\nu}_{3}|^{2}f(z_{2})/(4\pi^{2}M)$ which is roughly order of ${\cal O}(0.01)$ eV.
Once $m_{0}f(z_{2})$ is fixed as input, the parameters $y_{1}, y_{2}, \kappa$ and $\phi$ can be determined from the experimental results of mixing angles and mass-squared differences.  In addition, the CP phases $\delta_{CP}$ and $\varphi_{1,2}$ can be predicted after determining the model parameters.

\begin{figure}[t]
\begin{minipage}[t]{6.0cm}
\epsfig{figure=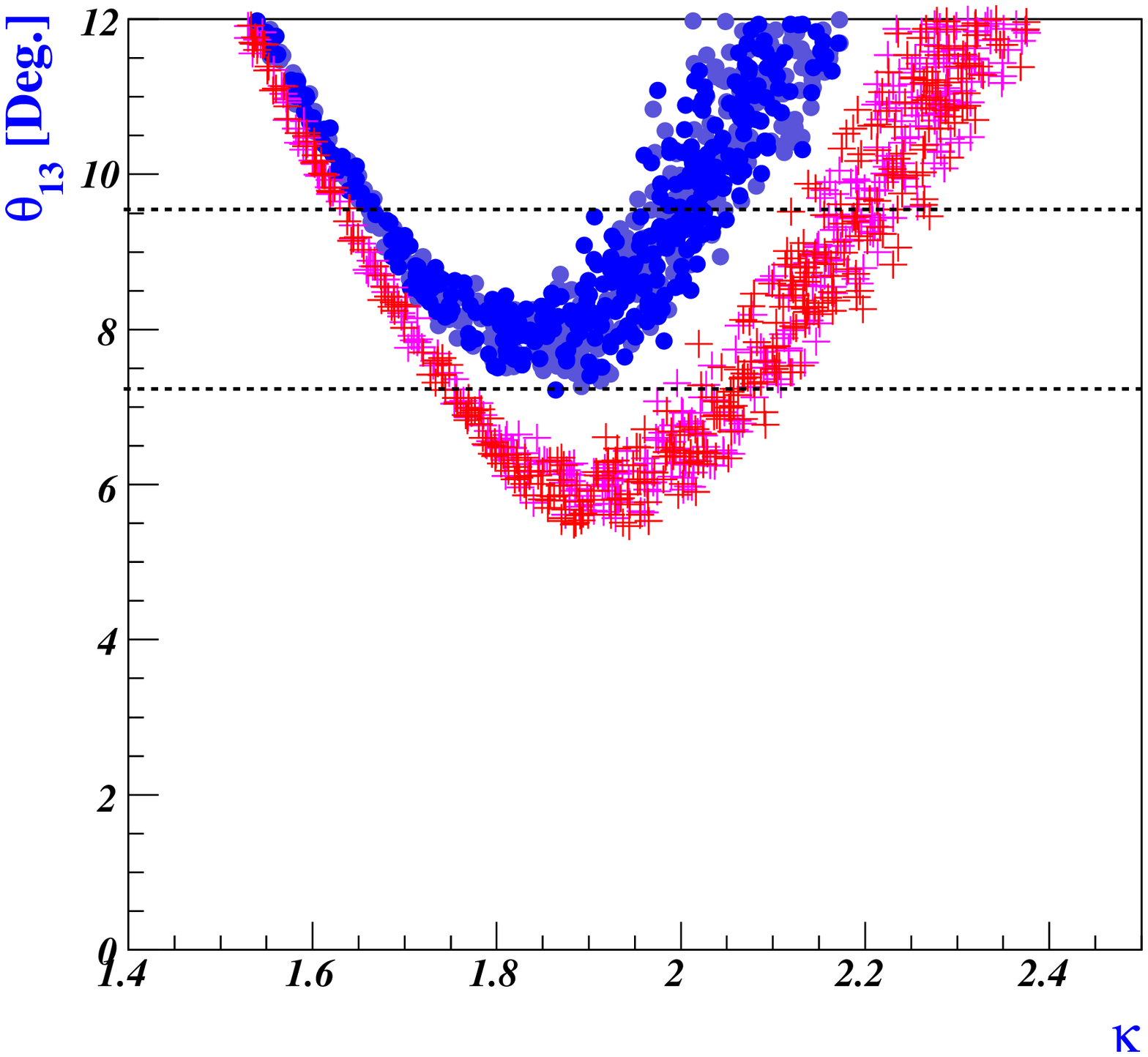,width=6.5cm,angle=0}
\end{minipage}
\hspace*{1.0cm}
\begin{minipage}[t]{6.0cm}
\epsfig{figure=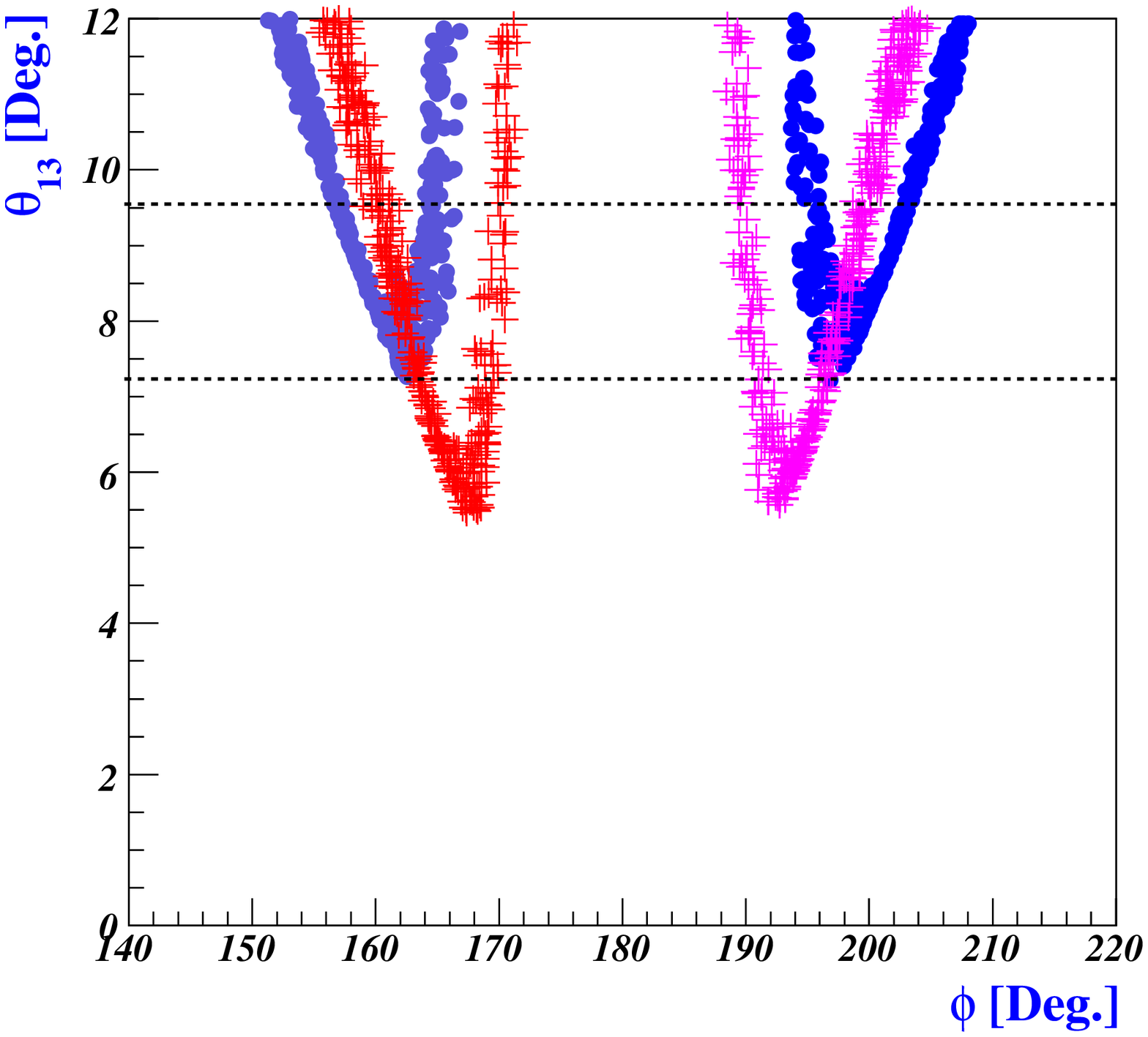,width=6.5cm,angle=0}
\end{minipage}\\
\begin{minipage}[t]{6.0cm}
\epsfig{figure=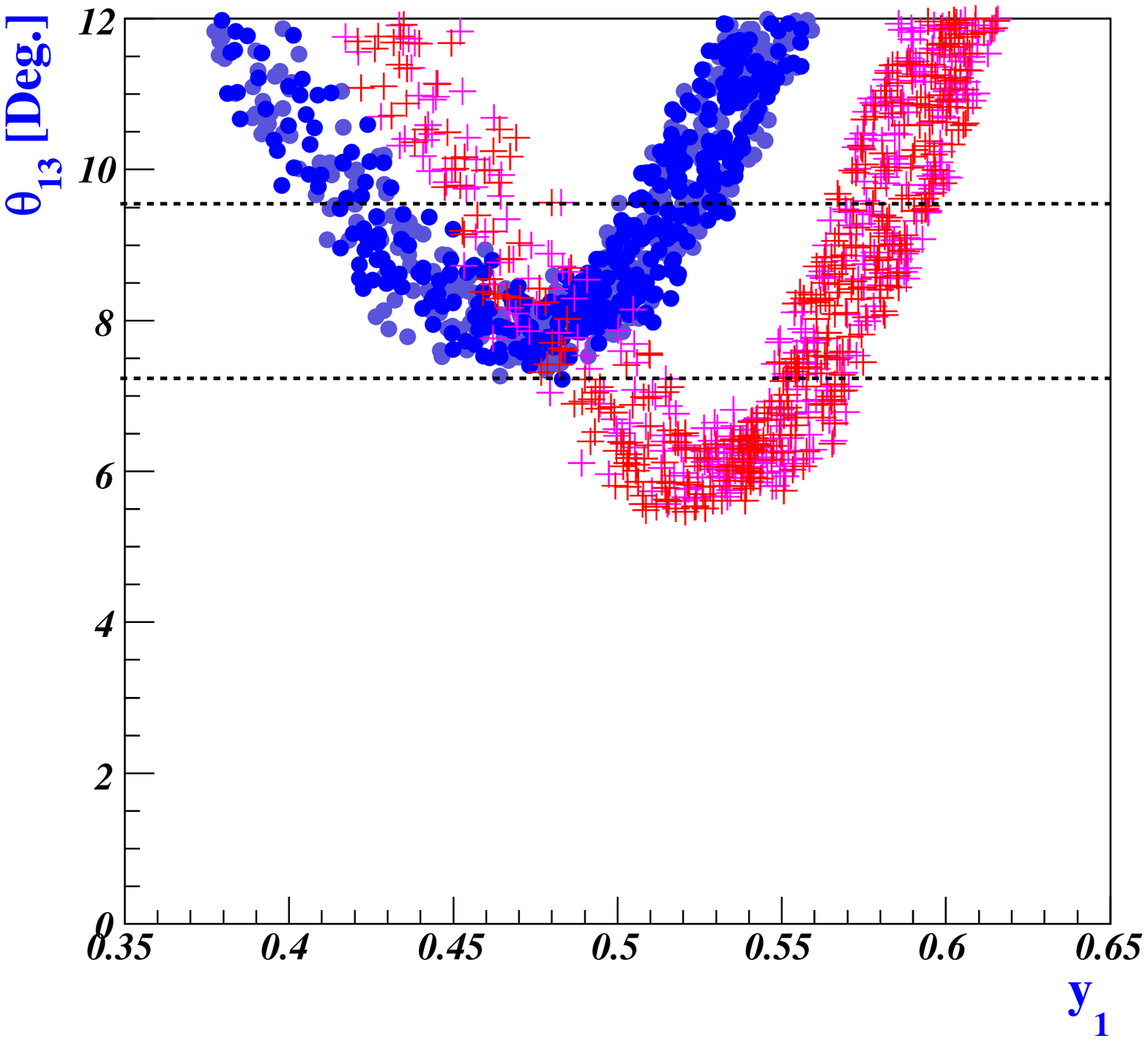,width=6.5cm,angle=0}
\end{minipage}
\hspace*{1.0cm}
\begin{minipage}[t]{6.0cm}
\epsfig{figure=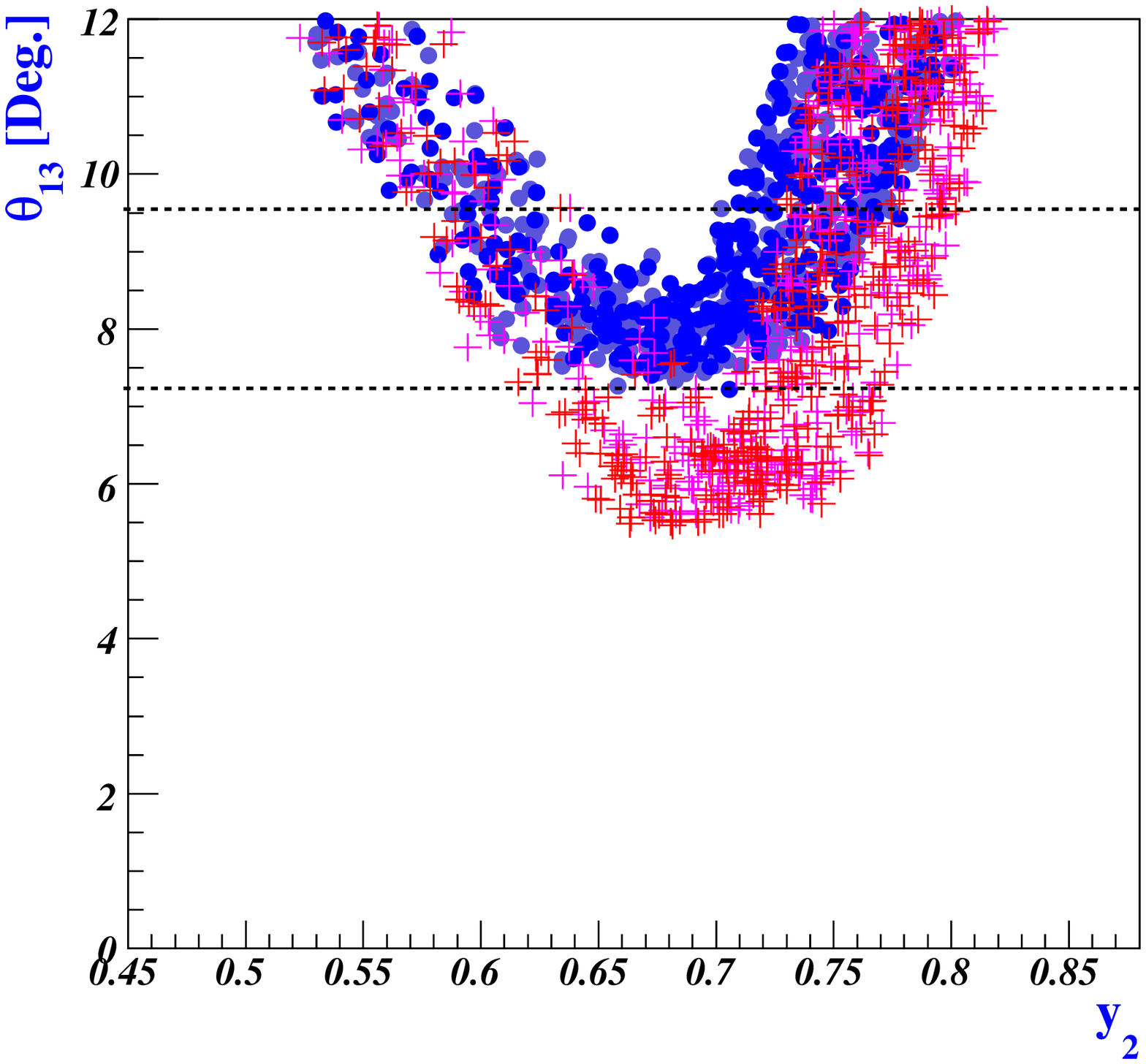,width=6.5cm,angle=0}
\end{minipage}
\caption{\label{FigA1}
Plots for NMH displaying the reactor mixing angle $\theta_{13}$
versus $\kappa$ (upper left panel) and versus $\phi_{1}$ (upper right panel), and the reactor angle $\theta_{13}$ versus $y_{1}$ (lower left panel) and versus $y_{2}$ (lower right panel). Here the blue-type dots and red-type crosses data points correspond to $\bar{m}_{\eta_i}=500$ GeV and $10^{8}$ GeV, respectively. The horizontal dotted lines in plots indicate the upper and lower bounds on $\theta_{13}$ given in Eq.~(\ref{expnu}) at $3\sigma$.}
\end{figure}

\subsection{Normal mass hierarchy}

Using the formulas for the neutrino mixing parameters and input values of $M$, $y^{\nu}_{3}$, $v_{\Phi}$, $\lambda^{\Phi\eta}_{3}, \bar{m}_{\eta_i}$ presented above, we obtain the allowed regions of the unknown model parameters. The results are given for the Case-I by
 \begin{eqnarray}
  &&0.5<\kappa<2.2~,\qquad0.37<y_{1}<0.56~,\qquad0.52< y_{2}\lesssim0.8~,\nonumber\\
  &&150^{\circ}<\phi<168^{\circ}~,\qquad192^{\circ}<\phi<208^{\circ}~,
  \label{input1}
 \end{eqnarray}
and for the case-II,
 \begin{eqnarray}
  &&1.5<\kappa<2.4~,\qquad0.41<y_{1}<0.62~,\qquad0.51< y_{2}<0.82~,\nonumber\\
  &&154^{\circ}<\phi<172^{\circ}~,\qquad188^{\circ}<\phi<206^{\circ}~.
  \label{input2}
 \end{eqnarray}
\begin{figure}[t]
\begin{minipage}[t]{6.0cm}
\epsfig{figure=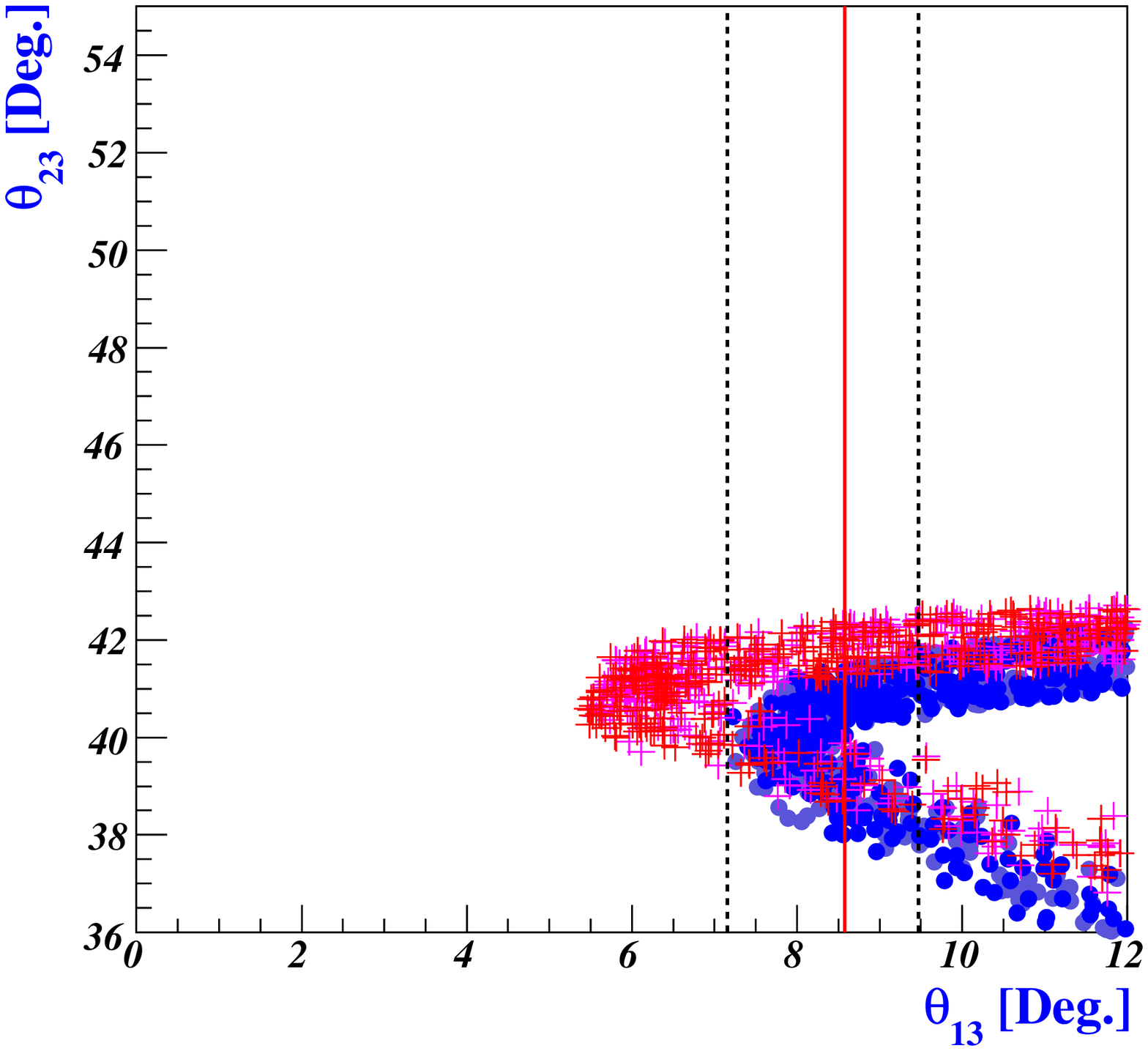,width=6.5cm,angle=0}
\end{minipage}
\hspace*{1.0cm}
\begin{minipage}[t]{6.0cm}
\epsfig{figure=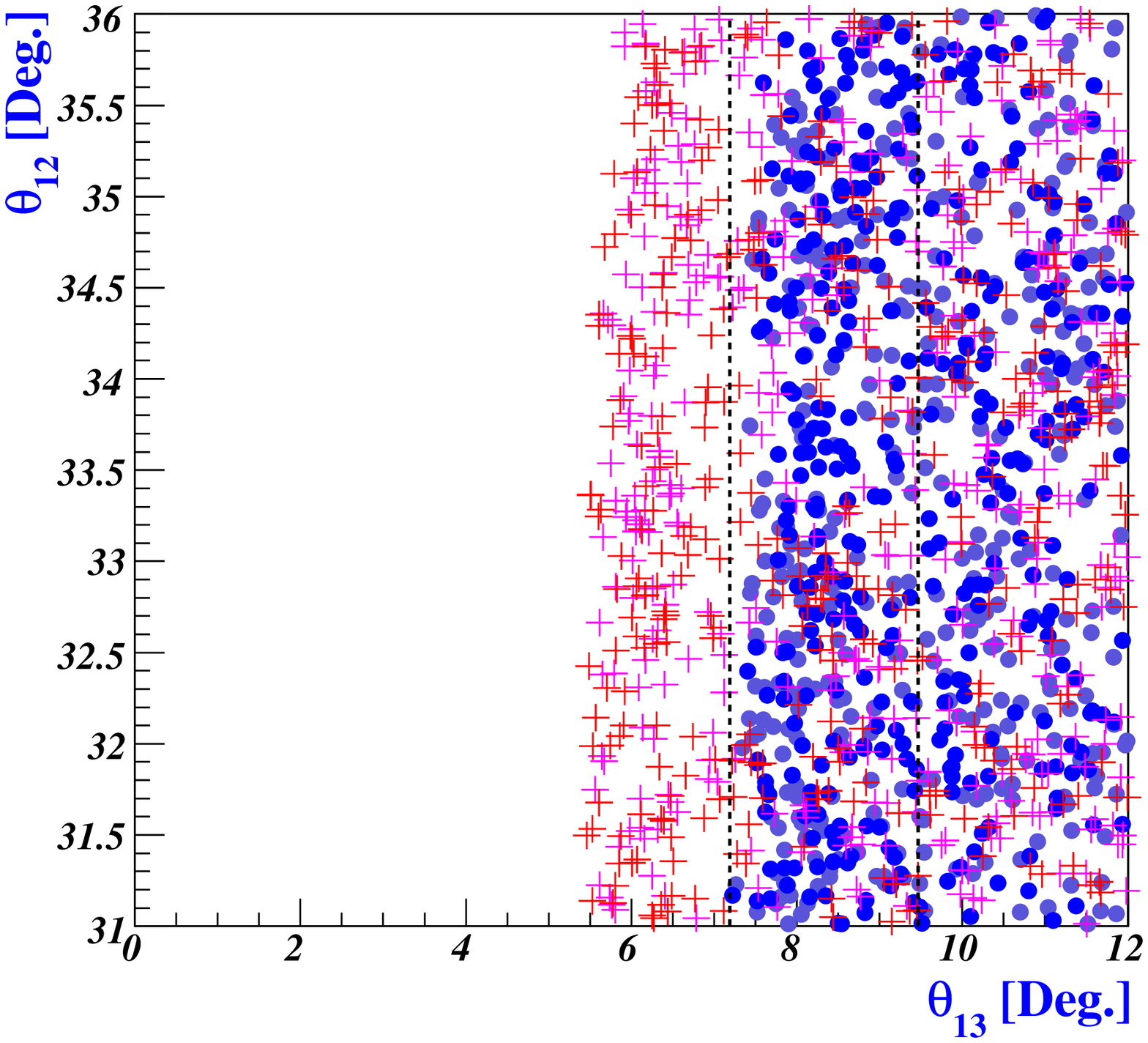,width=6.5cm,angle=0}
\end{minipage}
\caption{\label{FigA2}
Plots for NMH displaying the allowed values for the atmospheric mixing angle $\theta_{23}$ (left) and the solar mixing angle $\theta_{12}$ (right) versus the mixing angle $\theta_{13}$, respectively.
The thick line corresponds to $\theta_{13} = 8.6^{\circ}$ which is the best-fit value of Eq.~(\ref{expnu}). And the vertical dotted lines in plots indicate the upper and lower bounds on $\theta_{13}$ given in Eq.~(\ref{expnu}) at $3\sigma$}
\end{figure}
For those parameter regions, we investigate how a non-zero $\theta_{13}$, a deviation from maximality $\theta_{23}$ and a Dirac CP phase can be determined by the mass scale of $\eta$ for the normal mass hierarchy, after fixing a leptogenesis scale. In Figs.~\ref{FigA1}-\ref{FigA3}, the data points represented by dots and crosses indicate results for the different input scale of parameter $\bar{m}_{\eta_i}=500$ GeV and $10^{8}$ GeV, respectively.
For different ranges of $\phi$ as given in Eqs.(\ref{input1},\ref{input2}), we display the corresponding data points with different colors, blue and bright-blue dots, and red and hot-pink crosses, respectively.
In fact, the blue and bright blue points correspond to the case-I, whereas the red and hot pink crosses
correspond to the case-II.
The upper-left, upper-right, lower-left and lower-right plots in Fig.~\ref{FigA1} show how the mixing angle $\theta_{13}$ depends on the parameter $\kappa=\lambda^{s}_{\chi}v_\chi/M$, the CP-phase $\phi$, the parameter $y_{1}$, and $y_{2}$, respectively.
The points located between two dashed lines in the plots are in consistent with
the values of $\theta_{13}$ from the global fits including the Daya Bay and RENO experiments at $3\sigma$ C.L.
%

Fig.~\ref{FigA2} shows how the estimated values of $\theta_{13}$ depend on the mixing angles $\theta_{23}$ and $\theta_{12}$. The vertical lines corresponds to the experimental limits on the mixing angle $\theta_{13}$. As can be seen in the left plot of Fig.~\ref{FigA2}, $\theta_{23}$ compatible with the measured values of $\theta_{13}$ at $3\sigma$'s favors large deviations from maximality only to $\theta_{23}<45^{\circ}$. We see that the measured values of $\theta_{13}$ can be achieved for $37.5^{\circ}\lesssim\theta_{23}<42^{\circ}$ in the case-I as presented by blue dots, whereas for $38^{\circ}<\theta_{23}\lesssim40.5^{\circ}$ and $41^{\circ}\lesssim\theta_{23}\lesssim42.5^{\circ}$ in the Case-II as presented by red crosses, which are consistent with the experimental bounds at $1\sigma$ as can be seen in Eq.~(\ref{expnu}).
From the right plot of Fig.~\ref{FigA2}, we see that the
predictions for $\theta_{13}$ do not strongly depend on $\theta_{12}$ in the allowed region.

Leptonic CP violation can be detected through the neutrino oscillations which are sensitive to the Dirac CP phase $\delta_{CP}$, but insensitive to the Majorana phases in $U_{\rm PMNS}$~\cite{Branco:2002xf}.
To see how the parameters are correlated with low-energy CP violation observables measurable through neutrino oscillations, we consider the leptonic CP violation parameter defined by the
Jarlskog invariant~\cite{Jarlskog:1985ht}
\begin{align}
J_{CP}\equiv{\rm Im}[U_{e1}U_{\mu2}U^{\ast}_{e2}U^{\ast}_{\mu1}]
 =\frac{1}{8}\sin2\theta_{12}\sin2\theta_{23} \sin2\theta_{13}\cos\theta_{13}
 \sin\delta_{CP} .
\end{align}
The Jarlskog invariant $J_{CP}$ can be expressed in terms of the elements of the matrix $h=m_{\nu}m^{\dag}_{\nu}$~\cite{Branco:2002xf}:
 \begin{eqnarray}
  J_{CP}=-\frac{{\rm Im}\{h_{12}h_{23}h_{31}\}}{\Delta m^{2}_{21}\Delta m^{2}_{31}\Delta m^{2}_{32}}~.
  \label{JCP}
 \end{eqnarray}
The behaviors of $J_{CP}$ and $\delta_{CP}$ as a function of $\theta_{13}$ are plotted on the upper left and right panel of Fig.~\ref{FigA3}.
We see that the value of $J_{CP}$ lies in the range $-0.015\lesssim J_{CP}<0.025$ (blight blue) and $-0.026<J_{CP}<0.017$ (blue) for the Case-I, and $0.018\lesssim J_{CP}\lesssim0.03$ and $-0.026\lesssim J_{CP}\lesssim-0.006$ (hot-pink) and $0.008\lesssim J_{CP}\lesssim0.026$ and $-0.03\lesssim J_{CP}\lesssim-0.018$ (red) for the Case-II in the measured value of $\theta_{13}$ at $3\sigma$'s.  Also, in our model we have
 \begin{eqnarray}
  {\rm Im}\{h_{12}h_{23}h_{31}\}&=&m^{6}_{0}\frac{27y^{2}_{1}y^{2}_{2}(y^{2}_{2}-1)}{2}\Big(\sin(\psi_{1}-\psi_{2})\{....\}+\sin(2\psi_{1}-\psi_{2})\{.....\}\nonumber\\
  &+&\sin\psi_{2}\{....\}+\sin(\psi_{1}+\psi_{2})\{....\}\Big)~,
  \label{JCP1}
 \end{eqnarray}
in which $\{.....\}$ stands for a complicated lengthy function of $y_{1}$, $y_{2}$, $a$, $b$, $f(z_{1})$, $f(z_{2})$ and $f(z_{3})$. Clearly, Eq.~(\ref{JCP1}) indicates that in the limit of $y_{2}\rightarrow1$ the leptonic CP violation $J_{CP}$ goes to zero.
When $y_{2}\neq1$, i.e.\ for the normal hierarchy case, $J_{CP}$ could go to zero as cancelation among the terms composed of $\sin\psi_{12},\sin(\psi_{1}+\psi_{2}),\sin(2\psi_{1}-\psi_{2})$ and $\sin\psi_{2}$ multiplied by $y_{1,2},a,b,f(z_{1}),f(z_{2})$ and $f(z_{3})$ even if CP phases $\psi_{1,2}$ (or $\sin\phi$) are non zero.

\begin{figure}[b]
\begin{minipage}[b]{6.0cm}
\epsfig{figure=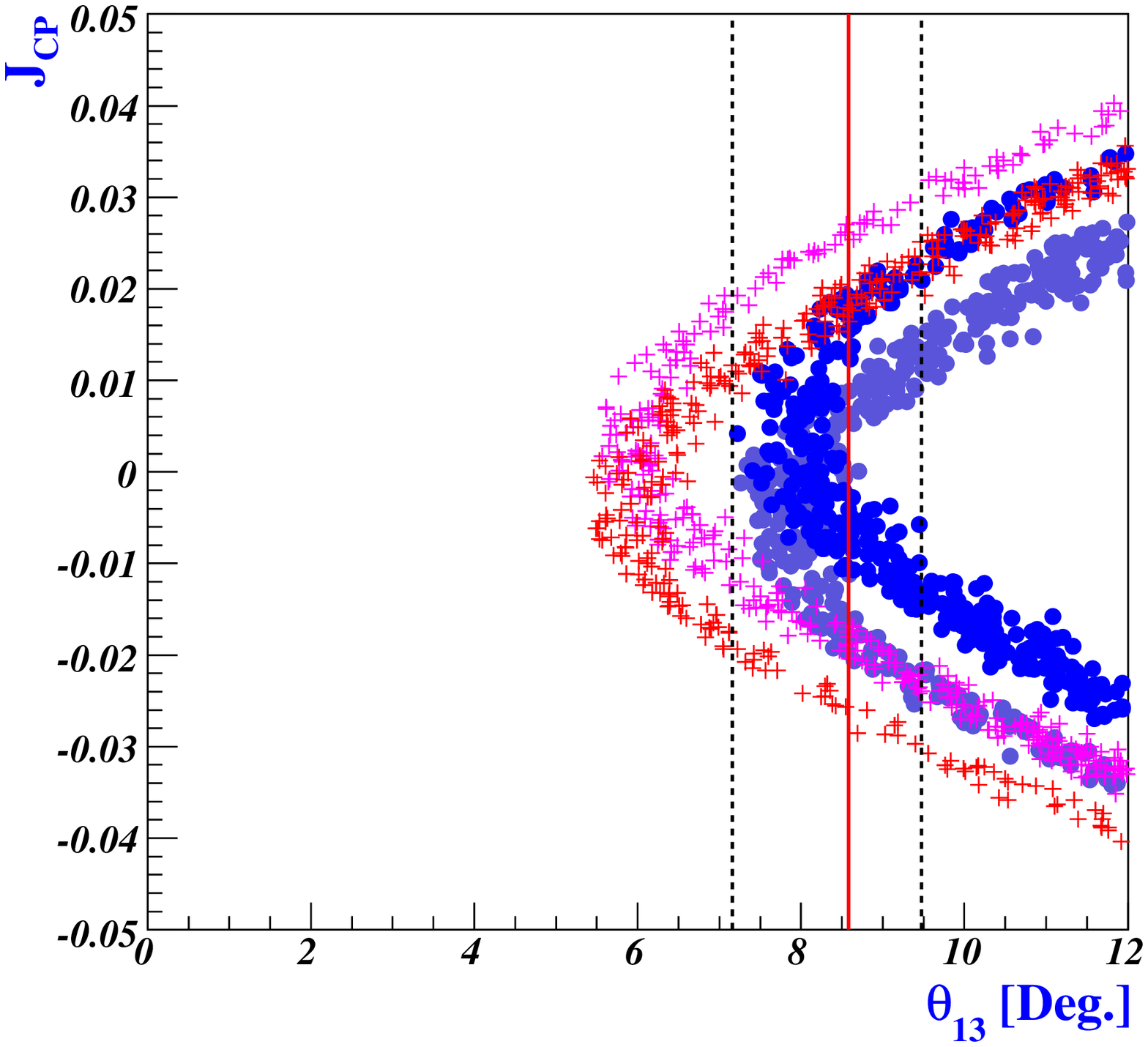,width=6.5cm,angle=0}
\end{minipage}
\hspace*{1.0cm}
\begin{minipage}[b]{6.0cm}
\epsfig{figure=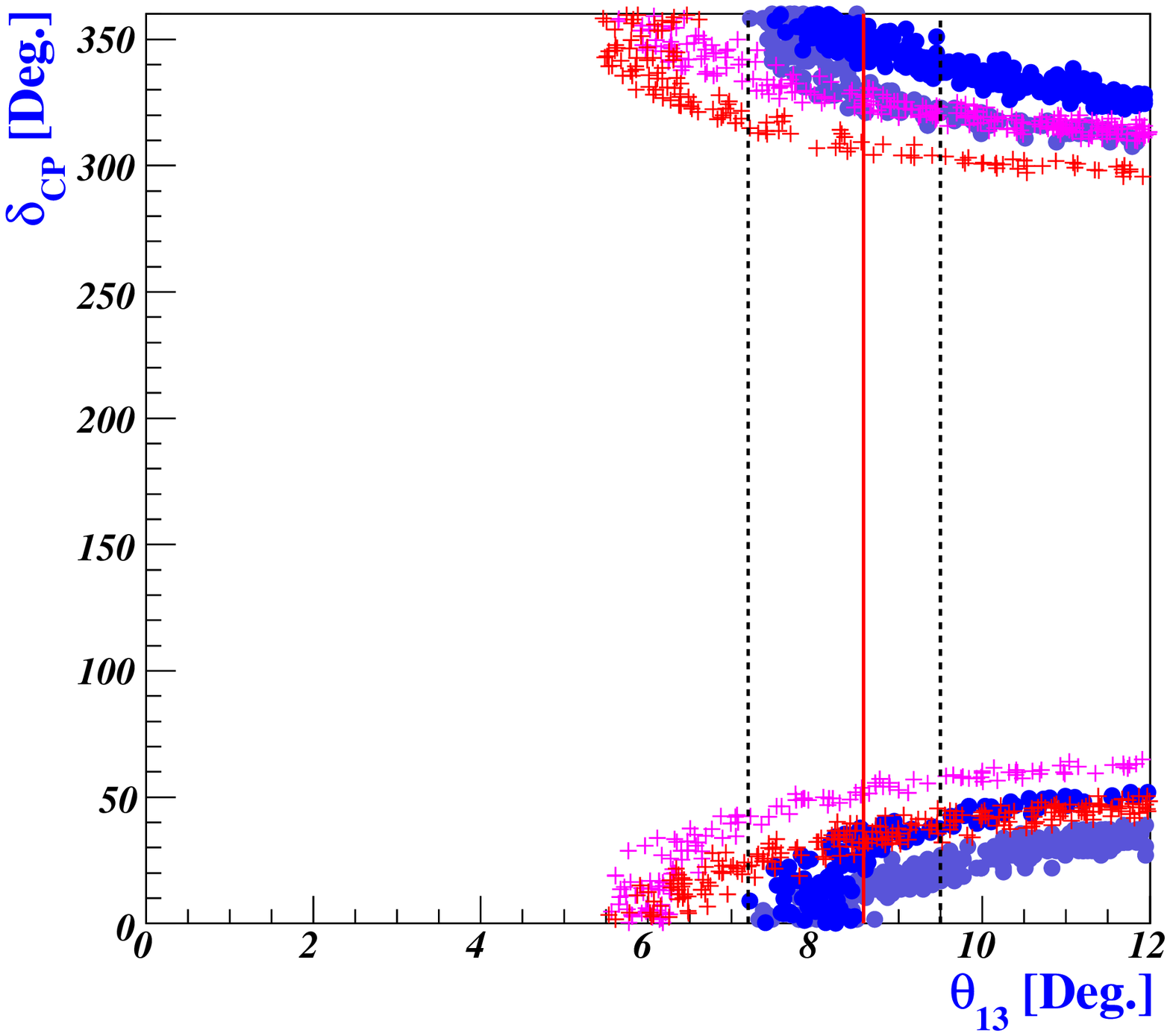,width=6.5cm,angle=0}
\end{minipage}
\caption{\label{FigA3}
Plots for NMH displaying the Jarlskog invariant $J_{CP}$ versus the reactor angle $\theta_{13}$ (left), and the Dirac CP phase $\delta_{CP}$ versus the reactor angle $\theta_{13}$ (right).
The thick line corresponds to $\theta_{13} = 8.6^{\circ}$ which is the best-fit value of Eq.~(\ref{expnu}). And the vertical dotted lines in plots indicate the upper and lower bounds on $\theta_{13}$ given in Eq.~(\ref{expnu}) at $3\sigma$}
\end{figure}

\begin{figure}[t]
\begin{minipage}[t]{6.0cm}
\epsfig{figure=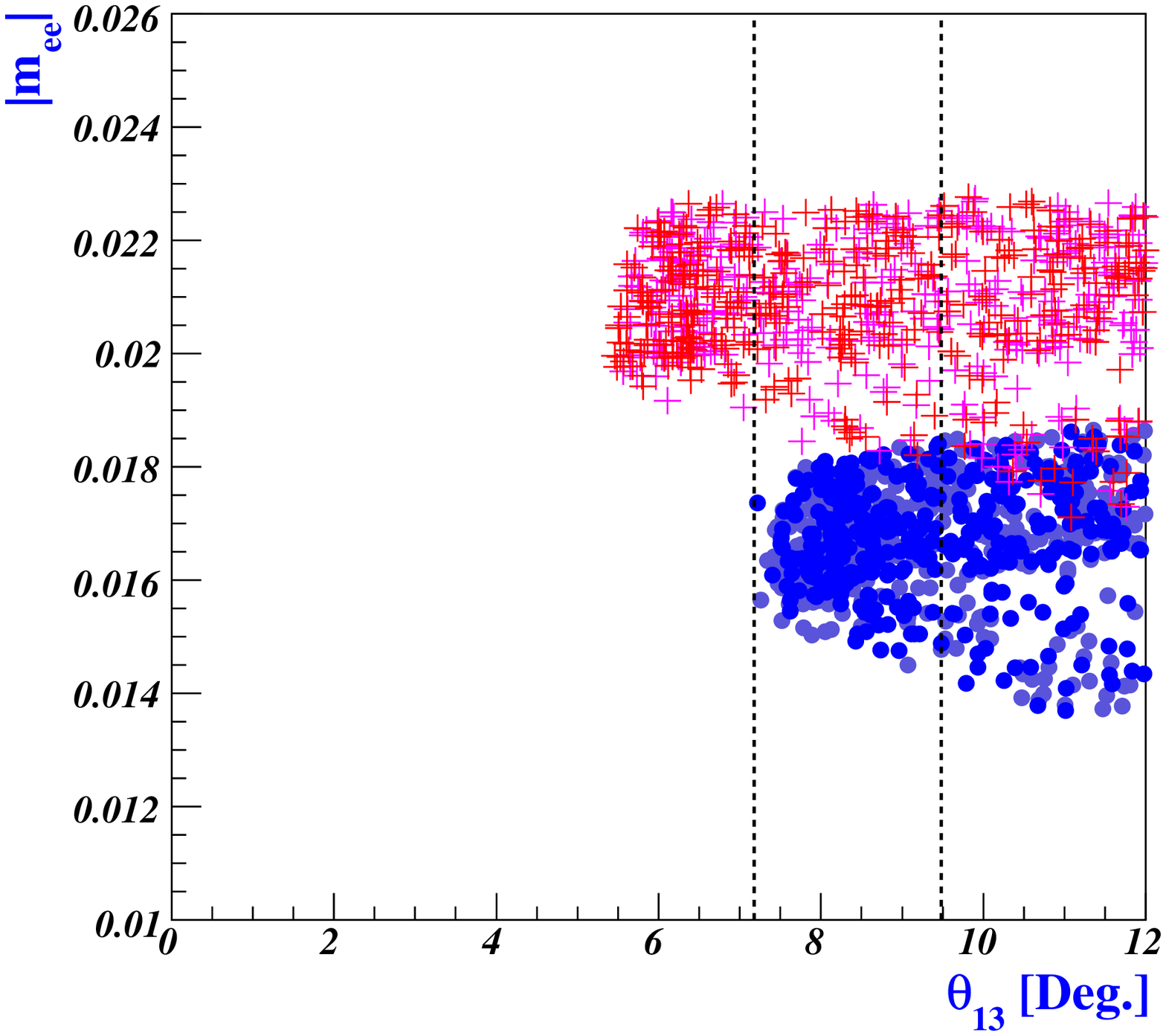,width=6.5cm,angle=0}
\end{minipage}
\hspace*{1.0cm}
\begin{minipage}[t]{6.0cm}
\epsfig{figure=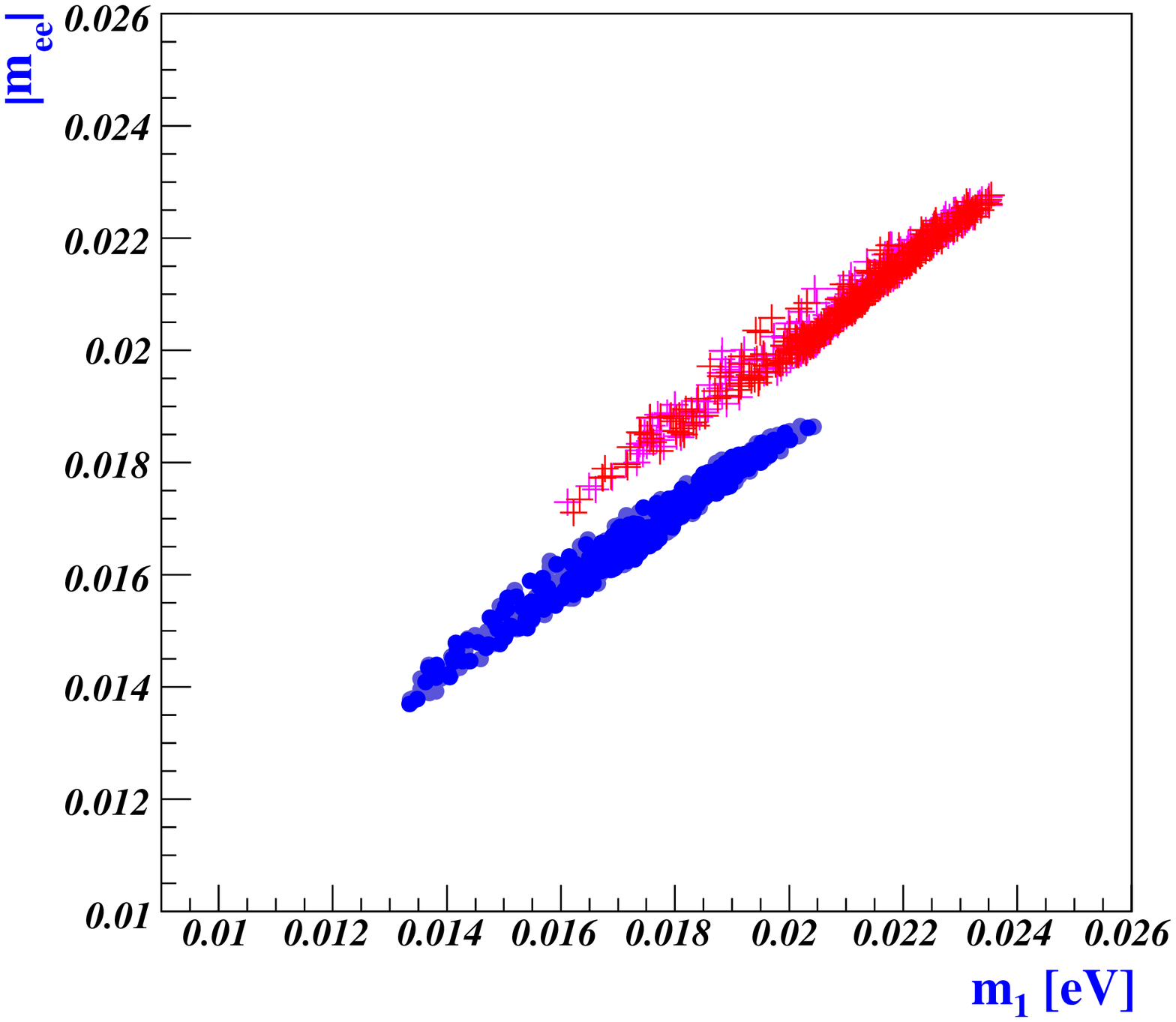,width=6.5cm,angle=0}
\end{minipage}
\caption{\label{FigA1p}
Plots for NMH displaying the effective neutrino mass $|m_{ee}|$ as a function of the mixing angle $\theta_{13}$ (left) and the lightest neutrino mass $m_{1}$ (right). The vertical dotted lines indicate the upper and lower bounds on $\theta_{13}$ given in Eq.~(\ref{expnu}) at $3\sigma$.}
\end{figure}

The right plot of Fig.~\ref{FigA3} shows the behavior of the Dirac CP phase $\delta_{CP}$ as a function of $\theta_{13}$. Interestingly enough, $\delta_{CP}$ for normal mass ordering favors values $0^{\circ}\leq\delta_{CP}\lesssim60^{\circ}$ and $300^{\circ}\lesssim\delta_{CP}\leq360^{\circ}$.
For each case, the blue and bright blue points correspond to the case-I, whereas the red and hot pint crosses
correspond to the case-II.

%
%
Since there is only one phase $\phi$ which is generated spontaneously in our Lagrangian, as will be shown in Sec. VI (see, Fig.~\ref{FigA4}), the right value of $\eta_{B}$ (baryon asymmetry of the Universe) will restrict the size of $\delta_{CP}$ and predict $1^{\circ}\lesssim\delta_{CP}\lesssim9^{\circ}$.

Moreover, we can straightforwardly obtain the effective neutrino mass $|m_{ee}|$ that characterizes the amplitude for neutrinoless double beta decay :
 \begin{eqnarray}
  |m_{ee}|\equiv \left|\sum_{i}(U_{\rm PMNS})^{2}_{ei}m_{i}\right|~,
  \label{mee}
 \end{eqnarray}
where $U_{\rm PMNS}$ is given in Eq.~(\ref{Unu}).
The left and right plots in Fig.~\ref{FigA1p} show the behavior of the effective neutrino mass $|m_{ee}|$ in terms of $\theta_{13}$ and the lightest neutrino mass $m_{1}$, respectively.
In the left plot of Fig.~\ref{FigA1p}, for the measured values of $\theta_{13}$ at $3\sigma$'s, the effective neutrino mass $|m_{ee}|$ can be in the range $0.0185\lesssim|m_{ee}|[{\rm eV}]<0.14$ (Case-I) and $0.018<|m_{ee}|[{\rm eV}]<0.023$ (Case-II). Our model predicts that the effective mass $|m_{ee}|$ is within the sensitivity of planned neutrinoless double-beta decay experiments.

\subsection{Inverted mass hierarchy}

\begin{figure}[t]
\begin{minipage}[t]{6.0cm}
\epsfig{figure=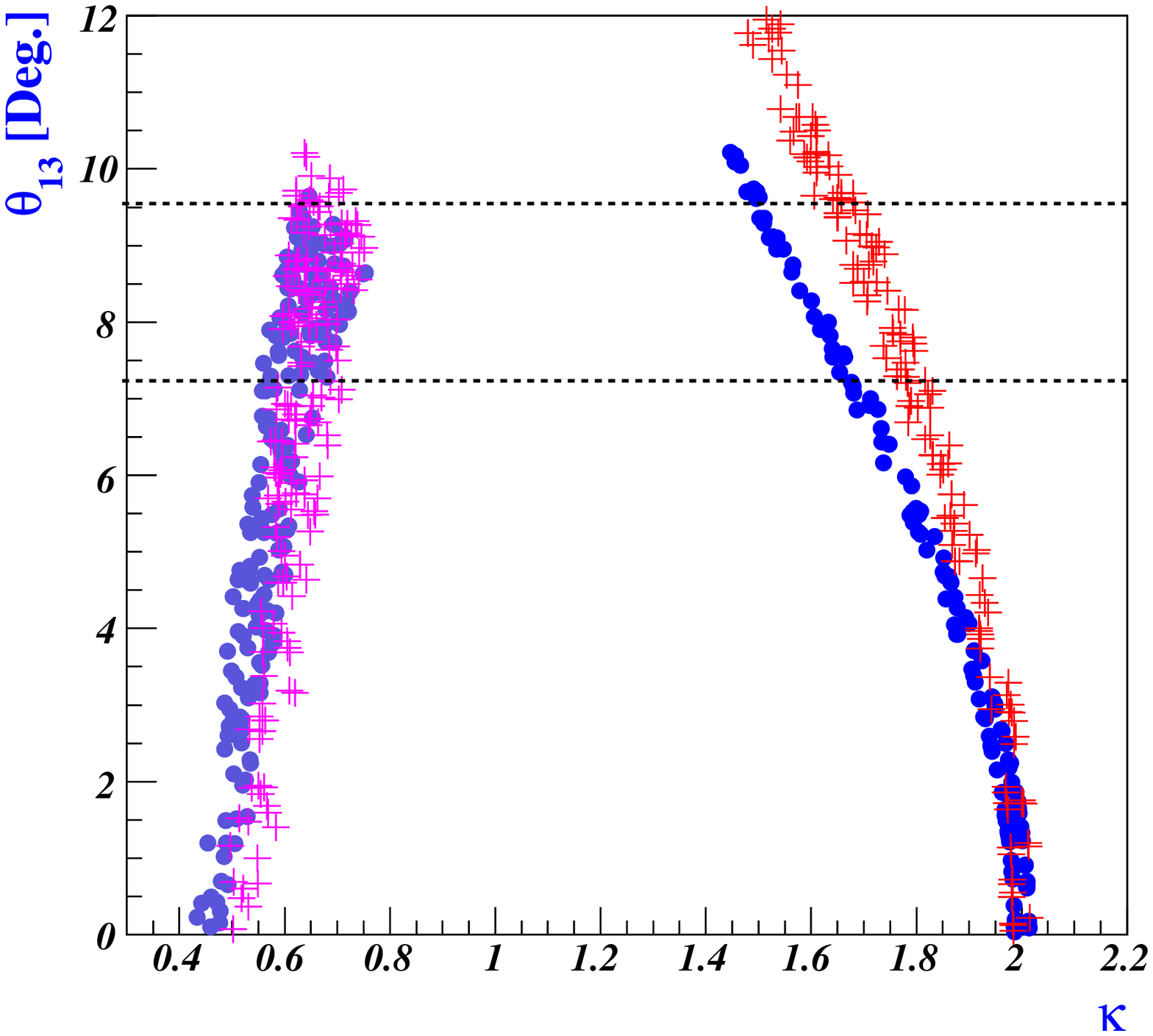,width=6.5cm,angle=0}
\end{minipage}
\hspace*{1.0cm}
\begin{minipage}[t]{6.0cm}
\epsfig{figure=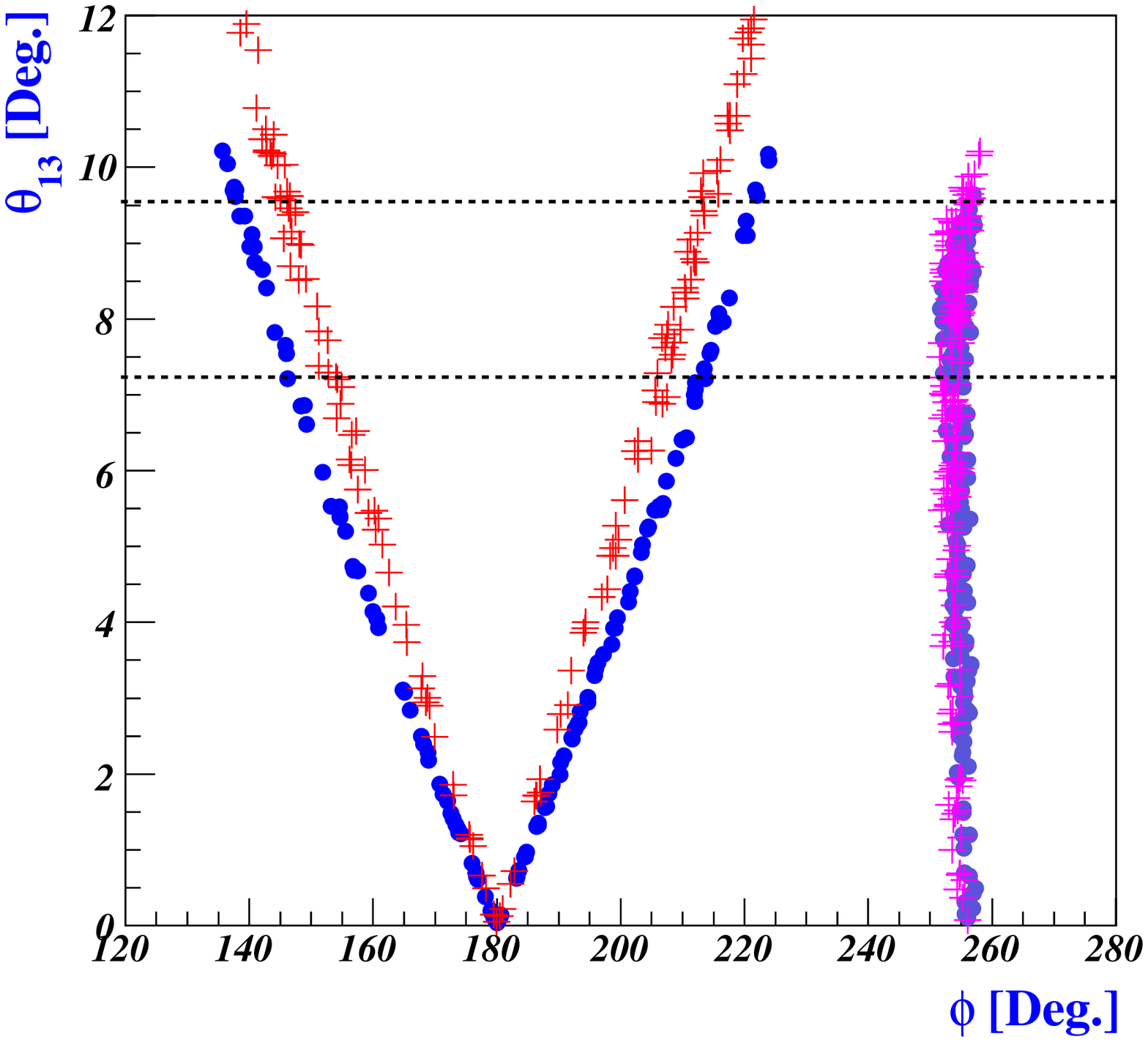,width=6.5cm,angle=0}
\end{minipage}\\
\begin{minipage}[t]{6.0cm}
\epsfig{figure=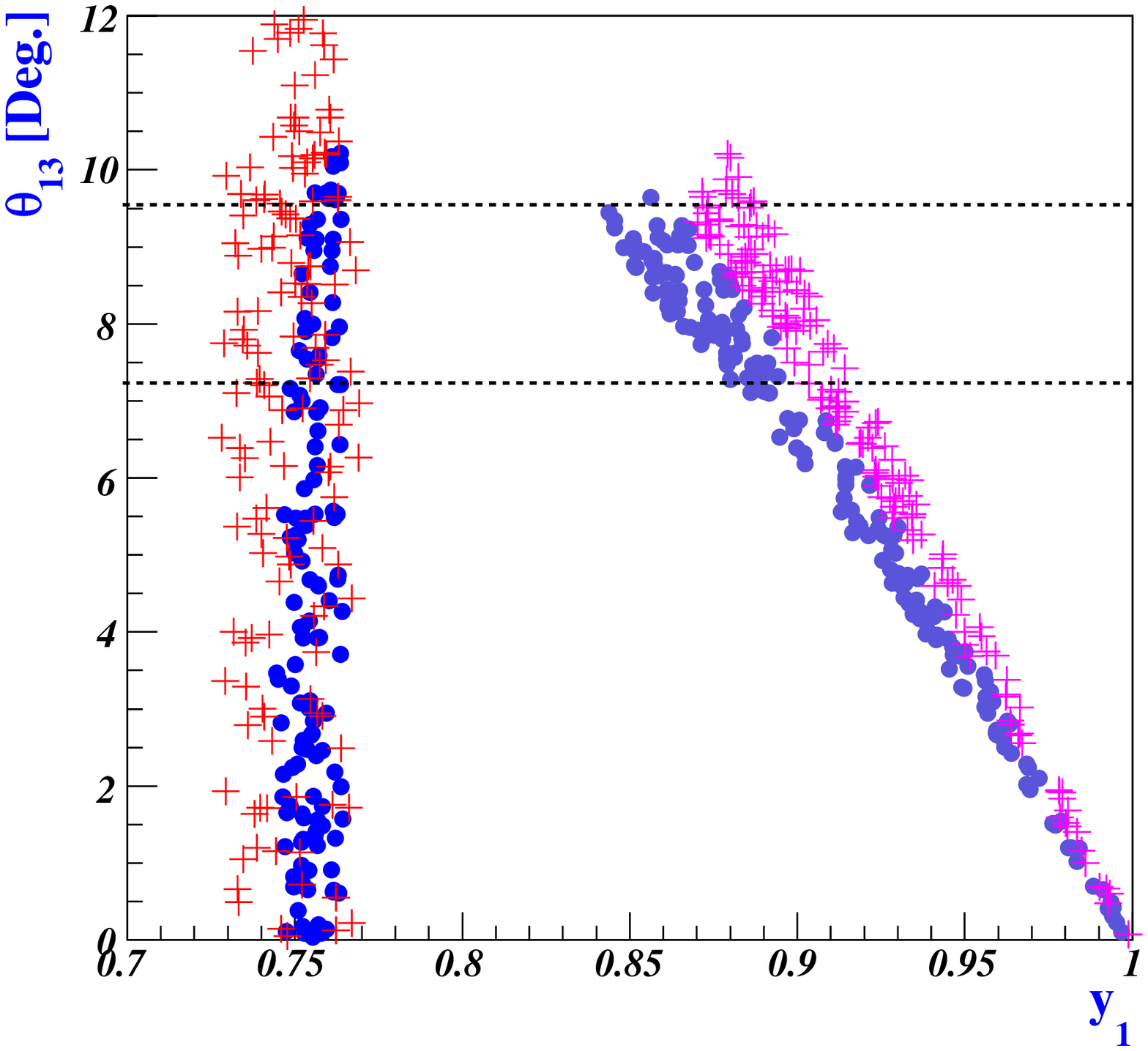,width=6.5cm,angle=0}
\end{minipage}
\hspace*{1.0cm}
\begin{minipage}[t]{6.0cm}
\epsfig{figure=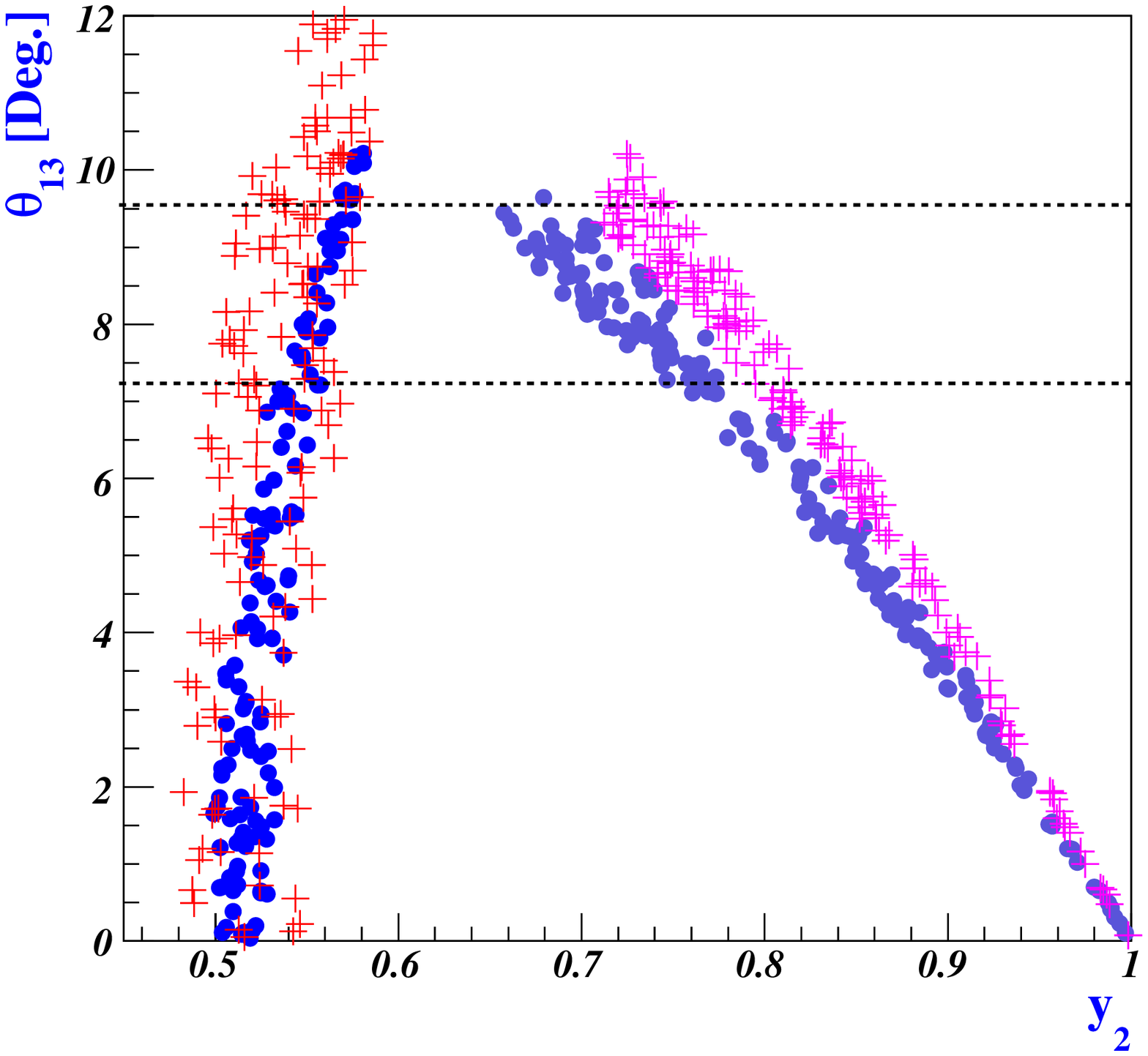,width=6.5cm,angle=0}
\end{minipage}
\caption{\label{FigB1}
Same as Fig.~\ref{FigA1} except for IMH, and $\bar{m}_{\eta_i}=100$ GeV and $10^{9}$ GeV correspond to the blue-type dots and red-type crosses data points, respectively.}
\end{figure}
\begin{figure}[t]
\begin{minipage}[t]{6.0cm}
\epsfig{figure=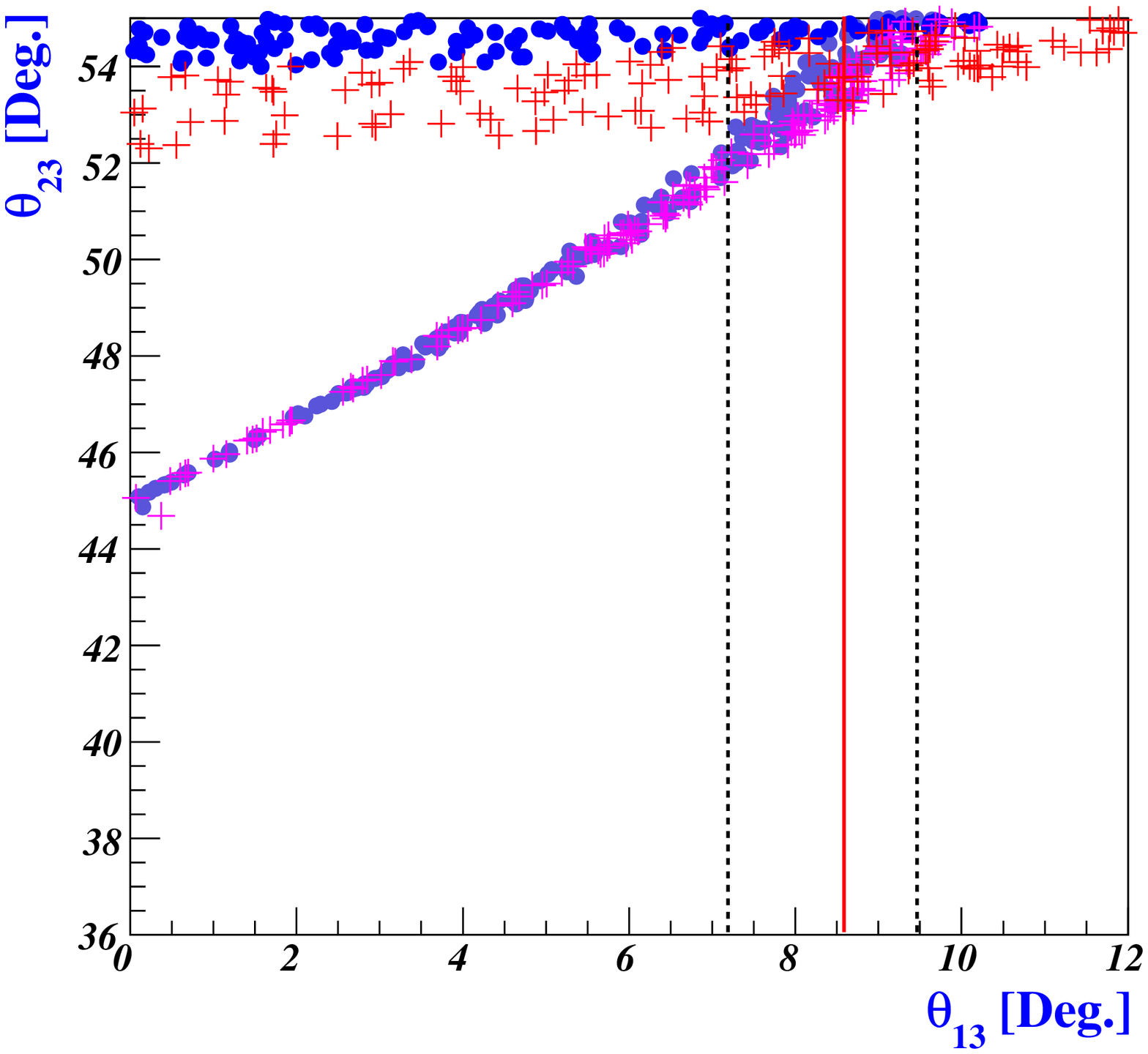,width=6.5cm,angle=0}
\end{minipage}
\hspace*{1.0cm}
\begin{minipage}[t]{6.0cm}
\epsfig{figure=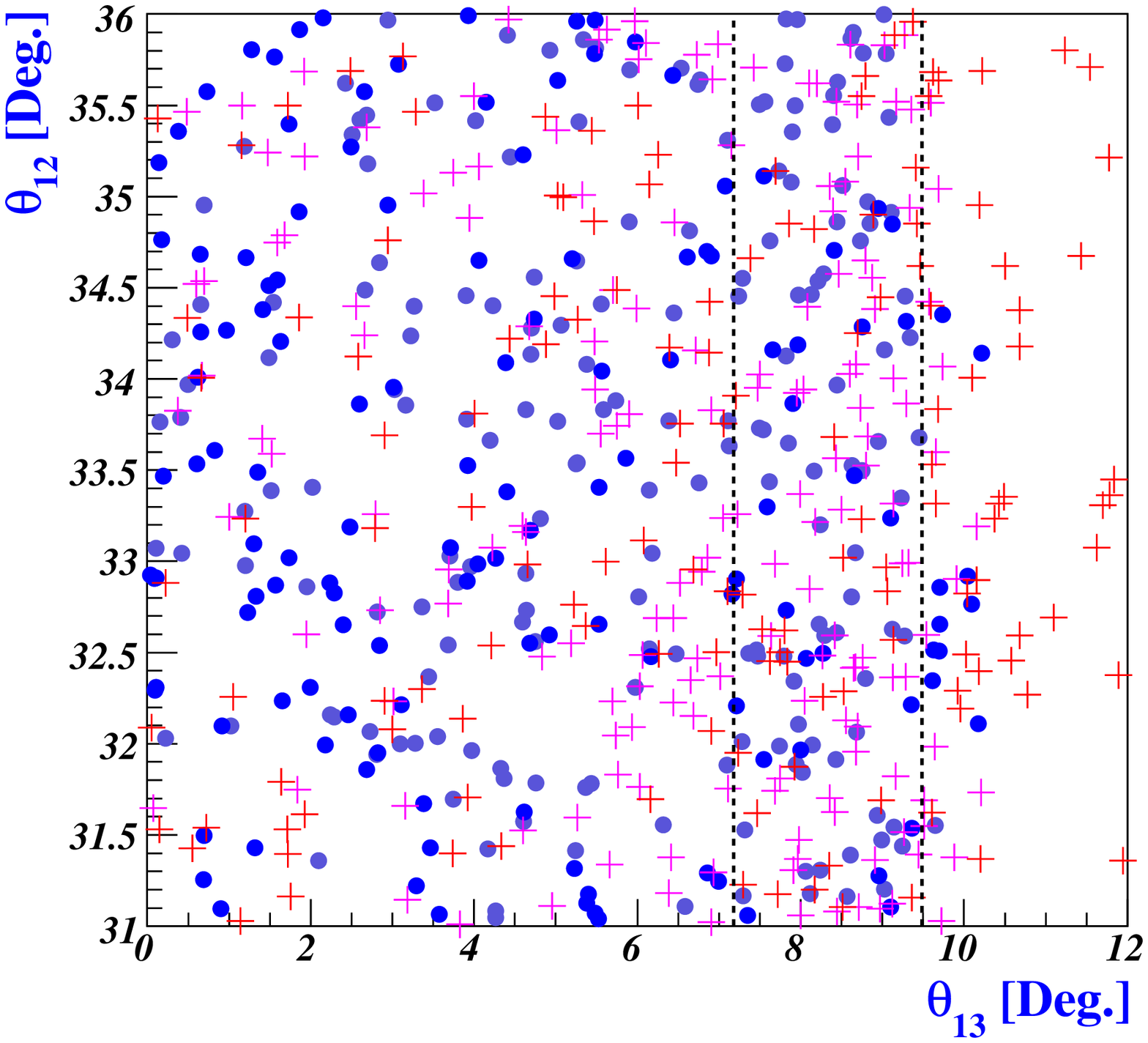,width=6.5cm,angle=0}
\end{minipage}
\caption{\label{FigB2}
Same as Fig.~\ref{FigA2} except for IMH.}
\end{figure}
\begin{figure}[b]
\begin{minipage}[t]{6.0cm}
\epsfig{figure=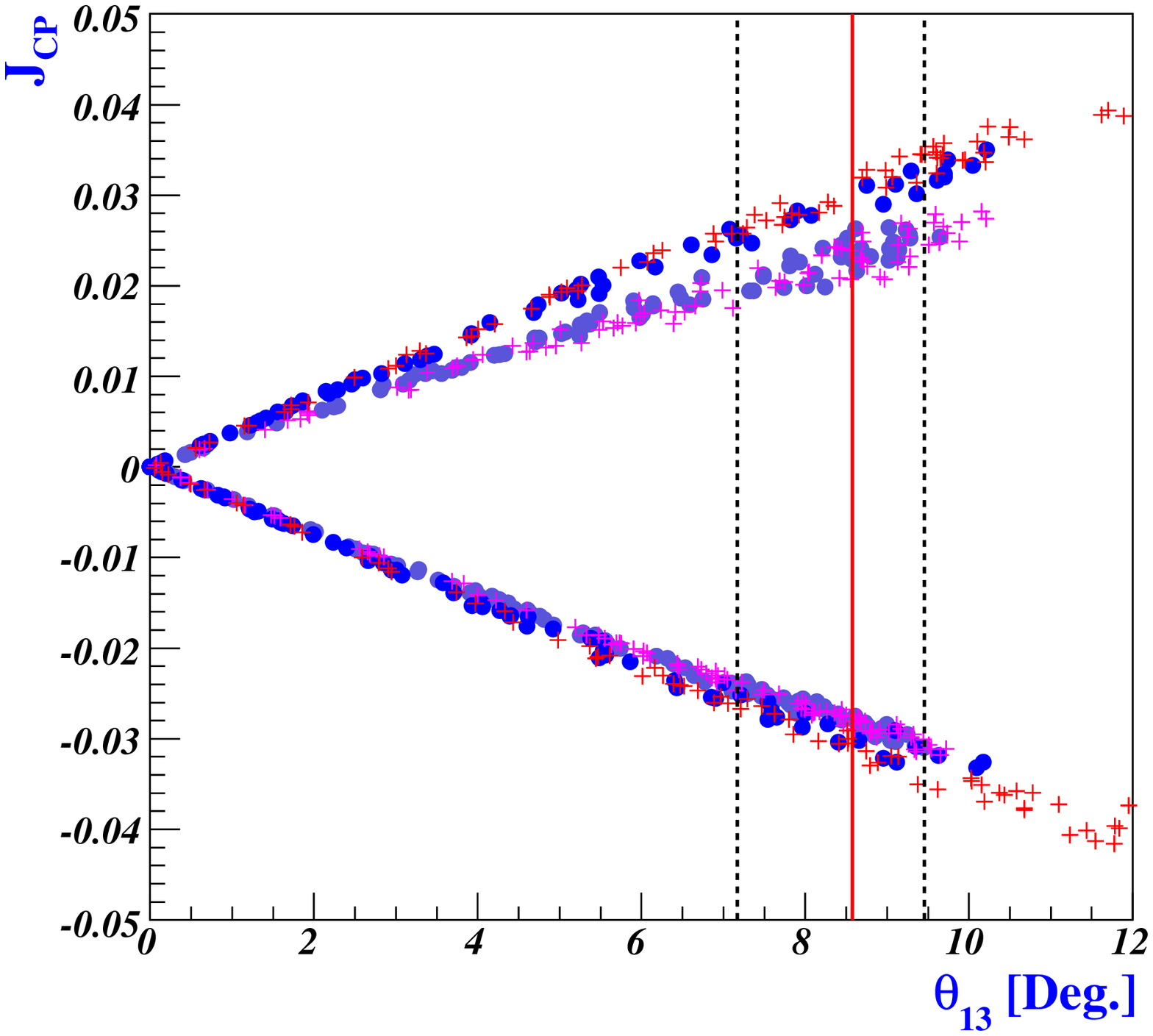,width=6.5cm,angle=0}
\end{minipage}
\hspace*{1.0cm}
\begin{minipage}[t]{6.0cm}
\epsfig{figure=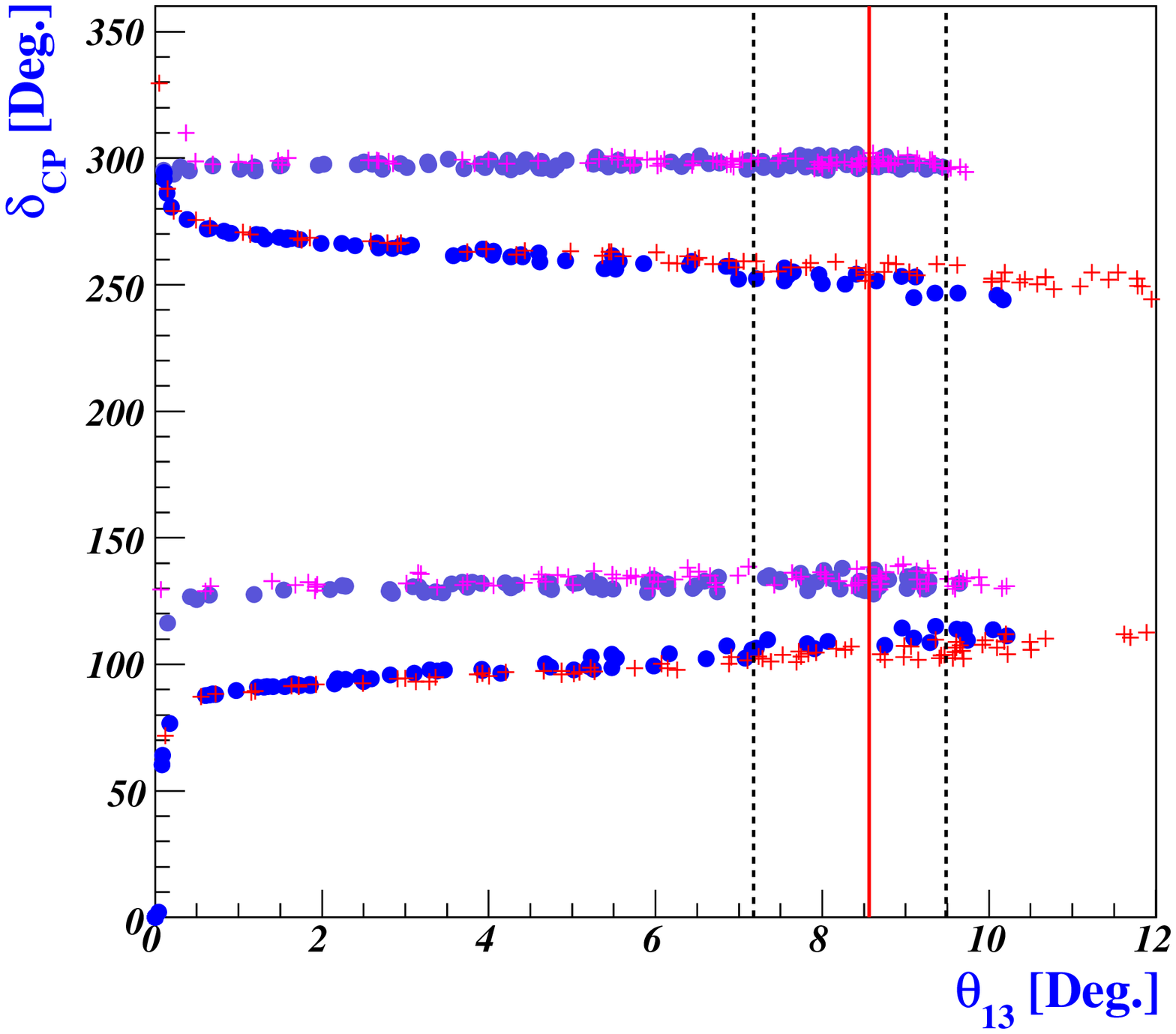,width=6.5cm,angle=0}
\end{minipage}
\caption{\label{FigB3}
Same as Fig.~\ref{FigA3} except for IMH.}
\end{figure}

Just as in NMH, using the formulas for the neutrino mixing parameters and our values of $M, y^{\nu}_{3}, v_{\Phi}, \lambda^{\Phi\eta}_{3}, \bar{m}_{\eta_i}$, we obtain the following allowed regions of the unknown model parameters: for  the case-I with $\bar{m}_{\eta_i}\simeq{\cal O}(v_{\Phi})$,
 \begin{eqnarray}
  &&0.4<\kappa<0.7~,\quad1.45<\kappa<2.05~,\quad0.74\lesssim y_{1}\lesssim0.77~,\quad0.84< y_{1}\lesssim1~,\nonumber\\
  &&0.5\lesssim y_{2}\lesssim0.57~,\quad0.66\lesssim y_{1}\lesssim1~,\quad135^{\circ}\lesssim\phi\lesssim220^{\circ}~,\quad250^{\circ}\lesssim\phi\lesssim260^{\circ}~,
  \label{input3}
 \end{eqnarray}
and for the case-II with $\bar{m}_{\eta_i}\simeq{\cal O}(v_{\chi})$,
 \begin{eqnarray}
  &&0.5<\kappa<0.75~,\quad1.45<\kappa\lesssim2~,\quad0.72<y_{1}\lesssim0.77~,\quad0.87\lesssim y_{1}\lesssim1~,\nonumber\\
  &&0.48\lesssim y_{2}\lesssim0.59~,\quad0.7\lesssim y_{1}\lesssim1~,\quad135^{\circ}\lesssim\phi\lesssim220^{\circ}~,\quad250^{\circ}\lesssim\phi\lesssim260^{\circ}~.
  \label{input4}
 \end{eqnarray}
For these parameter regions, we in turn investigate how the mixing angle $\theta_{13}$
depends on other parameters and how Dirac CP phase $\delta_{CP}$ can be determined
for the Case-I and II. Similar to NMH case, in Figs.~\ref{FigB1}-\ref{FigB3},
the data points represented by blue-type dots and red-type crosses indicate results
for the Case-I and Case-II, respectively.
The upper left and right panel in Fig.~\ref{FigB1} show how the mixing angle $\theta_{13}$ depends on the parameter $\kappa=\lambda^{s}_{\chi}v_{\chi}/M$ and the phase $\phi$, respectively; the lower left and right panel show how $\theta_{13}$ depends on the parameter $y_{1}$ and $y_{2}$, respectively.
The points located between two dashed lines in the plots are in consistent with
the values of $\theta_{13}$ from the global fits including the Daya Bay and RENO experiments at $3\sigma$ C.L.
%
Fig.~\ref{FigB2} shows how the estimated values of $\theta_{13}$ depend on the atmospheric and solar mixing angles, $\theta_{23}$ and $\theta_{12}$. In the left-plot of Fig.~\ref{FigB2}, we see that the measured range of $\theta_{13}$ at $3\sigma'$s can be achieved for $51^{\circ}\lesssim\theta_{23}\lesssim54^{\circ}$ (blue dots) and $54^{\circ}\lesssim\theta_{23}\lesssim55^{\circ}$ (bright-blue dots) for the Case-I, whereas
it can be achieved for  $52^{\circ}\lesssim\theta_{23}\lesssim55^{\circ}$ (red crosses) and hot-pink crosses $51^{\circ}\lesssim\theta_{23}\lesssim54^{\circ}$ (hot-pink crosses) for the Case-II.
Comparing two left-hand plots in Fig.~\ref{FigA2} and Fig.~\ref{FigB2}, we see that NMH prefers to
$\theta_{23}<45^{\circ}$, whereas IMH to $\theta_{23}>45^{\circ}$.
Thus, the type of mass hierarchy is strongly correlated with the octant of $\theta_{23}$ in our model.
Future determinations of the octant of $\theta_{23}$ and mass hierarchy would test our model.
The right-plot of Fig.~\ref{FigB2} shows that the predictions for $\theta_{13}$ do not strongly depend on $\theta_{12}$ in the allowed region.

We plot $J_{CP}$ as a function of the mixing angle
$\theta_{13}$ in the left-hand plot of Fig.~\ref{FigB3}. $J_{CP}$ has non-zero values for the measured range of $\theta_{13}$; $0.015\lesssim J_{CP}\lesssim0.035$ and $-0.036\lesssim J_{CP}\lesssim-0.022$, which could be tested in future experiments such as the upcoming long baseline neutrino oscillation ones, but it goes to zero for $\theta_{13}\rightarrow0$, which corresponds to $y_{2}\rightarrow1$ or $\phi\rightarrow\pi$ (or $\sin\psi_{1,2}\rightarrow0$), as can be seen from Eq.~(\ref{JCP1}). The right-hand plot of Fig.~\ref{FigB3} shows the behavior of the Dirac CP phase $\delta_{CP}$, where $\delta_{CP}$ can have discrete values around $100^{\circ}$, $135^{\circ}$, $255^{\circ}$ and $300^{\circ}$. As will be shown in Sec. VI (see, Fig.~\ref{FigB4}), the right magnitude of $\eta_{B}$ will restrict the information on $\delta_{CP}$ and it turns out that the values around $100^{\circ}$, $135^{\circ}$ and $300^{\circ}$ are consistent with leptogenesis.

\begin{figure}[t]
\begin{minipage}[t]{6.0cm}
\epsfig{figure=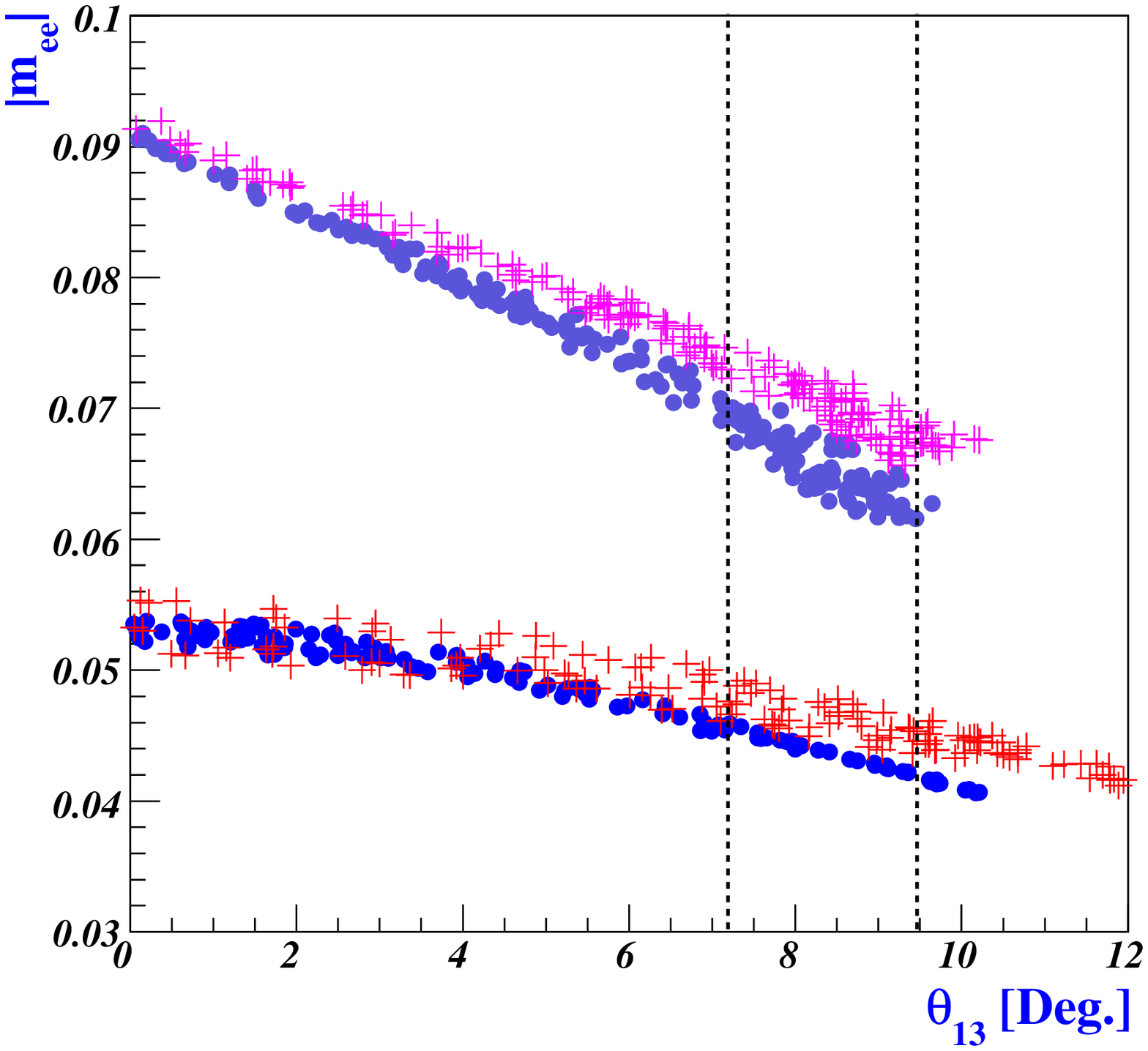,width=6.5cm,angle=0}
\end{minipage}
\hspace*{1.0cm}
\begin{minipage}[t]{6.0cm}
\epsfig{figure=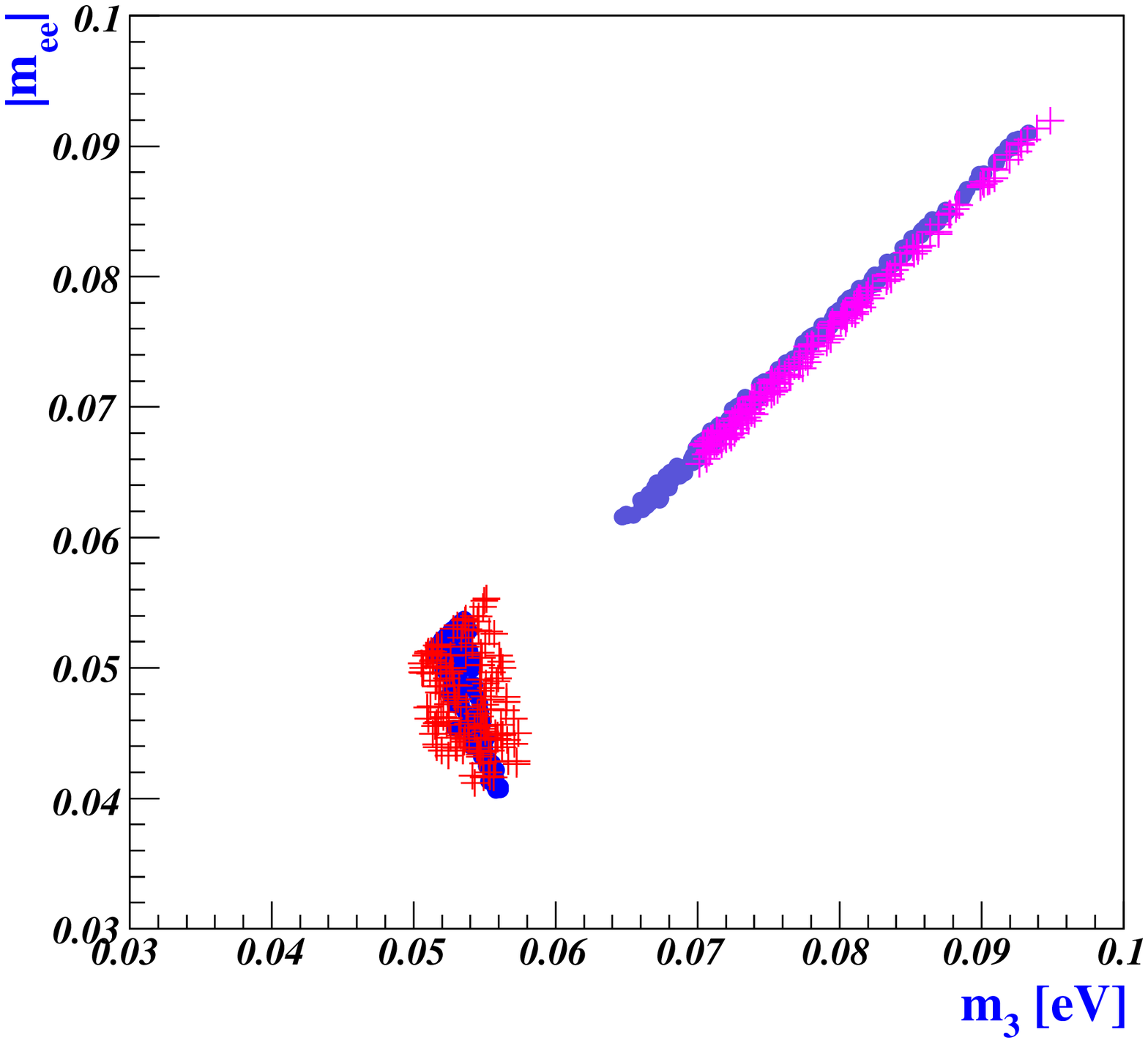,width=6.5cm,angle=0}
\end{minipage}
\caption{\label{FigB1p}
Same as Fig.~\ref{FigA1p} except for IMH  and the lightest neutrino mass $m_{3}$.}
\end{figure}
Similar to Fig.~\ref{FigA1p}, we plot the behavior of the effective neutrino mass $|m_{ee}|$ in terms of $\theta_{13}$ and the lightest neutrino mass $m_{3}$, respectively.
In the left plot of Fig.~\ref{FigB1p}, for the measured values of $\theta_{13}$ at $3\sigma$'s, the effective neutrino mass $|m_{ee}|$ can be in the ranges $0.042\lesssim|m_{ee}|[{\rm eV}]\lesssim0.048$ and $0.062\lesssim|m_{ee}|[{\rm eV}]\lesssim0.072$ (Case-I) and $0.044<|m_{ee}|[{\rm eV}]\lesssim0.05$ and $0.066<|m_{ee}|[{\rm eV}]\lesssim0.074$ (Case-II). The inverted mass hierarchy in our model predicts that the effective mass $|m_{ee}|$ is within the sensitivity of planned neutrinoless double-beta decay experiments.

\section{Leptogenesis and its link with low energy observables}

In addition to radiatively achieving the smallness of neutrino masses through one loop mediated by singlet heavy Majorana neutrinos,
in this model, the baryogenesis through so-called leptogenesis \cite{Fukugita:1986hr} can be realized from the decay of the singlet heavy Majorana neutrinos.
In early Universe, the decay of the right-handed heavy Majorana neutrino into a lepton and scalar boson is able to generate a nonzero lepton asymmetry, which in turn gets recycled into a baryon asymmetry through non-perturbative sphaleron processes.
We are in the energy scale\footnote{In order for baryogenesis via leptogenesis to be realized at around TeV scale, one needs either an enhancement of lepton asymmetry if the neutrino Yukawa coupling is very small~\cite{Pilaftsis:1997jf} or a dilution of washout factor if it is very large~\cite{Gu:2008yk}.} where $A_{4}$ symmetry~\footnote{There are some interesting papers on leptogenesis with flavor symmetry~\cite{A4lep}} is broken but the SM gauge group remains unbroken. So, both the charged and neutral scalars are physical.

The CP asymmetry generated through the interference between tree and one-loop diagrams for the decay of the heavy Majorana neutrino $N_{i}$ into $\eta$ and $(\nu,\ell_{\alpha})$ is given, for each lepton flavor $\alpha~(=e,\mu,\tau)$, by \cite{lepto2}
 \begin{eqnarray}\nonumber
  \varepsilon^{\alpha}_{i} &=& \frac{1}{8\pi(\tilde{Y}^{\dag}_{\nu}\tilde{Y}_{\nu})_{ii}}\sum_{j\neq i}{\rm
  Im}\Big\{(\tilde{Y}^{\dag}_{\nu}\tilde{Y}_{\nu})_{ij}(\tilde{Y}_{\nu})^{\ast}_{\alpha i}(\tilde{Y}_{\nu})_{\alpha j}\Big\}g\Big(\frac{M^{2}_{j}}{M^{2}_{i}}\Big),
 \label{cpasym1}
 \end{eqnarray}
where the function $g(x)$ is given by $g(x)= \sqrt{x}\Big[\frac{1}{1-x}+1-(1+x){\rm ln}\frac{1+x}{x}\Big]~.$
Here $i$ denotes a generation index and $\Gamma(N_i \rightarrow \cdot \cdot \cdot)$ stands for the decay width of the $i$th-generation right-handed neutrino. Another important ingredient carefully treated for successful leptogenesis is the wash-out factor $K^{\alpha}_{i}$ arisen mainly due to the inverse decay of the Majorana neutrino $N_{i}$ into the lepton flavor $\alpha(=e,\mu,\tau)$ \cite{Abada}. The explicit form of $K^{\alpha}_{i}$ is given by
 \begin{eqnarray}
  K^{\alpha}_{i} =\frac{\Gamma(N_{i}\rightarrow \eta \ell_{\alpha})}{H(M_{i})}
  =\frac{m_{\ast}}{M_{i}}(\tilde{Y}^{\ast}_{\nu})_{\alpha i}(\tilde{Y}_{\nu})_{\alpha i}~,
  \label{K-factor2}
 \end{eqnarray}
where $\Gamma(N_{i}\rightarrow \eta\ell_{\alpha})$ is the partial decay rate of the process $N_{i}\rightarrow\ell_{\alpha}+\eta$, $H(M_{i})=(4\pi^{3}g_{\ast}/45)^{\frac{1}{2}}M^{2}_{i}/M_{\rm Pl}$ with the Planck mass $M_{\rm Pl}=1.22\times10^{19}$ GeV is the Hubble parameter at temperature $T\simeq M_{i}$ and $m_{\ast}=\big(\frac{45}{2^{8}\pi^{5}g_{\ast}}\big)^{\frac{1}{2}}M_{\rm Pl}\simeq2.83\times10^{16}$ GeV with the effective number of degrees of freedom given by $g_{\ast}\simeq g_{\ast \rm SM}=106.75$.
And $\Gamma_{N_i}$ is a decay width of the process, $N_{i} \rightarrow \eta, \ell_{\alpha}$, defined as $\Gamma_{N_{i}}\equiv\sum_{\alpha}[\Gamma(N_{i}\rightarrow \ell_{\alpha}\eta)
  +\Gamma(N_{i}\rightarrow\overline{\ell}_{\alpha}\eta^{\dag})]=\frac{1}{8\pi}(\tilde{Y}^{\dag}_{\nu}\tilde{Y}_{\nu})_{ii}M_{i}$.

Since the factor $K^{\alpha}_{i}$ depends on both heavy right-handed neutrino mass $M_{i}$ and neutrino Yukawa coupling, the produced CP-asymmetries are strongly washed out for a rather large neutrino Yukawa coupling. In order for this enormously huge wash-out factor to be tolerated, we can consider higher scale leptogenesis. Assuming large and mild hierarchical neutrino Yukawa couplings, the lepton asymmetry and the wash-out factor are roughly given as $\varepsilon^{\alpha}_{i}\sim10^{-2}|y^{\nu}_{3}|^{2}$ and  $K^{\alpha}_{i}\sim m_{\ast}|y^{\nu}_{3}|/M$, respectively.
Finally, we get BAU  whose magnitude should be order of $10^{-10}$ from the product of $\varepsilon^{\alpha}_{i}$ and $K^{\alpha}_{i}$, and
can naively estimate the scale of $M$ by appropriately taking the magnitude of $y^{\nu}_{3}$,
for example,  $M\sim 10^{10}$ GeV for $|y^{\nu}_{3}|=1(0.01)$.
From our numerical analysis, we have found that it is impossible to reproduce the observed baryon asymmetry for $M_{i}\lesssim10^{9}$ GeV.
Therefore, it is necessary $M_{i}\gtrsim10^{9}$ GeV for successful leptogenesis, so that only the tau Yukawa interactions are supposed to be in thermal equilibrium.

Now, combining with Eqs.~(\ref{YnuT}), (\ref{YnuYnu}) and (\ref{cpasym1}), we get expressions for two flavored lepton asymmetries given by
 \begin{eqnarray}
  \varepsilon^{e\mu}_{1}&=&\frac{|y^{\nu}_{3}|^{2}}{12\pi}\left(\frac{(y^{2}_{1}-6y^{2}_{2})(1-2y^{2}_{1}+y^{2}_{2})}{3(1+4y^{2}_{1}+y^{2}_{2})}\sin\psi_{1}g(x_{12})-\frac{y^{2}_{2}(1-y^{2}_{2})}{4(1+4y^{2}_{1}+y^{2}_{2})}\sin\psi_{12}g(x_{13})\right)~,\nonumber\\
  \varepsilon^{\tau}_{1}&=&\frac{|y^{\nu}_{3}|^{2}}{48\pi}\left(-\frac{2(1-2y^{2}_{1}+y^{2}_{2})}{3(1+4y^{2}_{1}+y^{2}_{2})}\sin\psi_{1}g(x_{12})+\frac{1-y^{2}_{2}}{1+4y^{2}_{1}+y^{2}_{2}}\sin\psi_{12}g(x_{13})\right)~,\nonumber\\
  \varepsilon^{e\mu}_{2}&=&\frac{|y^{\nu}_{3}|^{2}}{48\pi}\left(\frac{(1-2y^{2}_{1}+y^{2}_{2})(y^{2}_{2}-2y^{2}_{1})}{3(1+y^{2}_{1}+y^{2}_{2})}\sin\psi_{1}g(x_{21})+\frac{y^{2}_{2}(1-y^{2}_{2})}{1+y^{2}_{1}+y^{2}_{2}}\sin\psi_{2}g(x_{23})\right)~,\nonumber\\
  \varepsilon^{\tau}_{2}&=&\frac{|y^{\nu}_{3}|^{2}}{48\pi}\left(\frac{1-2y^{2}_{1}+y^{2}_{2}}{3(1+y^{2}_{1}+y^{2}_{2})}\sin\psi_{1}g(x_{21})-\frac{1-y^{2}_{2}}{1+y^{2}_{1}+y^{2}_{2}}\sin\psi_{2}g(x_{23})\right)~,\nonumber\\
  \varepsilon^{e\mu}_{3}&=&-y^{2}_{2}\varepsilon^{\tau}_{3}=\frac{|y^{\nu}_{3}|^{2}y^{2}_{2}(1-y^{2}_{2})}{144\pi(1+y^{2}_{2})}\Big(\sin\psi_{12}g(x_{31})-2\sin\psi_{2}g(x_{32})\Big)~,
  \label{leptonasym01}
 \end{eqnarray}
where
 \begin{eqnarray}
  g(x_{12})&=&\frac{1}{a}\left[\frac{a^2}{a^{2}-1}+1-\frac{a^2+1}{a^2}\ln(a^2+1)\right]~,\nonumber\\
  g(x_{13})&=&\frac{b}{a}\left[\frac{a^2}{a^2-b^{2}}+1-\frac{a^2+b^2}{a^2}\ln\frac{a^2+b^2}{b^2}\right]~,\nonumber\\
  g(x_{21})&=&a\left[\frac{1}{1-a^{2}}+1-(1+a^2)\ln\frac{1+a^2}{a^2}\right]~,\nonumber\\
  g(x_{23})&=&b\left[\frac{1}{1-b^{2}}+1-(1+b^2)\ln\frac{1+b^2}{b^2}\right]~,\nonumber\\
  g(x_{31})&=&\frac{a}{b}\left[\frac{b^2}{b^{2}-a^{2}}+1-\frac{a^2+b^2}{b^2}\ln\frac{a^2+b^2}{a^2}\right]~,\nonumber\\
  g(x_{32})&=&\frac{1}{b}\left[\frac{b^2}{b^2-1}+1-\frac{b^2+1}{b^2}\ln(b^2+1)\right]~.
 \end{eqnarray}
As anticipated, in the limit of $y_{1,2}\rightarrow1$, the CP-asymmetries are going to vanish. Each CP asymmetry given in Eq.~(\ref{leptonasym01}) is weighted differently by the corresponding wash-out parameter given by Eq.~(\ref{K-factor2}), and thus expressed with different weight in the final form of the baryon asymmetry~\cite{Abada};
 \begin{eqnarray}
  \eta_{\emph{B}}&\simeq&
  -2\times10^{-2}\sum_{N_{i}}\Big[\varepsilon^{e\mu}_{i}\tilde{\kappa}\Big(\tiny{\frac{417}{589}}K^{e\mu}_{i}\Big)
  +\varepsilon^{\tau}_{i}\tilde{\kappa}\Big(\tiny\frac{390}{589}K^{\tau}_{i}\Big)\Big]~,
  \label{etaB}
 \end{eqnarray}
where $\varepsilon^{e\mu}_{i}=\varepsilon^{e}_{i}+\varepsilon^{\mu}_{i}$, $K^{e\mu}_{i}=K^{e}_{i}+K^{\mu}_{i}$ and the wash-out factor
 \begin{eqnarray}
  \tilde{\kappa}\simeq\Big(\frac{8.25}{K^{\alpha}_{i}}+\Big(\frac{K^{\alpha}_{i}}{0.2}\Big)^{1.16}\Big)^{-1}~.
 \end{eqnarray}
Here we have shown an expression for two flavored leptogenesis.  We note that $\psi_{1,2}$ and $g(x_{ij})$ in Eq.~(\ref{leptonasym01}) are the functions of the parameters $\phi$ and $\kappa$. While the values of parameters $y_{1,2}, \kappa$ and $\phi$ can be determined from the analysis as demonstrated in Sec.~IV and V, $y^{\nu}_{3}$ depends on the magnitude of $M$ through the relations defined in Eqs.~(\ref{K-factor2}) and (\ref{etaB}).

\begin{figure}[t]
\begin{minipage}[t]{6.0cm}
\epsfig{figure=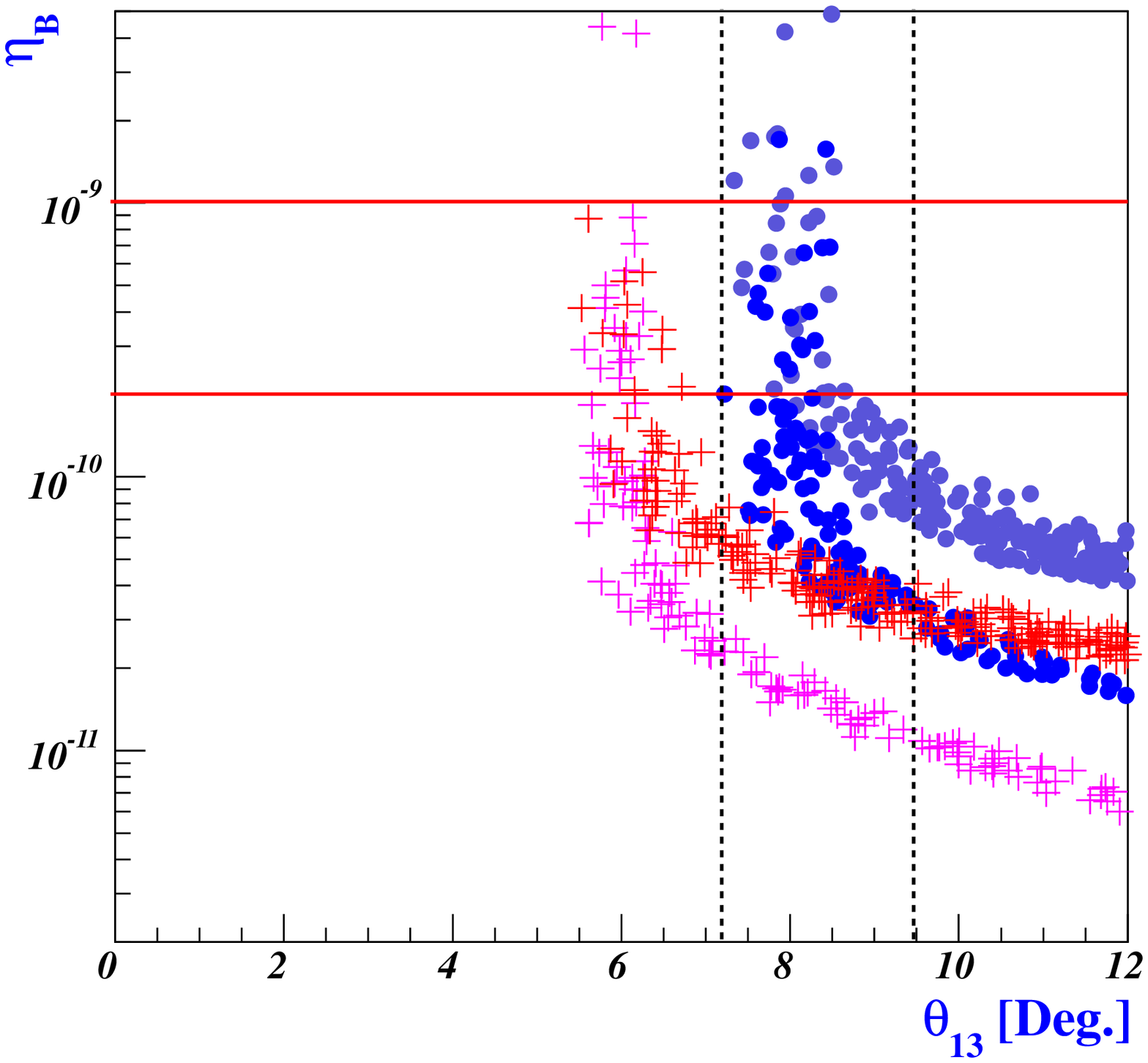,width=6.5cm,angle=0}
\end{minipage}
\hspace*{1.0cm}
\begin{minipage}[t]{6.0cm}
\epsfig{figure=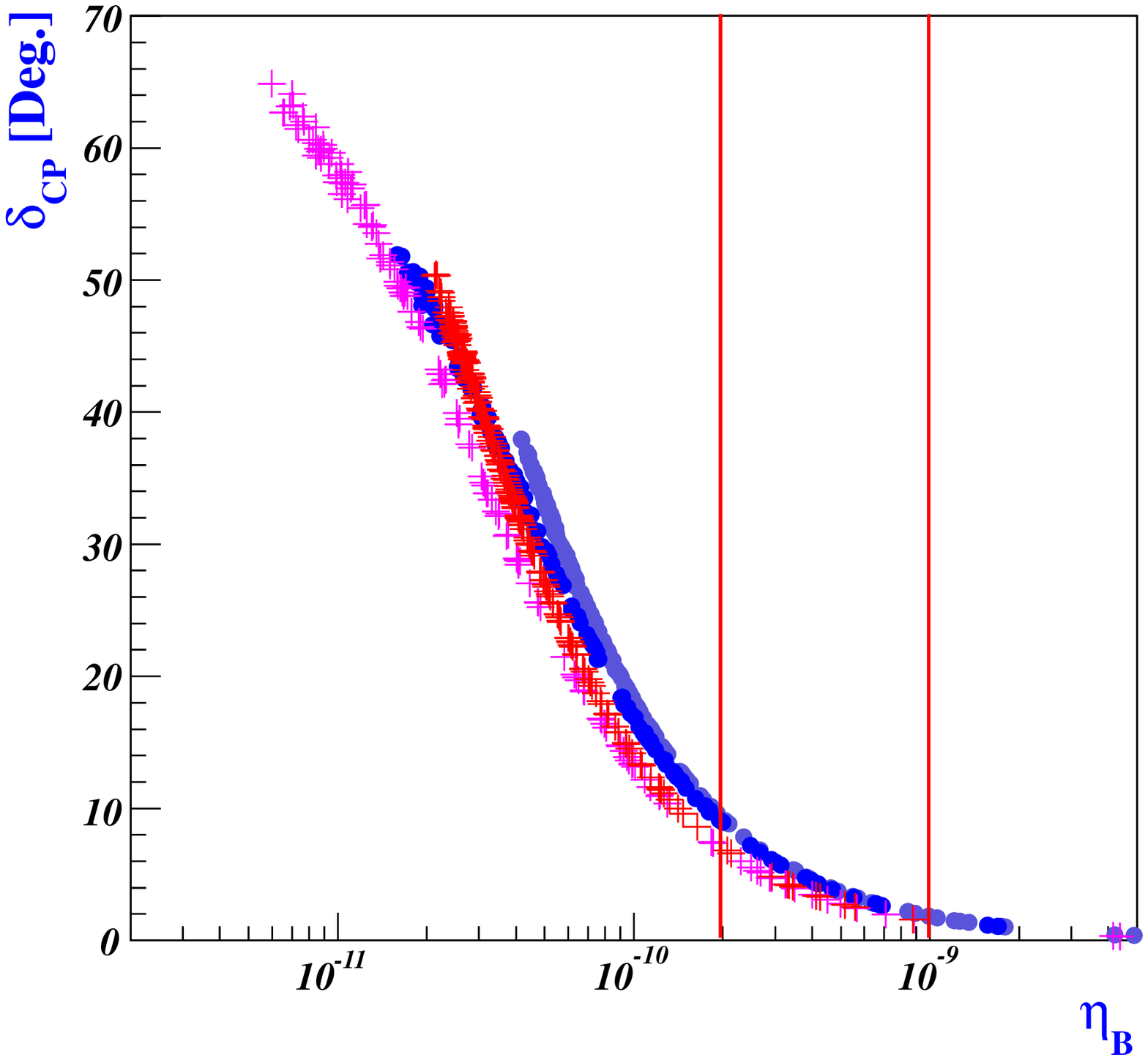,width=6.5cm,angle=0}
\end{minipage}
\caption{\label{FigA4}
 For NMH. Plot for the $\eta_{B}$ versus the mixing angle $\theta_{13}$ (left plot) and predictions for the Dirac CP phase $\delta_{CP}$ versus $\eta_{B}$ (right plot). Red-type crosses and blue-type dots data points correspond to $\bar{m}_{\eta_i}=10^{8}$ GeV and $500$ GeV, respectively.
The solid horizontal and vertical lines correspond to phenomenologically allowed regions $2\times10^{-10}\leq\eta_{B}\leq10^{-9}$, and the horizontal dotted lines correspond to the $3\sigma$ bounds given in Eq.~(\ref{expnu}).}
\end{figure}

For NMH, the predictions for $\eta_{B}$ as a function of $\theta_{13}$ (left plot) and for $\delta_{CP}$ as a function of $\eta_{B}$ (right plot) are shown, respectively, in Fig.~\ref{FigA4}.
As benchmarks, we take two parameter sets given in Table II.
The red crosses correspond to the former and blue dots to the latter. The solid horizontal and vertical lines correspond to experimentally allowed regions $2\times10^{-10}\leq\eta_{B}\leq10^{-9}$, and the horizontal dotted lines correspond to the $3\sigma$ bounds on neutrino data given in Eq.~(\ref{expnu}). The blue dots corresponding to $\bar{m}_{\eta_i}=500$ GeV satisfy the large $\theta_{13}$, and which in turn favor the Dirac CP phase ranged $1^{\circ}\lesssim\delta_{CP}\lesssim10^{\circ}$ (see the right plot in Fig.~\ref{FigA3}).

\begin{figure}[t]
\begin{minipage}[t]{6.0cm}
\epsfig{figure=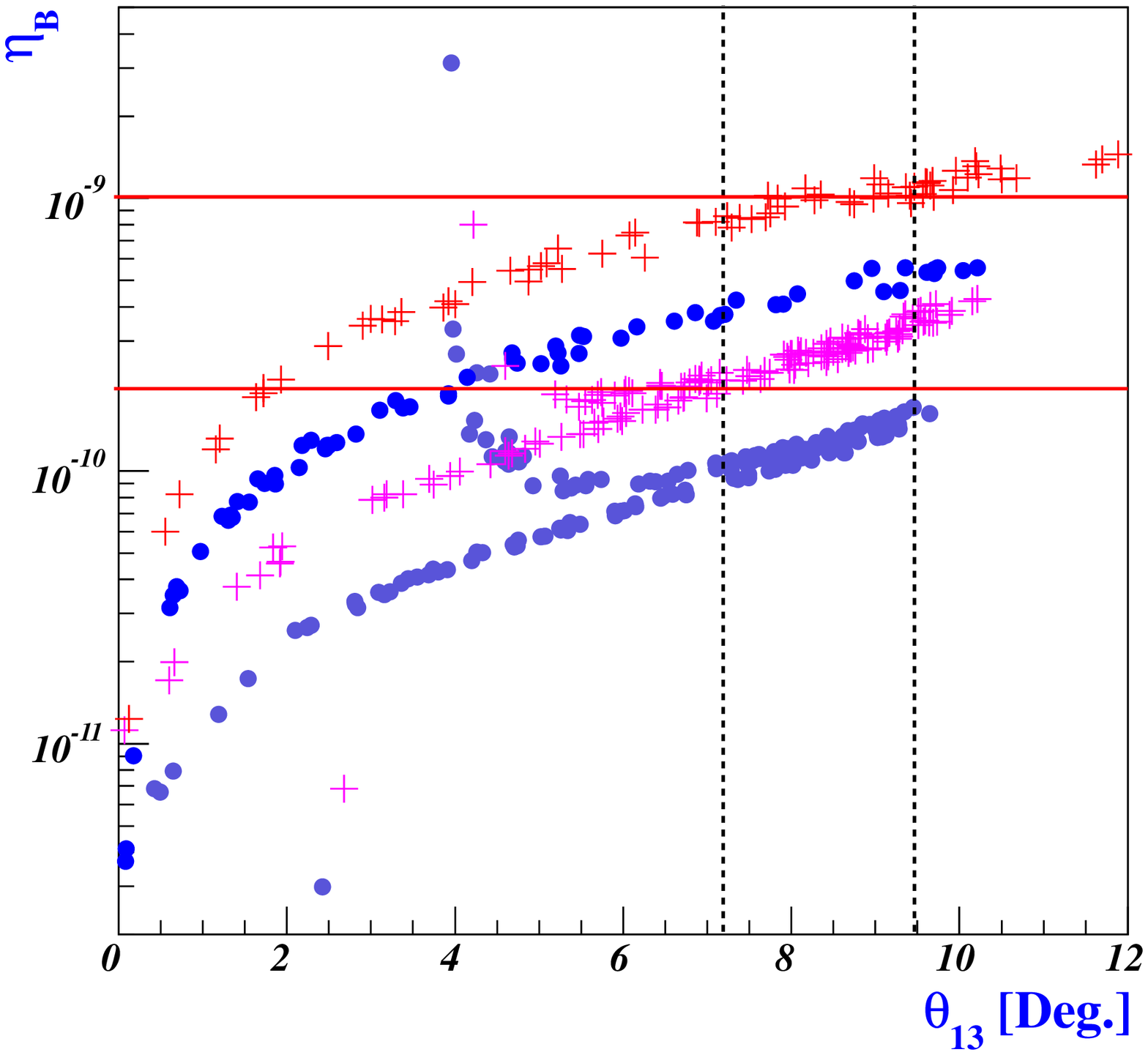,width=6.5cm,angle=0}
\end{minipage}
\hspace*{1.0cm}
\begin{minipage}[t]{6.0cm}
\epsfig{figure=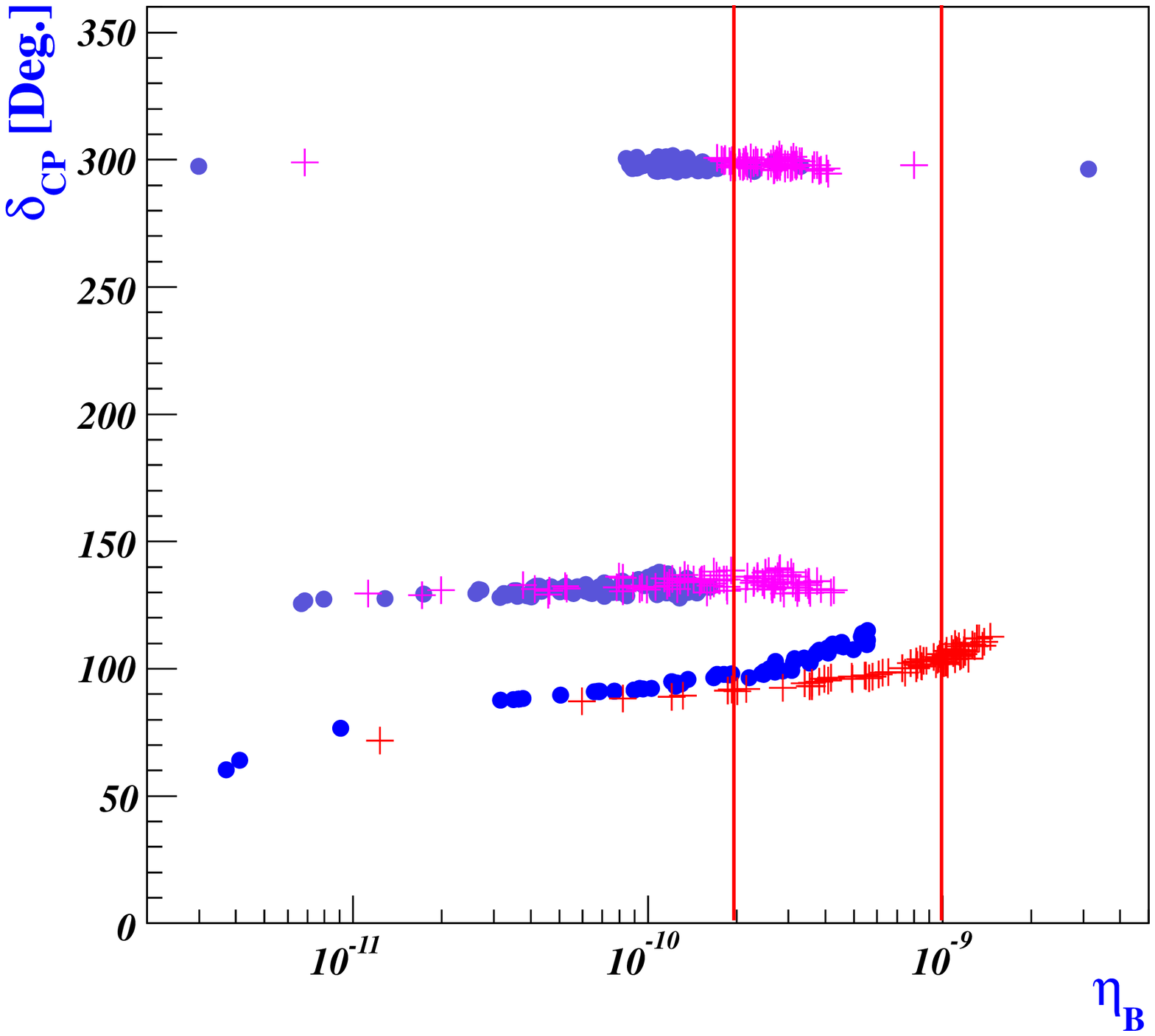,width=6.5cm,angle=0}
\end{minipage}
\caption{\label{FigB4}
 Same as Fig.~\ref{FigA4} except for IMH and $\bar{m}_{\eta_i}=10^{9}$ GeV and $100$ GeV correspond to the data points red-type crosses and blue-type dots, respectively.}
\end{figure}

For IMH, Fig.~\ref{FigB4} shows the predictions for $\eta_{B}$ as a function of $\theta_{13}$ (left plot) and for $\delta_{CP}$ as a function of $\eta_{B}$ (right plot), respectively.
As benchmarks,  we take two parameter sets given in Table II. The red crosses and blue dots correspond to the former and the latter, respectively. On the contrary to NMH, both the blue and red dots satisfy the large $\theta_{13}$,  which in turn favor the values of the Dirac CP phases around $\delta_{CP}\sim100^{\circ},135^{\circ},300^{\circ}$ (see the right plot in Fig.~\ref{FigB3}).

\section{Conclusion}

We have proposed a simple renormalizable model for the SCPV based on $SU(2)_{L}\times U(1)_{Y}\times A_{4}$ symmetry. We have introduced a right-handed neutrino $N_{R}$, a complex gauge singlet scalar $\chi$, and an $SU(2)_{L}$-doublet scalar $\eta$, all of which are $A_{4}$-triplets. In addition to the gauge and flavor symmetries, we have introduced an extra auxiliary $Z_{2}$ symmetry so that (i) a light neutrino mass could be generated though one loop diagram, (ii) vacuum alignment problem which occurs in the presence of two $A_{4}$-triplets could be naturally solved, and (iii) there could be a good dark matter candidate.
In our model CP is spontaneously broken at high energies, after breaking of flavor symmetry, by a complex vacuum expectation value of $A_{4}$-triplet and gauge singlet scalar field, leading to a natural source of low and high energy CP violation.
Then, we have investigated CP violation in the lepton sector and shown how the CP phases in PMNS could be arisen through spontaneous symmetry breaking mechanism.
And with compactified model parameters we have explained the smallness of neutrino masses and shown a mass texture displaying the mild hierarchy of neutrino mass. The light neutrino mass matrix is in the form of a deviated TBM generated through unequal neutrino Yukawa couplings, as can be seen in Figures~\ref{FigA1} and \ref{FigB1}. In the limiting case of equal active-neutrino Yukawa couplings, the mixing matrix recovers the exact TBM. In addition, we have shown that unequal neutrino Yukawa couplings can provide a source of high-energy CP violation, perhaps strong enough to be responsible for leptogenesis. Moreover, we have shown how to link between leptonic mixing and leptogenesis through the SCPV.

In a numerical example, where we have fixed the masses of dark matter,
first we have shown that the normal mass hierarchy favors relatively
large values of $\theta_{13}$, large deviations from maximality of $\theta_{23}<\pi/4$
and Dirac-CP phase $0^{\circ}\leq\delta_{CP}\leq50^{\circ}$
and $300^{\circ}\leq\delta_{CP}\leq360^{\circ}$, which is compatible with the global
analysis in $1\sigma$ experimental bounds. Second, we have shown that
within the measured values of $\theta_{13}$ the inverted case favors large deviations
from maximality of $\theta_{23}>\pi/4$ and Dirac-CP phase has discrete values around
$100^{\circ},135^{\circ},255^{\circ}$ and $300^{\circ}$. And in both cases we have
shown the effective neutrino mass $|m_{ee}|$ which is within the sensitivity of
planned neutrinoless double-beta decay experiments. Finally, with a successful
leptogenesis our numerical results give more predictive values on the Dirac CP phase:
for the normal mass hierarchy $1^{\circ}\lesssim\delta_{CP}\lesssim10^{\circ}$
and for inverted one $\delta_{CP}\sim100^{\circ},135^{\circ},300^{\circ}$.
Interestingly, future precise measurements of $\theta_{23}$, whether $\theta_{23}>45^{\circ}$ or $\theta_{23}<45^{\circ}$, will provide more information on $\delta_{CP}$ as well as the mass pattern for normal mass hierarchy or inverted one.

\acknowledgments{
The work of C.S.K. was supported in part by the National Research Foundation of Korea (NRF)
grant funded by Korea government of the Ministry of Education, Science and Technology (MEST)
(Grant No. 2011-0017430) and
(Grant No. 2011-0020333).
The work of S.K.K. was supported in part by NRF grant funded by Korean government of MEST (Grant No. 2011-0003287).}

\newpage

\appendix

\section{$A_{4}$}
Here we recall that $A_{4}$ is the symmetry group of the tetrahedron and the finite groups of the even permutation of four objects \cite{Altarelli:2005yp}.
The group $A_{4}$ has two generators $S$ and $T$, satisfying the relation $S^{2}=T^{3}=(ST)^{3}={\bf 1}$.
In the three-dimensional unitary representation, $S$ and $T$ are given by
 \begin{eqnarray}
 S={\left(\begin{array}{ccc}
 1 &  0 &  0 \\
 0 &  -1 & 0 \\
 0 &  0 &  -1
 \end{array}\right)}~,\qquad T={\left(\begin{array}{ccc}
 0 &  1 &  0 \\
 0 &  0 &  1 \\
 1 &  0 &  0
 \end{array}\right)}~.
 \label{generator}
 \end{eqnarray}
The group $A_{4}$ has four irreducible representations, one triplet ${\bf 3}$ and three singlets ${\bf 1}, {\bf 1}', {\bf 1}''$
with the multiplication rules ${\bf 3}\otimes{\bf 3}={\bf 3}_{s}\oplus{\bf 3}_{a}\oplus{\bf 1}\oplus{\bf 1}'\oplus{\bf 1}''$,
${\bf 1}'\otimes{\bf 1}''={\bf 1}$, ${\bf 1}'\otimes{\bf 1}'={\bf 1}''$
and ${\bf 1}''\otimes{\bf 1}''={\bf 1}'$.
Let's denote two $A_4$ triplets as $(a_{1}, a_{2}, a_{3})$ and $(b_{1}, b_{2}, b_{3})$,
then we have
 \begin{eqnarray}
  (a\otimes b)_{{\bf 3}_{\rm s}} &=& (a_{2}b_{3}+a_{3}b_{2}, a_{3}b_{1}+a_{1}b_{3}, a_{1}b_{2}+a_{2}b_{1})~,\nonumber\\
  (a\otimes b)_{{\bf 3}_{\rm a}} &=& (a_{2}b_{3}-a_{3}b_{2}, a_{3}b_{1}-a_{1}b_{3}, a_{1}b_{2}-a_{2}b_{1})~,\nonumber\\
  (a\otimes b)_{{\bf 1}} &=& a_{1}b_{1}+a_{2}b_{2}+a_{3}b_{3}~,\nonumber\\
  (a\otimes b)_{{\bf 1}'} &=& a_{1}b_{1}+\omega a_{2}b_{2}+\omega^{2}a_{3}b_{3}~,\nonumber\\
  (a\otimes b)_{{\bf 1}''} &=& a_{1}b_{1}+\omega^{2} a_{2}b_{2}+\omega a_{3}b_{3}~,
 \end{eqnarray}
where $\omega=e^{i2\pi/3}$ is a complex cubic-root of unity.

\section{The Higgs mass}

 Our model contains four Higgs doublets and three Higgs singlets.
And we can write, after the breaking of the flavor and electroweak symmetry,
 \begin{eqnarray}
  \Phi &=&
  {\left(\begin{array}{c}
  0 \\
  v+h
 \end{array}\right)}~,\qquad
 \eta_{j} =
  {\left(\begin{array}{c}
  \eta^{+}_{j} \\
  h_{j}+iA_{j}
 \end{array}\right)}~,~(j=1,2,3)\nonumber\\
 \chi_{1}&=& (v_{\chi}+\chi_{01})e^{i\phi}~,\chi_{2}=\chi_{02}~,\chi_{3}=\chi_{03}~,
  \label{Higgs}
 \end{eqnarray}
with the SM VEV $v=174$ GeV and $\eta^{+}_{j}\equiv(\eta^{-}_{j})^{\ast}$. Since the degree of freedom in $\Phi$ are eaten away by massive gauge bosons $W^{\pm}$ and $Z$, we can put $\varphi^{\pm}=0, A_{0}=0$, without loss of generality.
Here, we present the masses of physical scalar bosons, where the standard Higgs $h$ is mixed with $\chi_{0i}$, not with $h_{i}, A_{i}$. Since CP is conserved in our Lagrangian, then the couplings in the scalar potential given in Eq.~(\ref{potential}) are real.
The neutral Higgs boson mass matrix in the basis of $(h, \chi_{01}, \chi_{02}, \chi_{03}, h_{1}, A_{1}, h_{2}, h_{3}, A_{2}, A_{3})$ is block diagonalized due to $Z_{2}$ symmetry and CP conservation, which is given by
 \begin{eqnarray}
 {\emph{M}}^{2}_{\rm neutral}=
 {\left(\begin{array}{cccccccccc}
 m^{2}_{h} & m^{2}_{h\chi_1} & 0 & 0 & 0 & 0 & 0 & 0 & 0 & 0 \\
 m^{2}_{h\chi_1} & m^{2}_{\chi_1} & 0 & 0 & 0 & 0 & 0 & 0 & 0 & 0 \\
 0 & 0 & m^{2}_{\chi_2} & m^{2}_{\chi_2\chi_3} & 0 & 0 & 0 & 0 & 0 & 0 \\
 0 & 0 & m^{2}_{\chi_2\chi_3} & m^{2}_{\chi_3} & 0 & 0 & 0 & 0 & 0 & 0 \\
 0 & 0 & 0 & 0 & m^{2}_{h_1} & 0 & 0 & 0 & 0 & 0 \\
 0 & 0 & 0 & 0 & 0 &  m^{2}_{A_1} & 0 & 0 & 0 & 0 \\
 0 & 0 & 0 & 0 & 0 & 0 & m^{2}_{h_2} & m^{2}_{h_2h_3} & 0 & m^{2}_{h_2A_3} \\
 0 & 0 & 0 & 0 & 0 & 0 & m^{2}_{h_3h_2} & m^{2}_{h_3} & m^{2}_{h_3A_2} & 0 \\
 0 & 0 & 0 & 0 & 0 & 0 & 0 & m^{2}_{A_2h_3} & m^{2}_{A_2} & m^{2}_{A_2A_3} \\
 0 & 0 & 0 & 0 & 0 & 0 & m^{2}_{A_3h_2} & 0 & m^{2}_{A_3A_2} & m^{2}_{A_3}
 \end{array}\right)}~,
  \label{Higgsmass1}
 \end{eqnarray}
where the unprimed particles are not mass eigenstates, and mass parameters are given as
 \begin{eqnarray}
  m^{2}_{h}&=&4\lambda^{\Phi}v^{2}_{\Phi}~,\qquad m^{2}_{h\chi_1}=4v_{\Phi}v_{\chi}\lambda^{\Phi\chi}\cos2\phi~,\nonumber\\
  m^{2}_{\chi_1}&=&8v^{2}_{\chi}\left\{\tilde{\lambda}^{\chi}_{2}\cos2\phi+(\lambda^{\chi}_{1}+\lambda^{\chi}_{2})\cos4\phi\right\}~,\qquad m^{2}_{\chi_2\chi_3}=6v_{\chi}(\xi^{\chi}_{1}+\tilde{\xi}^{\chi}_{1})\cos\phi\nonumber\\
  m^{2}_{\chi_{2}}&=&m^{2}_{\chi}+v^{2}_{\chi}\Big(4\tilde{\lambda}^{\chi}_{3}+4\tilde{\lambda}^{\chi}_{4}-\tilde{\lambda}^{\chi}_{2}-(\tilde{\lambda}^{\chi}_{2}-4\tilde{\lambda}^{\chi}_{3}+4\tilde{\lambda}^{\chi}_{4}-4\lambda^{\chi}_{1}+2\lambda^{\chi}_{2}-8\lambda^{\chi}_{3})\cos2\phi\nonumber\\
  &-&\sqrt{3}\tilde{\lambda}^{\chi}_{2}\sin2\phi\Big)+2(v^{2}_{\Phi}\lambda^{\Phi\chi}+\mu^{2}_{\chi})~,\nonumber\\
  m^{2}_{\chi_{3}}&=&m^{2}_{\chi}+v^{2}_{\chi}\Big(4\tilde{\lambda}^{\chi}_{3}-4\tilde{\lambda}^{\chi}_{4}-\tilde{\lambda}^{\chi}_{2}-(\tilde{\lambda}^{\chi}_{2}-4\tilde{\lambda}^{\chi}_{3}-4\tilde{\lambda}^{\chi}_{4}-4\lambda^{\chi}_{1}+2\lambda^{\chi}_{2}-8\lambda^{\chi}_{3})\cos2\phi\nonumber\\
  &+&\sqrt{3}\tilde{\lambda}^{\chi}_{2}\sin2\phi\Big)+2(v^{2}_{\Phi}\lambda^{\Phi\chi}+\mu^{2}_{\chi})~,\nonumber\\
  m^{2}_{h_1}&=&v^{2}_{\Phi}(\lambda^{\eta\Phi}_{1}+\lambda^{\eta\Phi}_{2}+2\lambda^{\eta\Phi}_{3})+\mu^{2}_{\eta}+2v^{2}_{\chi}(\lambda^{\eta\chi}_{1}+\lambda^{\eta\chi}_{2})\cos2\phi~,\nonumber\\
  m^{2}_{A_1}&=&v^{2}_{\Phi}(\lambda^{\eta\Phi}_{1}+\lambda^{\eta\Phi}_{2}-2\lambda^{\eta\Phi}_{3})+\mu^{2}_{\eta}+2v^{2}_{\chi}(\lambda^{\eta\chi}_{1}+\lambda^{\eta\chi}_{2})\cos2\phi~,\nonumber\\
  m^{2}_{h_{2}}&=&v^{2}_{\Phi}(\lambda^{\eta\Phi}_{1}+\lambda^{\eta\Phi}_{2}+2\lambda^{\eta\Phi}_{3})+\mu^{2}_{\eta}+v^{2}_{\chi}\left((2\lambda^{\eta\chi}_{1}-\lambda^{\eta\chi}_{2})\cos2\phi-\sqrt{3}\lambda^{\eta\chi}_{2}\sin2\phi\right)~,\nonumber\\
  m^{2}_{h_{3}}&=&v^{2}_{\Phi}(\lambda^{\eta\Phi}_{1}+\lambda^{\eta\Phi}_{2}+2\lambda^{\eta\Phi}_{3})+\mu^{2}_{\eta}+v^{2}_{\chi}\left((2\lambda^{\eta\chi}_{1}-\lambda^{\eta\chi}_{2})\cos2\phi+\sqrt{3}\lambda^{\eta\chi}_{2}\sin2\phi\right)~,\nonumber\\
  m^{2}_{A_{2}}&=&v^{2}_{\Phi}(\lambda^{\eta\Phi}_{1}+\lambda^{\eta\Phi}_{2}-2\lambda^{\eta\Phi}_{3})+\mu^{2}_{\eta}+v^{2}_{\chi}\left((2\lambda^{\eta\chi}_{1}-\lambda^{\eta\chi}_{2})\cos2\phi-\sqrt{3}\lambda^{\eta\chi}_{2}\sin2\phi\right)~,\nonumber\\
  m^{2}_{A_{3}}&=&v^{2}_{\Phi}(\lambda^{\eta\Phi}_{1}+\lambda^{\eta\Phi}_{2}-2\lambda^{\eta\Phi}_{3})+\mu^{2}_{\eta}+v^{2}_{\chi}\left((2\lambda^{\eta\chi}_{1}-\lambda^{\eta\chi}_{2})\cos2\phi+\sqrt{3}\lambda^{\eta\chi}_{2}\sin2\phi\right)~,\nonumber\\
  m^{2}_{h_2h_3}&=& m^{2}_{A_2A_3}=2v_{\chi}\xi^{\eta\chi}_{1}\cos\phi~,\qquad m^{2}_{h_2A_3}= -2v_{\chi}\xi^{\eta\chi}_{2}\sin\phi=-m^{2}_{A_2h_3}~.
 \end{eqnarray}
Since the matrix in Eq.~(\ref{Higgsmass1}) is block diagonalized, it is easy to obtain the mass spectrum given as follows;
 \begin{eqnarray}
 m^{2}_{h'}&=& \frac{1}{2}\Big\{m^{2}_{h}+m^{2}_{\chi_1}-\sqrt{(m^{2}_{h}-m^{2}_{\chi_1})^{2}+4(m^{2}_{h\chi_1})^2}\Big\}~,\nonumber\\ m^{2}_{\chi'_1}&=&  \frac{1}{2}\Big\{m^{2}_{h}+m^{2}_{\chi_1}+\sqrt{(m^{2}_{h}-m^{2}_{\chi_1})^{2}+4(m^{2}_{h\chi_1})^2}\Big\}~,\nonumber\\ m^{2}_{\chi'_2}&=& m^{2}_{\chi_2}-m^{2}_{\chi_2\chi_3}~,\quad m^{2}_{\chi'_3}= m^{2}_{\chi_2}+m^{2}_{\chi_2\chi_3}~,\nonumber\\
 m^{2}_{h'_1}&=&m^{2}_{h_1}~,\qquad m^{2}_{A'_1}=m^{2}_{A_1} ~,\nonumber\\
 m^{2}_{h'_2}&=& v^{2}_{\Phi}\lambda^{\eta\Phi}_{12}+\mu^{2}_{\eta}+v^{2}_{\chi}(2\lambda^{\eta\chi}_{1}-\lambda^{\eta\chi}_{2})\cos2\phi-\Big\{4v^{4}_{\Phi}\lambda^{\eta\Phi2}_{3}\nonumber\\
 &+&v^{2}_{\chi}\left\{4\xi^{\eta\chi2}_{1}\cos^{2}\phi+4\xi^{\eta\chi2}_{2}\sin^{2}\phi+3v^{2}_{\chi}\lambda^{\eta\chi2}_{2}\sin^{2}2\phi\right\}-4v_{\chi}\lambda^{\eta\Phi}_{3}\sqrt{\Upsilon}\Big\}^{\frac{1}{2}}~,\nonumber\\
 m^{2}_{A'_2}&=&  v^{2}_{\Phi}\lambda^{\eta\Phi}_{12}+\mu^{2}_{\eta}+v^{2}_{\chi}(2\lambda^{\eta\chi}_{1}-\lambda^{\eta\chi}_{2})\cos2\phi+\Big\{4v^{4}_{\Phi}\lambda^{\eta\Phi2}_{3}\nonumber\\
 &+&v^{2}_{\chi}\left\{4\xi^{\eta\chi2}_{1}\cos^{2}\phi+4\xi^{\eta\chi2}_{2}\sin^{2}\phi+3v^{2}_{\chi}\lambda^{\eta\chi2}_{2}\sin^{2}2\phi\right\}-4v_{\chi}\lambda^{\eta\Phi}_{3}\sqrt{\Upsilon}\Big\}^{\frac{1}{2}}~,\nonumber\\
 m^{2}_{h'_3}&=&  v^{2}_{\Phi}\lambda^{\eta\Phi}_{12}+\mu^{2}_{\eta}+v^{2}_{\chi}(2\lambda^{\eta\chi}_{1}-\lambda^{\eta\chi}_{2})\cos2\phi-\Big\{4v^{4}_{\Phi}\lambda^{\eta\Phi2}_{3}\nonumber\\
 &+&v^{2}_{\chi}\left\{4\xi^{\eta\chi2}_{1}\cos^{2}\phi+4\xi^{\eta\chi2}_{2}\sin^{2}\phi+3v^{2}_{\chi}\lambda^{\eta\chi2}_{2}\sin^{2}2\phi\right\}+4v_{\chi}\lambda^{\eta\Phi}_{3}\sqrt{\Upsilon}\Big\}^{\frac{1}{2}}~,\nonumber\\
 m^{2}_{A'_3}&=&  v^{2}_{\Phi}\lambda^{\eta\Phi}_{12}+\mu^{2}_{\eta}+v^{2}_{\chi}(2\lambda^{\eta\chi}_{1}-\lambda^{\eta\chi}_{2})\cos2\phi+\Big\{4v^{4}_{\Phi}\lambda^{\eta\Phi2}_{3}\nonumber\\
 &+&v^{2}_{\chi}\left\{4\xi^{\eta\chi2}_{1}\cos^{2}\phi+4\xi^{\eta\chi2}_{2}\sin^{2}\phi+3v^{2}_{\chi}\lambda^{\eta\chi2}_{2}\sin^{2}2\phi\right\}+4v_{\chi}\lambda^{\eta\Phi}_{3}\sqrt{\Upsilon}\Big\}^{\frac{1}{2}}~,
  \label{Higgsmass2}
 \end{eqnarray}
 where $\lambda^{\eta\Phi}_{12}\equiv \lambda^{\eta\Phi}_{1} +\lambda^{\eta\Phi}_{2}$ and $\Upsilon=4\xi^{\eta\chi2}_{1}\cos^{2}\phi+3v^{2}_{\chi}\lambda^{\eta\chi2}_{2}\sin^{2}2\phi$.
Note here that the primed particles denote mass eigenstates.
And the charged Higgs boson mass matrix in the basis of $(\eta^{\pm}_{1} ,\eta^{\pm}_{2}, \eta^{\pm}_{3})$ is given as
 \begin{eqnarray}
 m^{2}_{\rm charged}=
 {\left(\begin{array}{ccc}
 m^{2}_{\eta^{\pm}_{1}} &  0 & 0  \\
 0 & m^{2}_{\eta^{\pm}_{2}} & 0  \\
 0 & 0 & m^{2}_{\eta^{\pm}_{3}}
 \end{array}\right)}~,
  \label{Higgsmass3}
 \end{eqnarray}
where
 \begin{eqnarray}
  m^{2}_{\eta^{\pm}_{1}}&=&\mu^{2}_{\eta}+v^{2}_{\Phi}\lambda^{\eta\Phi}_{1}+2v^{2}_{\chi}\left(\lambda^{\eta\chi}_{1}+\lambda^{\eta\chi}_{2}\right)\cos2\phi~,\nonumber\\
  m^{2}_{\eta^{\pm}_{2}}&=&\mu^{2}_{\eta}+v^{2}_{\Phi}\lambda^{\eta\Phi}_{1}+v^{2}_{\chi}\left\{(2\lambda^{\eta\chi}_{1}-\lambda^{\eta\chi}_{2})\cos2\phi-\sqrt{3}\lambda^{\eta\chi}_{2}\sin2\phi\right\}~,\nonumber\\
  m^{2}_{\eta^{\pm}_{3}}&=&\mu^{2}_{\eta}+v^{2}_{\Phi}\lambda^{\eta\Phi}_{1}+v^{2}_{\chi}\left\{(2\lambda^{\eta\chi}_{1}-\lambda^{\eta\chi}_{2})\cos2\phi+\sqrt{3}\lambda^{\eta\chi}_{2}\sin2\phi\right\}~.
  \label{chargedeta}
 \end{eqnarray}
Note here that since there is no mixing the unprimed particles denote mass eigenstates.

Using $m^{2}_{h_i}$, $m^{2}_{A_i}$ in Eq.~(\ref{Higgsmass1}), the expressions for $\bar{m}^{2}_{\eta_{i}}$ appeared in Eq.~(\ref{lownu1}) are
 \begin{eqnarray}
  \bar{m}^{2}_{\eta_{1}}&=&\mu^{2}_{\eta}+v^{2}_{\Phi}\lambda^{\eta\Phi}_{12}+v^{2}_{\chi}\left(2\lambda^{\eta\chi}_{1}-\lambda^{\eta\chi}_{2}\right)\cos2\phi
  =m^{2}_{\eta^{\pm}_{1}}+v^{2}_{\Phi}\lambda^{\eta\Phi}_{2}~,\nonumber\\
  \bar{m}^{2}_{\eta_{2}}&=&\mu^{2}_{\eta}+v^{2}_{\Phi}\lambda^{\eta\Phi}_{12}+v^{2}_{\chi}\left(2\lambda^{\eta\chi}_{1}-\lambda^{\eta\chi}_{2}\right)\cos2\phi-\sqrt{3}v^{2}_{\chi}\lambda^{\eta\chi}_{2}\sin2\phi=m^{2}_{\eta^{\pm}_{2}}+v^{2}_{\Phi}\lambda^{\eta\Phi}_{2}~,\nonumber\\
  \bar{m}^{2}_{\eta_{3}}&=&\mu^{2}_{\eta}+v^{2}_{\Phi}\lambda^{\eta\Phi}_{12}+v^{2}_{\chi}\left(2\lambda^{\eta\chi}_{1}-\lambda^{\eta\chi}_{2}\right)\cos2\phi+\sqrt{3}v^{2}_{\chi}\lambda^{\eta\chi}_{2}\sin2\phi=m^{2}_{\eta^{\pm}_{3}}+v^{2}_{\Phi}\lambda^{\eta\Phi}_{2}~.
  \label{Higgsmass4}
 \end{eqnarray}

\section{Parametrization of the neutrino mass matrix}

We parameterize the hermitian matrix $m_{\nu}m^{\dag}_{\nu}$ as follows:
 \begin{eqnarray}
 m_{\nu}m^{\dag}_{\nu}= m^{2}_{0}\left(\begin{array}{ccc}
  \tilde{A}y^{2}_{1} & y_{1}y_{2}\left(\frac{P-Q}{2}-i\frac{3(R+S)}{2}\right) & y_{1}\left(\frac{Q+P}{2}-i\frac{3(R-S)}{2}\right) \\
  y_{1}y_{2}\left(\frac{P-Q}{2}+i\frac{3(R+S)}{2}\right) & y^{2}_{2}\frac{F+G+K}{4} & y_{2}\left(\frac{F-G}{4}-i\frac{3D}{2}\right) \\
  y_{1}\left(\frac{Q+P}{2}+i\frac{3(R-S)}{2}\right) & y_{2}\left(\frac{F-G}{4}+i\frac{3D}{2}\right) & \frac{F+G-K}{4}
  \end{array}\right).~\nonumber
 \end{eqnarray}
 All parameters appearing here are real, and equal to
 \begin{eqnarray}
  \tilde{A}&=&(1+y^{2}_{1}+y^{2}_{2})f(z_{2})^{2}+(1+4y^{2}_{1}+y^{2}_{2})\left(\frac{f(z_{1})}{a^{2}}\right)^{2}-\frac{2(1-2y^{2}_{1}+y^{2}_{2})f(z_{1})f(z_{2})\cos\psi_{1}}{a}~,\nonumber\\
  F&=&4(1+y^{2}_{1}+y^{2}_{2})f(z_{2})^{2}+(1+4y^{2}_{1}+y^{2}_{2})\left(\frac{f(z_{1})}{a^{2}}\right)^{2}+4\frac{(1-2y^{2}_{1}+y^{2}_{2})f(z_{1})f(z_{2})\cos\psi_{1}}{a}~,\nonumber\\
  P&=&2(1+y^{2}_{1}+y^{2}_{2})f(z_{2})^{2}-(1+4y^{2}_{1}+y^{2}_{2})\left(\frac{f(z_{1})}{a^{2}}\right)^{2}-\frac{(1-2y^{2}_{1}+y^{2}_{2})f(z_{1})f(z_{2})\cos\psi_{1}}{a}~,\nonumber\\
  G&=& 9(1+y^{2}_{2})\left(\frac{f(z_{3})}{b}\right)^{2}~,\qquad \qquad R=(1-2y^{2}_{1}+y^{2}_{2})\frac{f(z_{1})f(z_{2})\sin\psi_{1}}{a}~,\nonumber\\
  K&=&6(1-y^{2}_{2})\frac{f(z_{3})}{b}\left\{\frac{f(z_{1})}{a}\cos\psi_{12}+2f(z_{2})\cos\psi_{2}\right\}~,\nonumber\\
  Q&=&3(1-y^{2}_{2})\frac{f(z_{3})}{b}\left\{\frac{f(z_{1})}{a}\cos\psi_{12}-f(z_{2})\cos\psi_{2}\right\}~,\nonumber\\
  D&=&(1-y^{2}_{2})\frac{f(z_{3})}{b}\left\{\frac{f(z_{1})}{a}\sin\psi_{12}-2f(z_{2})\sin\psi_{2}\right\}~,\nonumber\\
  S&=&(1-y^{2}_{2})\frac{f(z_{3})}{b}\left\{\frac{f(z_{1})}{a}\sin\psi_{12}+f(z_{2})\sin\psi_{2}\right\}~,
 \label{MMele}
 \end{eqnarray}
where $\psi_{ij}\equiv\psi_{i}-\psi_{j}$.
In Eq.~(\ref{sol1}) the parameters $\Psi_{1},\Psi_{2},\Psi_{3}$ are defined by
 \begin{eqnarray}
  \Psi_{1}&=&c^{2}_{13}y^{2}_{1}\tilde{A}+s^{2}_{13}\Psi_{3}
  -\frac{y_{1}\sin2\theta_{13}}{2}\Big[c_{23}\left((Q+P)\cos\delta_{CP}+3(R-S)\sin\delta_{CP}\right)\nonumber\\
    &+&s_{23}y_{2}\left((P-Q)\cos\delta_{CP}+3(R+S)\sin\delta_{CP}\right)\Big]\Big\}\nonumber\\
  \Psi_{2}&=&\frac{1}{4}\left\{y^{2}_{2}c^{2}_{23}(F+G+K)+s^{2}_{23}(F+G-K)-y_{2}(F-G)\sin2\theta_{23}\right\}~,\nonumber\\
  \Psi_{3}&=&\frac{1}{4}\left\{y^{2}_{2}s^{2}_{23}(F+G+K)+c^{2}_{23}(F+G-K)+y_{2}(F-G)\sin2\theta_{23}\right\}~.
 \label{para1}
 \end{eqnarray}



\begin{thebibliography}{99}
\def\plb#1#2#3{Phys.\ Lett.\       {\bf B#1}, (#3) #2}
\def\npb#1#2#3{Nucl.\ Phys.\       {\bf B#1}, (#3) #2}
\def\prd#1#2#3{Phys.\ Rev.\        {\bf D#1}, (#3) #2}
\def\prl#1#2#3{Phys.\ Rev.\ Lett.\ {\bf #1},  (#3) #2}
\def\mpl#1#2#3{Mod.\ Phys.\ Lett.\ {\bf A#1}, (#3) #2}
\def\rep#1#2#3{Phys.\ Rep.\        {\bf #1},  (#3) #2}
\def\sci#1#2#3{Science             {\bf #1},  (#3) #2}
\def\astro#1#2#3{Astrophys.\ J.\   {\bf #1},  (#3) #2}
\def\epj#1#2#3{Eur.\ Phys.\ J.  {\bf C#1},  (#3) #2}
\def\jhep#1#2#3{JHEP               {\bf #1},  (#3) #2}
\def\jpg#1#2#3{J.\ Phys.\        {\bf G#1},  (#3) #2}
\def\ijmp#1#2#3{Int.\ J.\ Mod.\ Phys.\ {\bf #1},  (#3) #2}
\def\ptp#1#2#3{Prog.\ Theor.\ Phys.\ {\bf #1},  (#3) #2}

\bibitem{Farrar:1993hn}
  G.~R.~Farrar and M.~E.~Shaposhnikov,
  Phys.\ Rev.\ D {\bf 50}, 774 (1994)  [hep-ph/9305275];  
  M.~B.~Gavela, P.~Hernandez, J.~Orloff, O.~Pene and C.~Quimbay,
  Nucl.\ Phys.\ B {\bf 430}, 382 (1994)  [hep-ph/9406289];  
  P.~Huet and E.~Sather,
  Phys.\ Rev.\ D {\bf 51}, 379 (1995)  [hep-ph/9404302].  


\bibitem{CKM}
N. Cabibbo, Phys. Rev. Lett. {\bf 10}, 531 (1963);
  M.~Kobayashi and T.~Maskawa,
   Prog.\ Theor.\ Phys.\  {\bf 49}, 652 (1973).

\bibitem{Lee:1973iz}
  T.~D.~Lee,
  Phys.\ Rev.\ D {\bf 8}, 1226 (1973);  
T.~D.~Lee,
  Phys.\ Rept.\  {\bf 9}, 143 (1974).  

\bibitem{Branco:1999fs}
  G.~C.~Branco, L.~Lavoura and J.~P.~Silva,
  \emph{CP Violation},  Int.\ Ser.\ Monogr.\ Phys.\  {\bf 103} (Oxford University Press, 1999);
G.~C.~Branco, R.~G.~Felipe and F.~R.~Joaquim,
  Rev.\ Mod.\ Phys.\  {\bf 84}, 515 (2012)  [arXiv:1111.5332 [hep-ph]];
G.~C.~Branco, R.~Gonzalez Felipe, F.~R.~Joaquim and H.~Serodio,
  Phys.\ Rev.\ D {\bf 86}, 076008 (2012)  [arXiv:1203.2646 [hep-ph]].

\bibitem{Ma:2001dn}
  E.~Ma and G.~Rajasekaran, Phys.\ Rev.\  D {\bf 64}, 113012 (2001) [arXiv:hep-ph/0106291].

\bibitem{mutau}
 T. Fukuyama and H. Nishiura, arXiv:hep-ph/9702253;
 R. N. Mohapatra and S. Nussinov, Phys. Rev. {\bf D 60}, 013002 (1999);
 E. Ma and M. Raidal, Phys. Rev. Lett. {\bf 87}, 011802 (2001);
 C. S. Lam, [arXiv:hep-ph/0104116]; T. Kitabayashi and M. Yasue, Phys.Rev. D {\bf 67} 015006 (2003);
 W. Grimus and L. Lavoura, arXiv:hep-ph/0305046; 0309050;W. Grimus and L. Lavoura, Phys.\ Lett.\ B {\bf 572}, 189 (2003);
 Y. Koide, Phys.Rev. D {\bf 69}, 093001 (2004); A. Ghosal, hep-ph/0304090;
 W.~Grimus and L.~Lavoura, J.\ Phys.\ G {\bf 30}, 73 (2004);
 R.~N.~Mohapatra and W.~Rodejohann,
  Phys.\ Rev.\ D {\bf 72}, 053001 (2005) [hep-ph/0507312];
 Y.~H.~Ahn, S.~K.~Kang, C.~S.~Kim and J.~Lee,
  Phys.\ Rev.\ D {\bf 73}, 093005 (2006)
  [hep-ph/0602160];
 Y.~H.~Ahn, S.~K.~Kang, C.~S.~Kim and  J.~Lee,
  Phys.\ Rev.\ D {\bf 75}, 013012 (2007)
  [hep-ph/0610007].

\bibitem{loopnu}
  E.~Ma,
  Mod.\ Phys.\ Lett.\  A {\bf 21}, 1777 (2006)
  [arXiv:hep-ph/0605180];
P.~Fileviez Perez and M.~B.~Wise,
  Phys.\ Rev.\ D {\bf 80}, 053006 (2009)  [arXiv:0906.2950 [hep-ph]].

\bibitem{PDG}
  J. Beringer et al. (Particle Data Group), Phys. Rev. D86, 010001 (2012).

\bibitem{Ahn:2012tv}
  Y.~H.~Ahn, S.~K.~Kang,
  Phys.\ Rev.\ D {\bf 86}, 093003 (2012)  [arXiv:1203.4185 [hep-ph]].  

\bibitem{Ahn:2011yj}
  Y.~H.~Ahn, H.~-Y.~Cheng and S.~Oh,
  Phys.\ Rev.\ D {\bf 83}, 076012 (2011)
  [arXiv:1102.0879 [hep-ph]].

\bibitem{A4}
  X.~G.~He, Y.~Y.~Keum and R.~R.~Volkas, JHEP {\bf 0604}, 039 (2006) [arXiv:hep-ph/0601001].

\bibitem{vacuum}
  G.~Altarelli and F.~Feruglio,
  Nucl.\ Phys.\  B {\bf 741}, 215 (2006)
  [arXiv:hep-ph/0512103];
  I.~de Medeiros Varzielas, S.~F.~King and G.~G.~Ross,
  Phys.\ Lett.\  B {\bf 644}, 153 (2007)
  [arXiv:hep-ph/0512313];
  G.~Altarelli, F.~Feruglio and Y.~Lin,
  Nucl.\ Phys.\  B {\bf 775}, 31 (2007)
  [arXiv:hep-ph/0610165].

\bibitem{Holthausen:2012dk}
  M.~Holthausen, M.~Lindner and M.~A.~Schmidt,
  JHEP {\bf 1304}, 122 (2013)  [arXiv:1211.6953 [hep-ph]].  

\bibitem{review}
  M.~Fukugita and T.~Yanagida, Phys.\ Lett.\  B {\bf 174}, 45 (1986);
  G.~F.~Giudice {\it et al.}, Nucl.\ Phys.\ B {\bf 685}, 89 (2004) [arXiv:hep-ph/0310123];
  W.~Buchmuller, P.~Di Bari and M.~Plumacher, Annals Phys.\  {\bf 315}, 305 (2005)  [arXiv:hep-ph/0401240];
  A.~Pilaftsis and T.~E.~J.~Underwood, Phys.\ Rev.\  D {\bf 72}, 113001 (2005)  [arXiv:hep-ph/0506107].

\bibitem{type1_seesaw}
  P.~Minkowski,
  Phys.\ Lett.\ B {\bf 67}, 421 (1977);
  T.~Yanagida,
  in {\it Workshop on Unified Theories}, KEK report 79-18 p.95 (1979);
  M.~Gell-Mann, P.~Ramond and R.~Slansky,
  in {\it Supergravity} (North Holland, Amsterdam, 1979)
  eds. P.~van~Nieuwenhuizen, D.~Freedman, p.315;
  S.~L.~Glashow,
  NATO Adv.\ Study Inst.\ Ser.\ B Phys.\  {\bf 59}, 687 (1980);
  R.~Barbieri, D.~V.~Nanopoulos, G.~Morchio and F.~Strocchi,
  Phys.\ Lett.\ B {\bf 90}, 91 (1980);
  R.~N.~Mohapatra and G.~Senjanovic,
  Phys.\ Rev.\ Lett.\  {\bf 44}, 912 (1980);
  G.~Lazarides, Q.~Shafi and C.~Wetterich,
  Nucl.\ Phys.\  B {\bf 181}, 287 (1981).


\bibitem{type3_seesaw}
  R.~Foot, H.~Lew, X.~G.~He and G.~C.~Joshi,
  Z.\ Phys.\  C {\bf 44}, 441 (1989).

\bibitem{Data}
 F.~P.~An {\it et al.}  [DAYA-BAY Collaboration],
  Phys.\ Rev.\ Lett.\  {\bf 108}, 171803 (2012)  [arXiv:1203.1669 [hep-ex]];  
  J.~K.~Ahn {\it et al.}  [RENO Collaboration],
  Phys.\ Rev.\ Lett.\  {\bf 108}, 191802 (2012)  [arXiv:1204.0626 [hep-ex]];  
 K.~Abe {\it et al.} [T2K Collaboration],
  Phys.\ Rev.\ Lett.\ \ {\bf 107}, 041801  (2011)  [arXiv:1106.2822 [hep-ex]];
  P.~Adamson {\it et al.} [MINOS Collaboration],
  Phys.\ Rev.\ Lett.\ \ {\bf 107}, 181802  (2011)  [arXiv:1108.0015 [hep-ex]];
 H.~De Kerret {\it et al.} [Double Chooz Collaboration], talk presented at the Sixth International Workshop on Low Energy Neutrino Physics, November 9-11, 2011 (Seoul, Korea).


\bibitem{Ahn:2010cc}
  Y.~H.~Ahn and C.~-S.~Chen,
  Phys.\ Rev.\ D {\bf 81}, 105013 (2010)  [arXiv:1001.2869 [hep-ph]];  
  Y.~H.~Ahn and H.~Okada,
  Phys.\ Rev.\ D {\bf 85}, 073010 (2012)  [arXiv:1201.4436 [hep-ph]].  


\bibitem{Ahn:2012cg}
  Y.~H.~Ahn, S.~Baek and P.~Gondolo,
  Phys.\ Rev.\ D {\bf 86}, 053004 (2012)
  [arXiv:1207.1229 [hep-ph]].



\bibitem{Antusch:2005gp}
  S.~Antusch, J.~Kersten, M.~Lindner, M.~Ratz and M.~A.~Schmidt,
  JHEP {\bf 0503}, 024 (2005)  [hep-ph/0501272].  

\bibitem{GonzalezGarcia:2012sz}
  M.~C.~Gonzalez-Garcia, M.~Maltoni, J.~Salvado and T.~Schwetz,
  arXiv:1209.3023 [hep-ph].  

\bibitem{Branco:2002xf}
  G.~C.~Branco, R.~Gonzalez Felipe, F.~R.~Joaquim, I.~Masina, M.~N.~Rebelo and C.~A.~Savoy,
  Phys.\ Rev.\  D {\bf 67}, 073025 (2003)
  [arXiv:hep-ph/0211001].

\bibitem{Jarlskog:1985ht}
  C.~Jarlskog,
  Phys.\ Rev.\ Lett.\  {\bf 55}, 1039 (1985);
  D.~d.~Wu,
  Phys.\ Rev.\  D {\bf 33}, 860 (1986).


\bibitem{Fukugita:1986hr}
  M.~Fukugita and T.~Yanagida,
  Phys.\ Lett.\  B {\bf 174}, 45 (1986).

\bibitem{A4lep}
D.~Aristizabal Sierra, F.~Bazzocchi, I.~de Medeiros Varzielas, L.~Merlo, S.~Morisi and, Nucl.\ Phys.\ B {\bf 827}, 34 (2010) [arXiv:0908.0907 [hep-ph]];
F.~Riva, Phys.\ Lett.\ B {\bf 690} (2010) 443
[arXiv:1004.1177 [hep-ph]];
D.~Aristizabal Sierra, I.~de Medeiros Varzielas, arXiv:1205.6134 [hep-ph].
  I.~K.~Cooper, S.~F.~King and C.~Luhn,
   Nucl.\ Phys.\ B {\bf 859}, 159 (2012)  [arXiv:1110.5676 [hep-ph]].

\bibitem{Pilaftsis:1997jf}
  A.~Pilaftsis,
  Phys.\ Rev.\ D {\bf 56}, 5431 (1997)  [hep-ph/9707235].  


\bibitem{Gu:2008yk}
  P.~-H.~Gu and U.~Sarkar,
  Mod.\ Phys.\ Lett.\ A {\bf 25}, 501 (2010)  [arXiv:0811.0956 [hep-ph]].  

\bibitem{lepto2}
 L. Covi, E. Roulet and F. Vissani, Phys. Lett. {\bf B384}, (1996) 169; A.~Pilaftsis, Int.\ J.\ Mod.\ Phys.\ A {\bf 14}, (1999) 1811 [arXiv:hep-ph/9812256].

\bibitem{Abada}
  A.~Abada, S.~Davidson, F.~X.~Josse-Michaux, M.~Losada and A.~Riotto,  JCAP {\bf 0604},  (2006) 004 [arXiv:hep-ph/0601083];
   S.~Antusch, S.~F.~King and A.~Riotto,  JCAP {\bf 0611}, (2006)  011  [arXiv:hep-ph/0609038].


\bibitem{Altarelli:2005yp}
E. Ma and G. Rajasekaran, Phys. Rev. D 64 (2001) 113012 ;
K. S. Babu, E. Ma and J. W. F. Valle, Phys. Lett. B 552 (2003) 207;
M. Hirsch, J. C. Romao, S. Skadhauge, J. W. F. Valle and A. Villanova del Moral,
arXiv:hep-ph/0312244; M. Hirsch, J. C. Romao, S. Skadhauge, J. W. F. Valle and
A. Villanova del Moral, Phys. Rev. D69 (2004) 093006; E. Ma, Phys. Rev. D 70 (2004) 031901; E. Ma arXiv:hep-ph/0409075; E. Ma, New
J. Phys. 6 (2004) 104;
  G.~Altarelli and F.~Feruglio,
  Nucl.\ Phys.\  B {\bf 720}, 64 (2005);
 S.~Baek and M.~C.~Oh,
Phys.\ Lett.\ B {\bf 690}, 29 (2010)  [arXiv:0812.2704 [hep-ph]];  




\end{thebibliography}
\end{document}